%% file: ws-ijmpc.tex
\documentclass[a4paper,10pt,twoside,onecolumn,final]{ws-ijmpc} %\documentclass{ws-ijmpc}
\usepackage[super]{cite} % other options `nocompress`, `nosort`

\input{defs.tex}
\begin{document}

\markboth{W.A. Ahmad \& M. Sutcliffe}
{Dynamic T-decomposition for classical simulation of quantum circuits}

%%%%%%%%%%%%%%%%%%%%% Publisher's Area please ignore %%%%%%%%%%%%%%%
%\catchline{}{}{}{}{}
%%%%%%%%%%%%%%%%%%%%%%%%%%%%%%%%%%%%%%%%%%%%%%%%%%%%%%%%%%%%%%%%%%%%

\title{Dynamic T-decomposition for classical simulation of quantum circuits}

\author{Wira Azmoon Ahmad\footnote{Permanent address: wira.azmoon@u.nus.edu}}

\address{Department of Computer Science, University of Oxford, 7 Parks Road\\
Oxford, Oxfordshire, OX1 3QG, United Kingdom.\\
wira.ahmad@cs.ox.ac.uk}

\author{Matthew Sutcliffe\footnote{Permanent address: mjsutcliffe1999@gmail.com}}

\address{Department of Computer Science, University of Oxford, 7 Parks Road\\
Oxford, Oxfordshire, OX1 3QG, United Kingdom.\\
matthew.sutcliffe@cs.ox.ac.uk}

\maketitle

%\begin{history}
%\received{Day Month Year}
%\revised{Day Month Year}
%\end{history}
\vspace{1cm} %TEMP

\begin{abstract}
It is known that a quantum circuit may be simulated with classical hardware via stabilizer state (T-)decomposition in $O(2^{\alpha t})$ time, given $t$ non-Clifford gates and a decomposition efficiency $\alpha$. The past years have seen a number of papers presenting new decompositions of lower $\alpha$ to reduce this runtime and enable simulation of ever larger circuits. More recently, it has been demonstrated that well placed applications of apparently weaker (higher $\alpha$) decompositions can in fact result in better overall efficiency when paired with the circuit simplification strategies of ZX-calculus.

In this work, we take the most generalized T-decomposition (namely vertex cutting), which achieves a poor efficiency of $\alpha=1$, and identify common structures to which applying this can, after simplification via ZX-calculus rewriting, yield very strong effective efficiencies $\alpha_{\text{eff}}\ll1$. By taking into account this broader scope of the ZX-diagram and incorporating the simplification facilitated by the well-motivated cuts, we derive a handful of efficient T-decompositions whose applicabilities are relatively frequent. In benchmarking these new `\textit{dynamic}' decompositions against the existing alternatives, we observe a significant reduction in overall $\alpha$ and hence overall runtime for classical simulation, particularly for certain common circuit classes.

\end{abstract}

\ccode{PACS Nos.: 03.67.Lx, 03.67.Ac, 89.70.Eg}

\vspace{1cm} %TEMP

\section{Introduction}

Simulating quantum circuits with classical hardware is necessarily an inefficient endeavor. Without the quantum advantage, the runtime complexity of this task grows exponentially with the size or complexity of the circuit. For instance, via a T-decomposition approach \cite{Bravyi-2016bss}, the runtime grows as $O(2^{\alpha t})$, given an initial circuit with $t$ non-Clifford gates and a T-decomposition of efficiency $\alpha$. Nevertheless, in the present and near-term, where quantum hardware is very limited by size and noise \cite{Nielsen-Chuang-2010}, classical simulation is vital for verifying the behavior of quantum algorithms and quantum computers.

The graphical language of ZX-calculus was initially applied to the problem of classical simulation, in a meaningful capacity, with the work of Kissinger and van de Wetering \cite{Kissinger-2022}. This work demonstrated how the problem (of both strong and weak simulation) could be formulated in terms of ZX-calculus and benefit, quite dramatically, from its rewriting strategies.

Subsequent research \cite{Kissinger2022cat} extended this work by introducing more efficient decompositions that reduce the overall computational runtime of solving the problem. This enabled even larger and more complex quantum circuits to be classically simulated. As well as advancing the state of the art for a generally applicable decomposition, this work also --- along with a handful of other publications \cite{Codsi2022,Laakkonen2022} --- offered decompositions which depend upon more specific structures but which, when applicable, are even more efficient.

With a few exceptions \cite{sutcliffe2024-paramzx,sutcliffe2024part}, most related research has followed the implicit assumption that the means to improve overall efficiency is to discover new decompositions of lower $\alpha$. Recently, however, this assumption has been challenged, with new work (implicitly alluded to in Codsi \cite{Codsi2022} and then more formally addressed in Sutcliffe and Kissinger \cite{sutcliffe2024proccut}) demonstrating how apparently weaker (higher $\alpha$) decompositions can actually perform better overall when the broader neighborhood of the circuit, along with ZX-calculus rewriting, is taken into account. A recent proof of concept for this type of `\textit{dynamic decomposition}' approach \cite{sutcliffe2024proccut} showed a specific example of a common structure (namely `\textit{CNOT sandwiches}') where this reasoning applies, with a weighting heuristic to optimize for it.

This is the context into which this present paper originates. In this paper, we outline a family of heuristics for deciding when and where to apply the general `\textit{vertex cutting}' decomposition to maximize the ZX-simplification it facilitates. Ultimately, this can be expressed as a set of T-decompositions which are dynamic and map (scalably) to patterns that are inherently common to quantum circuits (as ZX-diagrams after initial simplification).

Finally, bringing these new dynamic T-decompositions together with a meta-heuristic for deciding which to apply at each step, we benchmark our approach against the preceding ZX-based method and show that, for certain classes of circuit especially, this can reduce the computational runtime by orders of magnitude, allowing significantly larger and more complex quantum circuits to be simulated classically.

\section{Background}

\subsection{The ZX-calculus}

Popularly, quantum circuits are expressed in circuit notation \cite{Nielsen-Chuang-2010}, composed of quantum logic gates. However, for reasoning with such circuits, the powerful alternative notation of \textit{ZX-calculus} \cite{Coecke-2011,coecke2017picturing,weteringWorking,KissingerWetering2024Book} has proven extremely useful. This provides a graphical language with which to describe and manipulate quantum circuits and has found uses in many related areas such as circuit optimization \cite{KissingerTcount,Duncan2020graphtheoretic,Kissinger2020}, compilation \cite{cowtan2019phase,khesin2023graphical}, and --- most relevant to this paper --- classical simulation \cite{Kissinger-2022,Kissinger2022cat,koch2023,sutcliffe2024proccut}.

In brief terms, any quantum circuit may be expressed as a ZX-diagram, being a linear map composed of green Z-spiders and red X-spiders, each with an associated phase $[0,2\pi)$, and both normal and Hadamard edges:

\begin{figure}[h!]
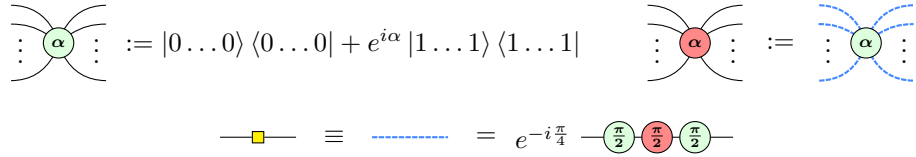

\ctikzfig{figures/introZX}
\caption{The basic components of a ZX-diagram, namely Z- and X-spiders, as well as Hadamard edges (being a yellow box or a blue dashed edge) and normal edges.}
\label{fig:introZX}
\end{figure}

\begin{figure}[h!]
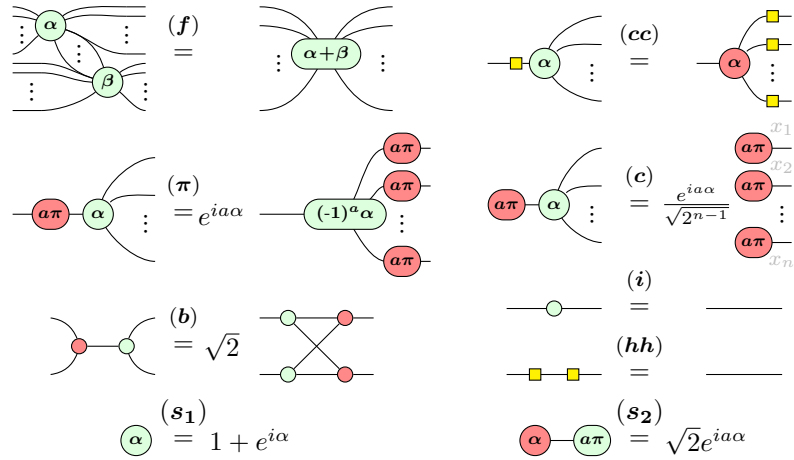

\ctikzfig{figures/ZX-rules}
\caption{The basic rewriting rules \cite{weteringWorking} of ZX-calculus, where $a\in\{0,1\}$ and $\alpha,\beta\in[0,2\pi)$. These rules remain valid with all colors inverted.}
\label{fig:zxrules}
\end{figure}

Such ZX-diagrams may then be simplified via the rewriting rules, as outlined in Figure \ref{fig:zxrules}, which serve to reduce the number of spiders (i.e. gates) and offer a simpler equivalent construction of the same circuit.

\subsection{Classical simulation of quantum circuits}

Given the limitations of today's noisy intermediate-scale quantum (NISQ) hardware \cite{Nielsen-Chuang-2010}, verifying the behavior of quantum algorithms can be a tricky task. The ability to simulate such algorithms with classical hardware, therefore, is of great utility. In particular, classical simulation of quantum circuits may refer to one of two things \cite{Kissinger-2022}:
\begin{itemize}
    \item To \textbf{strongly} classically simulate a quantum circuit is to determine (possibly up to some allowed error margin) the probability of a particular measurement outcome.
    \item To \textbf{weakly} classically simulate is to probabilistically sample a circuit's output distribution.
\end{itemize}

Weak simulation methods largely rely upon applications of strong simulation, hence finding more efficient techniques to achieve the latter may in turn improve the efficiency of the former. Given this, it is to the task of strong simulation that this paper is focused.

Utilizing ZX-calculus to this end \cite{Kissinger-2022}, an $n$-qubit quantum circuit may be expressed as a ZX-diagram with its inputs and outputs closed off according to the desired initial state, $\ket{0}^{\otimes n}$, and measurement bitstring, $\bra{x_1,x_2,\ldots,x_n}$ where $x_i\in\{0,1\}\forall i$:

%\begin{figure}[h!]
    \[\input{amplitude.tikz} \;\in \C\]
%    \caption{A scalar circuit involving Clifford+T diagram $V$.}
%    \label{fig:amplitude}
%\end{figure}

Quantum circuits restricted to the Clifford \cite{weteringWorking} gateset manifest as ZX-diagrams with phases limited to $\frac{m\pi}{2}$, where $m\in\{0,1,2,3\}$. Such circuits are known to be efficiently classically simulable \cite{gottesman1998}, with the outcome measurement probability, $P(x_1,x_2,\ldots,x_n)$, easily determined by reducing the diagram to a scalar, $A$, via the rewriting rules of Figure \ref{fig:zxrules}, where $P(x_1,x_2,\ldots,x_n)=|A|^2$.

\begin{figure}[h!]
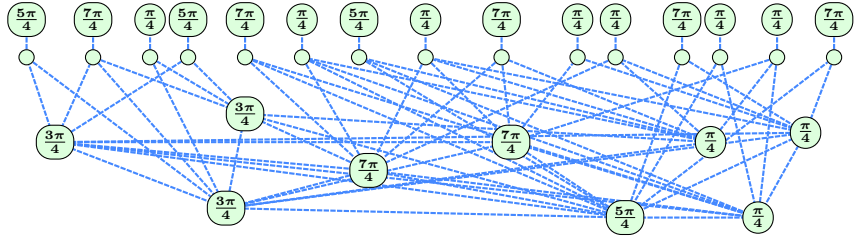

\ctikzfig{figures/example-scalar}
\caption{A ZX-diagram in reduced gadget form \cite{Kissinger-2022}.}
\label{fig:example-scalar}
\end{figure}

Conversely, classically simulating circuits of the broader Clifford+T gateset 
(required for approximate universality \cite{Nielsen-Chuang-2010}) is notoriously inefficient. 
Such circuits, expressed as ZX-diagrams, allow phases of $\frac{m\pi}{4}$, where $m\in\{0,1,\ldots,7\}$. 
The rewriting rules alone are generally insufficient to fully reduce these diagrams to scalars, 
rather simplifying them to their \textit{reduced gadget form} akin to Figure \ref{fig:example-scalar}.

From here, \textit{stabilizer state decompositions} (or \textit{T-decompositions}) may be employed to replace the reduced form Clifford+T ZX-diagram for a sum of simulable Clifford ZX-diagrams \cite{Kissinger-2022}. The simplest and most trivial such decomposition is the \textit{vertex cutting} decomposition \cite{Codsi2022}:

\begin{equation}
    \input{figures/cut.tikz}
    \label{fig:cut}
\end{equation}

Applying this to each of the non-Clifford `\textit{T-spiders}' (spiders of phase $\frac{m\pi}{4}$ for odd $m$) allows the overall ZX-diagram to be exchanged for a sum of $2^{t}$ efficiently simulable Clifford terms. Due to this exponential relationship, circuits with larger T-count (number of T-spiders) are significantly slower to simulate classically. Hence, there has been much research into finding more efficient decompositions, with the current state of the art for a generally applicable decomposition being that of Equation \ref{eq:magic-5-decomp} in \ref{app:cats} \cite{Kissinger2022cat}.

Each application of this decomposition effectively trades $4$ T-spiders for a sum of $3$ locally Clifford terms, hence it could be used to translate a $t$ T-count circuit into a sum of $3^{t/4}\approx2^{0.396t}$ terms. Thus, for a general $2^{\alpha t}$ terms, this decomposition achieves an efficiency of $\alpha\approx0.396$.

Other known decompositions are able to achieve a lower $\alpha$ but are only applicable provided specific structures, such as the `\textit{cats}' family of decompositions \cite{Kissinger2022cat}, outlined in \ref{app:cats}, which can achieve efficiencies as low as $\alpha=0.25$. However, the $\alpha$ values achieved by these non-universal decompositions do not imply an equivalent \textit{overall} efficiency, as it is not generally possible to remove all the T-spiders using only these efficient, but structure-specific, decompositions. Indeed, the overall effective efficiency of a collective decomposition strategy is measured, for a particular circuit, as:

\begin{equation}
    \alpha := \frac{\log_2{n}}{t}
\end{equation}

where $n$ is the final number of Clifford terms and $t$ is the initial T-count (after initial Clifford simplification).

Lastly, it should be noted that after each application of a decomposition, an extra round of Clifford simplification (via the rewriting rules of Figure \ref{fig:zxrules}) may subsequently lead to a further reduction in T-count, making the \textit{effective} efficiency $\alpha_{\text{eff}}$ of a particular decomposition potentially much lower than its \textit{apparent} efficiency: $\alpha_{\text{eff}}\leq\alpha$. This all depends on when and where the decomposition is applied and how much simplification it may facilitate. Generally, however, this is not taken into account as checking the $\alpha_{\text{eff}}$ for every available decomposition, for every applicable subgraph, at every step is computationally infeasible. Consequently, most decomposition strategies, at each step, simply apply the applicable decomposition with the lowest $\alpha$.

\section{Methods}

The main limitation we recognize about the existing literature is this assumption that, at each step, selecting the applicable decomposition with the lowest $\alpha$ is the optimal decision. In fact, as emphasized in the work of Sutcliffe and Kissinger \cite{sutcliffe2024proccut}, it is possible that selecting weaker (higher $\alpha$) decompositions can achieve the strongest effective efficiency (lowest $\alpha_{\text{eff}}$) by leading to more simplification.

Determining the \textit{most} optimal decomposition (and where to apply it) at each step would be \#P-hard \cite{movassagh2023hardness}. Nevertheless, using heuristics to motivate the choice at each step can lead to better results than that achieved by the assumption above. Sutcliffe and Kissinger \cite{sutcliffe2024proccut} provided a proof of concept of this, and more recently experimental work on applying reinforcement learning here has yielded promising results \cite{koziell-pipe2024}.

\subsection{Dynamic decompositions}
\label{sec:decomps}

In analyzing the structure of reduced form Clifford+T ZX-diagrams (such as that of Figure \ref{fig:example-scalar}), we identify a number of common patterns to which a simple well-placed cutting decomposition (Equation \ref{fig:cut}) can result in significant subsequent simplification via the rewriting rules. These patterns, as well as being common to such diagrams, are flexible in that we allow for variability and scalability, leading to structure-specific decompositions that are nevertheless very dynamic. These are as follows:

\begin{decom}[Lone phase decomposition]
    \label{decom:lonephase}
    The following decomposition:
    \begin{equation}
        \label{eq:lone-phase}
        \input{lone-phase.tikz}
    \end{equation}
    is valid and achieves:
    \begin{equation}
        \alpha=\frac{1}{n}
    \end{equation}
\end{decom}

\hfill
\hfill

\begin{decom}[Multi-$\ket{\text{cat}_\text{3}}$ decomposition \cite{Codsi2022}]
    \label{decom:multicat3}
    The following decomposition:
    \begin{equation}
        \label{eq:cat-3-cut}
        \input{multi-cat3.tikz}
    \end{equation}
    is valid and achieves:
    \begin{equation}
        \alpha=\frac{1}{2n+1}
    \end{equation}
\end{decom}

\hfill
\hfill

\begin{decom}
    \label{decom:d3}
    The following decomposition:
    \begin{equation}
        \label{eq:pair-decomp}
        \input{pair-decomp.tikz}
    \end{equation}
    is valid and achieves:
    \begin{equation}
        \alpha=\frac{1}{2n+2}
    \end{equation}
\end{decom}

\hfill
\hfill

\begin{decom}
    \label{decom:d4}
    The following decomposition:
    \begin{equation}
        \label{eq:pair-phase-decomp}
        \input{pair-phase-decomp.tikz}
    \end{equation}
    is valid and achieves:
    \begin{equation}
        \alpha=\frac{1}{n+2}
    \end{equation}
\end{decom}

In each case, the derivation (as well as variations and broader generalizations) may be found in \ref{app:derivs}.

Combining these with the baseline decompositions of Kissinger et al \cite{Kissinger2022cat} (see \ref{app:cats}), we can outline a decomposition strategy, expressed in Algorithm \ref{alg:greedy-cut}, to decide which decomposition to use at each step.

\begin{algorithm}
\caption{Greedy Algorithm with Heuristic}
\label{alg:greedy-cut}
\begin{algorithmic}[1]
\State \textbf{Input:} $\alpha$ values for $\catstate{4}$, $\catstate{6}$, $\catstate{5}$, $\catstate{3}$, 
and $\cbox{\magicspider}^{\otimes 5}$, and the graph $G$
\State \textbf{Output:} Chosen state and corresponding decomposition

\State Let $\alpha_{4} = 0.25, \alpha_{6} = 0.264,\alpha_{5} = 0.317, \alpha_{3} = 0.333, 
\alpha_{\otimes 5} = 0.396$
\State Compute the best catstate $\catstate{\text{best}}$ in $G$ with the lowest $\alpha$ value
\State Let $\alpha_{\text{best}}$ be the $\alpha$ value of $\catstate{\text{best}}$, 
or of $\cbox{\magicspider}^{\otimes 5}$ if no $\catstate{\text{best}}$ is found

\State Compute best available $\alpha_h$ on $G$ among our new decompositions

\If {$\alpha_h < \alpha_{\text{best}}$}
    \State Use the decomposition corresponding to $\alpha_h$
\Else
    \State Use the decomposition corresponding to $\catstate{\text{best}}$ or $\cbox{\magicspider}^{\otimes 5}$
\EndIf

\end{algorithmic}
\end{algorithm}

\section{Results}
\label{sec:results}

We benchmarked our method on randomly generated circuits of four distinct classes:
\begin{itemize}
    \item Pauli exponentials circuits
    \item Controlled-CZ (CCZ) circuits
    \item Modified hidden shift circuits
    \item Instantaneous Quantum Polynomial (IQP) circuits
\end{itemize}

Details of these circuit classes, and the parameters used for their generation, can be found in \ref{app:classes}.

\begin{figure}[ht]
    \centering
    % Top subplot spanning the width
    \begin{subfigure}[t]{\textwidth}
        \centering
        \includegraphics[width=\textwidth]{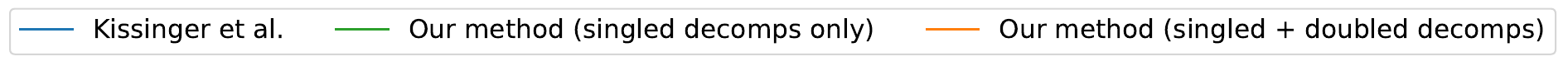}
    \end{subfigure}
    
    \vspace{0.2cm} % Adjust vertical space between rows
    
    % Side-by-side subfigures below the top
    \begin{subfigure}[b]{0.48\textwidth} % Adjust width as needed
        \centering
        \includegraphics[width=\textwidth]{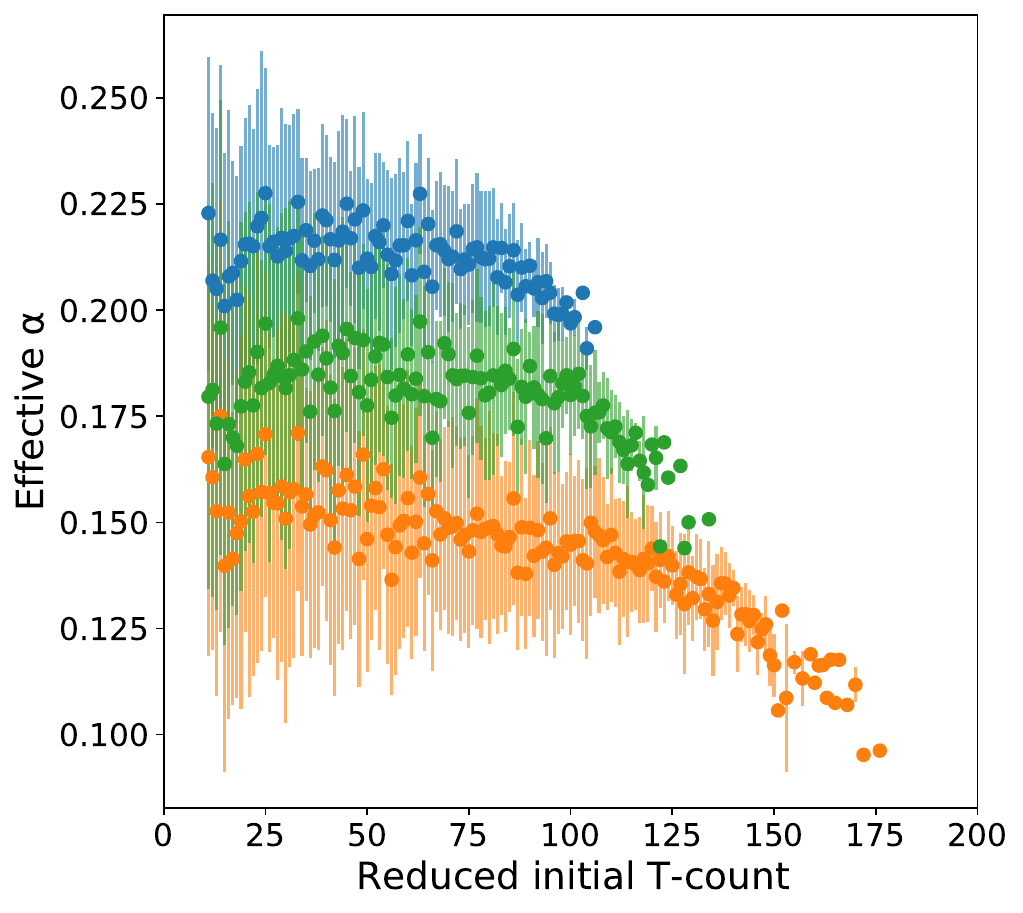}
        \caption{Overall $\alpha_{\text{eff}}$ versus T-count.}
        \label{fig:ccz:alpha}
    \end{subfigure}
    \hfill % Optional space between subfigures
    \begin{subfigure}[b]{0.48\textwidth}
        \centering
        \includegraphics[width=\textwidth]{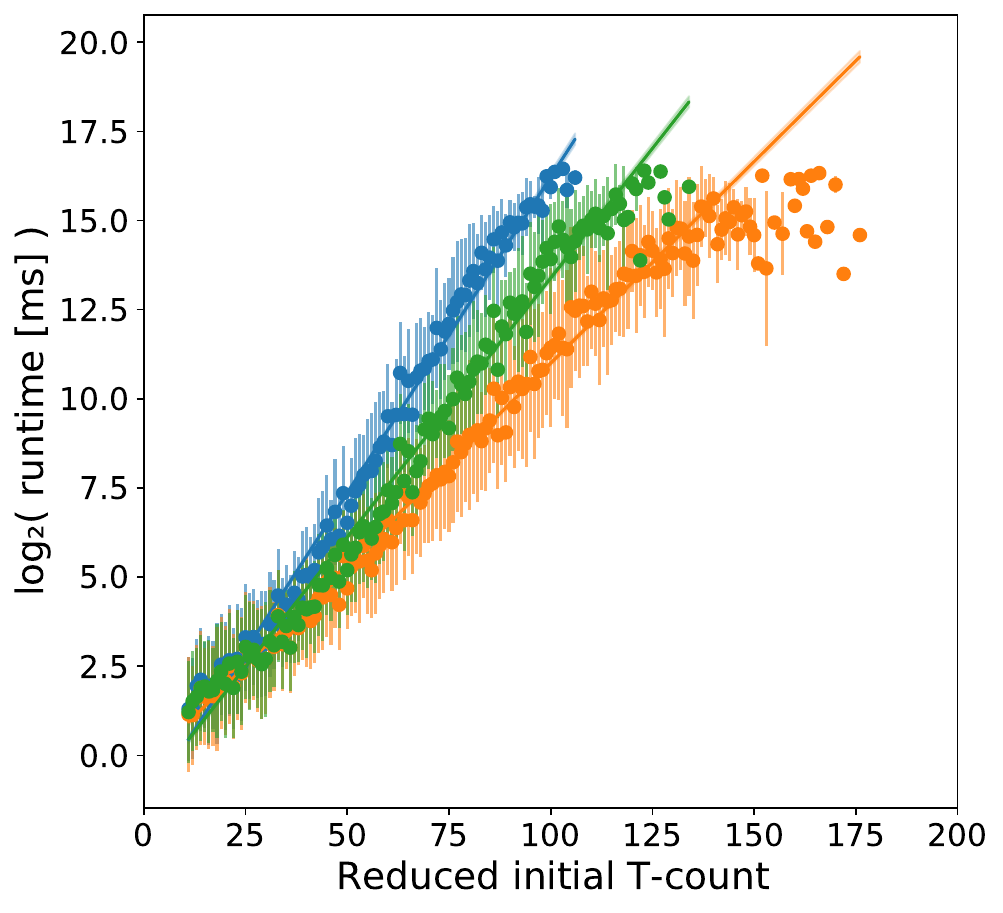}
        \caption{The ($\log_2$) runtimes versus T-count.}
        \label{fig:ccz:time}
    \end{subfigure}

    %\captionsetup{justification=centering}
    \caption{The measured results for classically simulating \textbf{CCZ} circuits, versus the method of Kissinger et al. \cite{Kissinger2022cat}}
    \label{fig:ccz}
\end{figure}

\begin{figure}[ht]
    \centering
    % Top subplot spanning the width
    \begin{subfigure}[t]{\textwidth}
        \centering
        \includegraphics[width=\textwidth]{figures/legend.pdf}
    \end{subfigure}
    
    \vspace{0.2cm} % Adjust vertical space between rows
    
    % Side-by-side subfigures below the top
    \begin{subfigure}[b]{0.48\textwidth} % Adjust width as needed
        \centering
        \includegraphics[width=\textwidth]{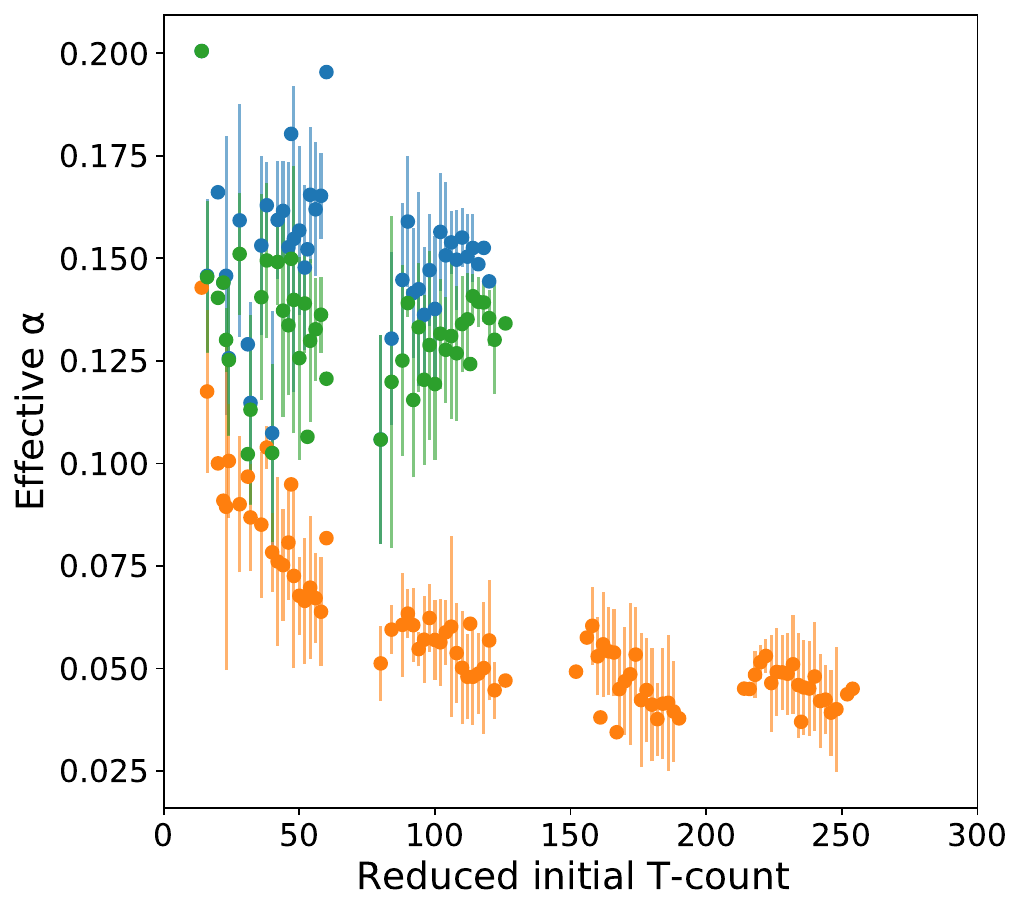}
        \caption{Overall $\alpha_{\text{eff}}$ versus T-count.}
        \label{fig:hs:alpha}
    \end{subfigure}
    \hfill % Optional space between subfigures
    \begin{subfigure}[b]{0.48\textwidth}
        \centering
        \includegraphics[width=\textwidth]{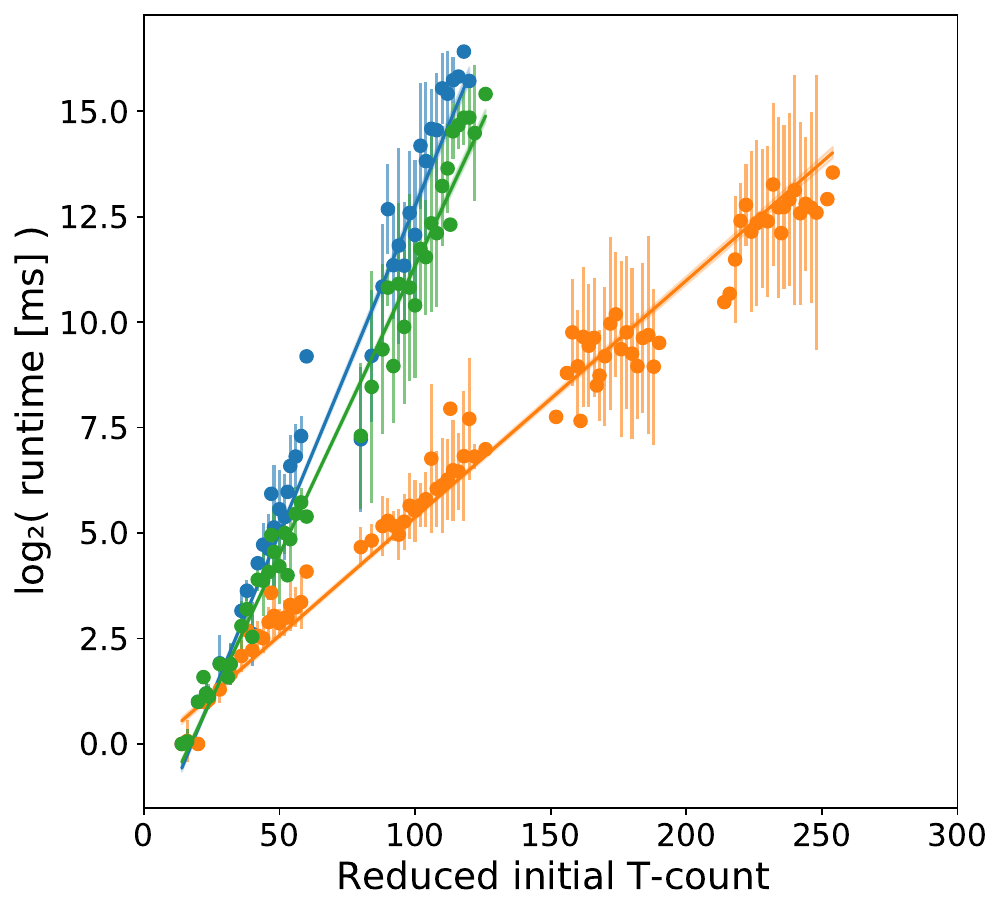}
        \caption{The ($\log_2$) runtimes versus T-count.}
        \label{fig:hs:time}
    \end{subfigure}
    
    %\captionsetup{justification=centering}
    \caption{The measured results for classically simulating \textbf{modified hidden shift} circuits, versus the method of Kissinger et al. \cite{Kissinger2022cat}}
    \label{fig:hs}
\end{figure}

\begin{figure}[ht]
    \centering
    % Top subplot spanning the width
    \begin{subfigure}[t]{\textwidth}
        \centering
        \includegraphics[width=\textwidth]{figures/legend.pdf}
    \end{subfigure}
    
    \vspace{0.2cm} % Adjust vertical space between rows
    
    % Side-by-side subfigures below the top
    \begin{subfigure}[b]{0.48\textwidth} % Adjust width as needed
        \centering
        \includegraphics[width=\textwidth]{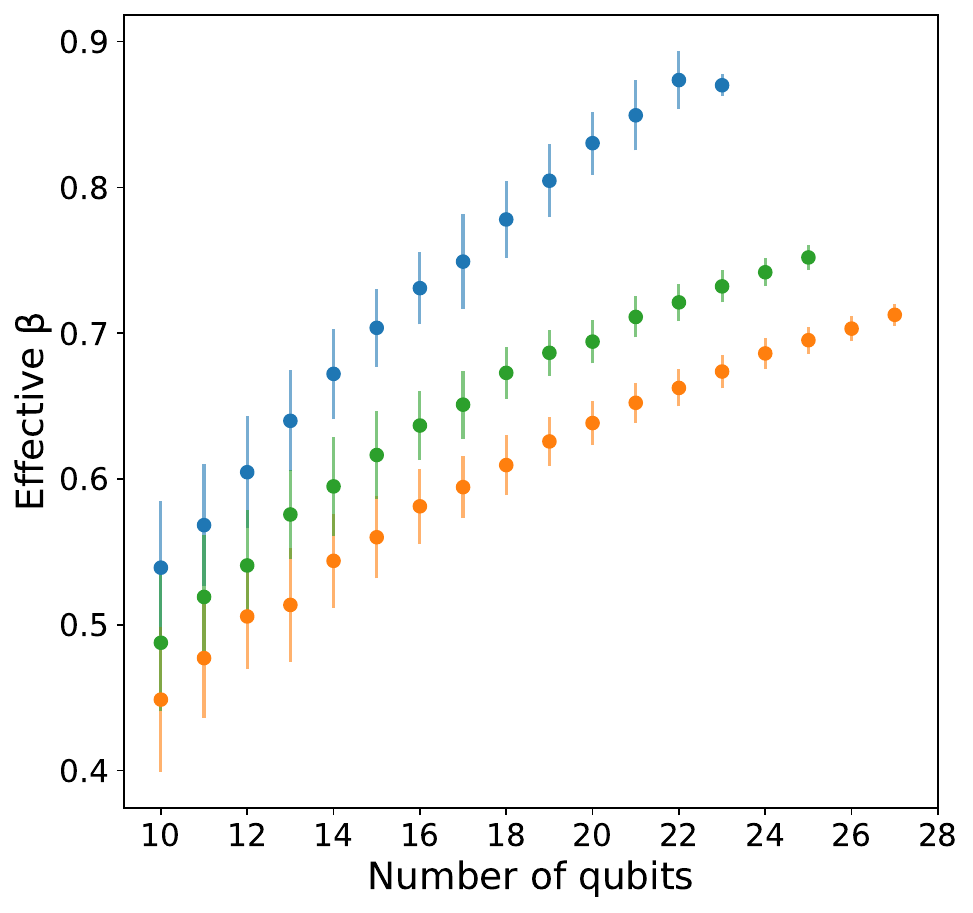}
        \caption{Overall $\beta_{\text{eff}}$ versus qubit count.}
        \label{fig:iqp:beta}
    \end{subfigure}
    \hfill % Optional space between subfigures
    \begin{subfigure}[b]{0.48\textwidth}
        \centering
        \includegraphics[width=\textwidth]{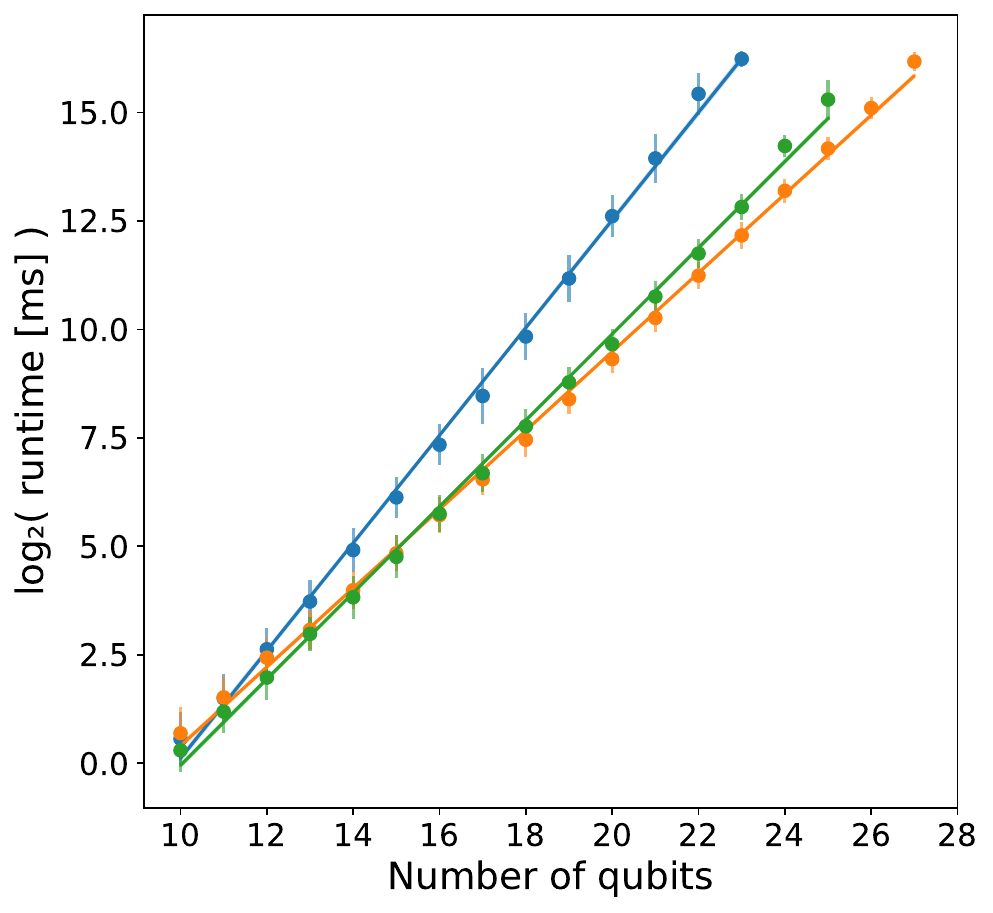}
        \caption{The ($\log_2$) runtimes versus qubit count.}
        \label{fig:iqp:time}
    \end{subfigure}
    
    %\captionsetup{justification=centering}
    \caption{The measured results for classically simulating \textbf{IQP} circuits, versus the method of Kissinger et al. \cite{Kissinger2022cat}}
    \label{fig:iqp}
\end{figure}

\begin{figure}[ht]
    \centering
    % Top subplot spanning the width
    \begin{subfigure}[t]{\textwidth}
        \centering
        \includegraphics[width=\textwidth]{figures/legend.pdf}
    \end{subfigure}
    
    \vspace{0.2cm} % Adjust vertical space between rows
    
    % Side-by-side subfigures below the top
    \begin{subfigure}[b]{0.48\textwidth} % Adjust width as needed
        \centering
        \includegraphics[width=\textwidth]{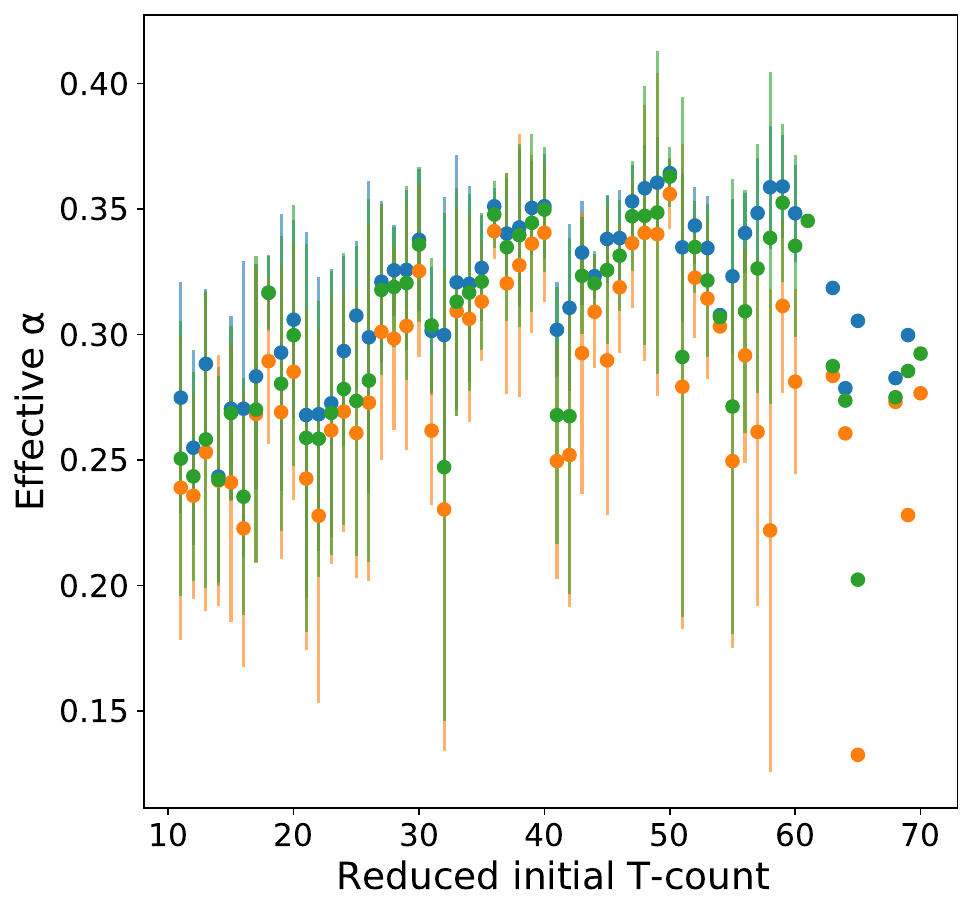}
        \caption{Overall $\alpha_{\text{eff}}$ versus T-count.}
        \label{fig:pauli:alpha}
    \end{subfigure}
    \hfill % Optional space between subfigures
    \begin{subfigure}[b]{0.48\textwidth}
        \centering
        \includegraphics[width=\textwidth]{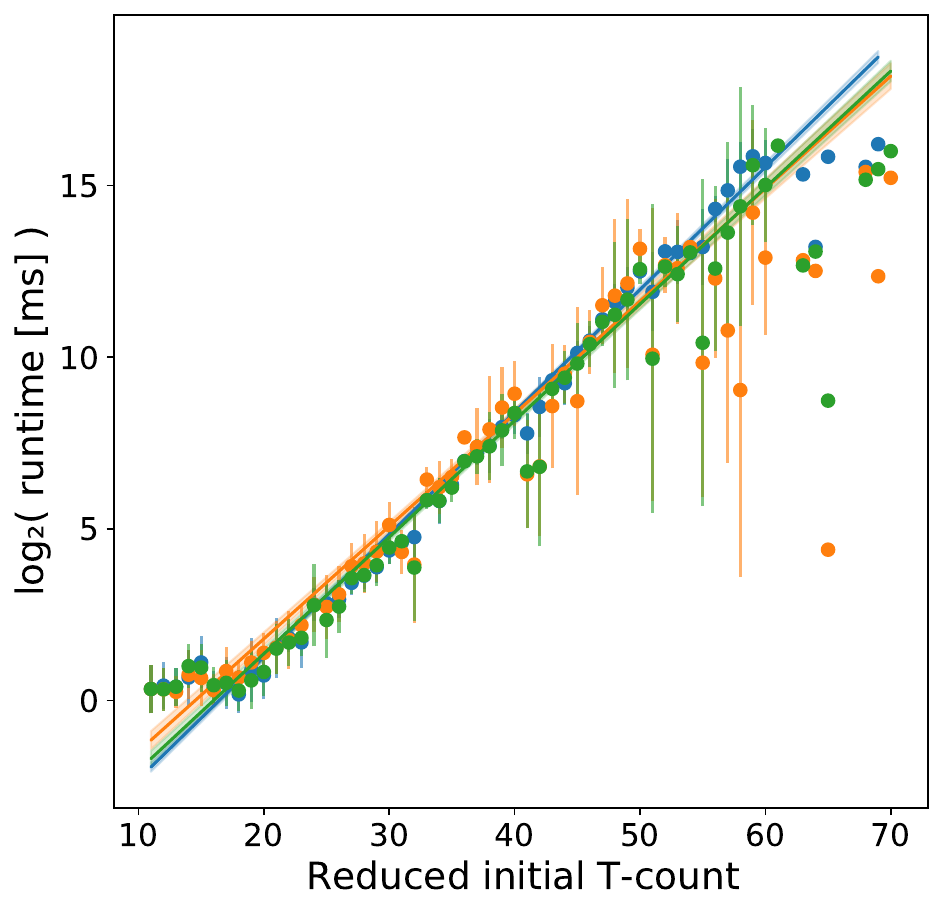}
        \caption{The ($\log_2$) runtimes versus T-count.}
        \label{fig:pauli:time}
    \end{subfigure}
    
    %\captionsetup{justification=centering}
    \caption{The measured results for classically simulating \textbf{random Pauli} circuits, versus the method of Kissinger et al. \cite{Kissinger2022cat}}
    \label{fig:pauli}
\end{figure}

The measured results are displayed in Figures \ref{fig:ccz} to \ref{fig:pauli}, showing how both the overall effective decomposition efficiency and the corresponding runtime of strong simulation varies with the initial T-count using this method for each circuit class. These results are grouped into the `\textit{singled}' cases (Decompositions \ref{decom:lonephase} and \ref{decom:multicat3}) and the `\textit{doubled}' cases (Decompositions \ref{decom:d3} and \ref{decom:d4}) and are compared against the method of Kissinger et al. \cite{Kissinger2022cat}

\vspace{2cm} %temp?

Evidently, for CCZ circuits, employing the singled decompositions resulted in a noticeable improvement to the efficiency, with the inclusion of the doubled decompositions providing a comparable additional improvement. Modified hidden shift circuits, meanwhile, were improved only marginally by the singled decompositions, but quite substantially given the doubled decompositions. Conversely, the singled decompositions offered a significant improvement to IQP circuits, with the doubled decompositions making only a minor difference. Lastly, on fully random Pauli exponential circuits, neither set of decompositions resulted in meaningful improvements versus the existing methods. (Note that the trailing off visible in Figure \ref{fig:ccz} and the gaps in Figure \ref{fig:hs} are artifacts of the parameters used for the random circuit generation.)

These results show that the decompositions presented in Section \ref{sec:decomps} are each more effective on certain classes of circuits than others. This makes sense as they rely upon specific structures which may be more common in some classes. Altogether, the inclusion of these decompositions led to very significant improvements for three of the four circuit classes (and arguably the more `real-world' classes), allowing such circuits to be classically simulated more rapidly, in turn enabling ever larger circuits to be simulated within feasible timeframes.

More thorough benchmarking and analysis of these decompositions can be found in the master's thesis of the first author \cite{ahmad2024}, with a review of the alternative ZX-based methods soon to be available in the PhD thesis \cite{sutcliffePhd} and review paper \cite{sutcliffeReview} of the second author.

\section{Conclusions}

In this work, we identified common circuit structures and patterns to which well-placed vertex cuts lead to significant reductions in T-count after ZX-calculus simplification along each branch. We formalized these patterns as dynamically structured T-decompositions with efficiency $\alpha$ calculated as a function of the scale of the pattern. Lastly, we benchmarked these decompositions against the current state of the art \cite{Kissinger2022cat} and demonstrated significant reductions in runtime for the task of classical simulation of quantum circuits, particularly for three common circuit classes.

\section*{Acknowledgments}
The authors would like to thank Professor Aleks Kissinger for his supervision and contribution to discussions on this work.

\appendix

\section{Derived ZX-Rules}

\begin{figure}[h!]
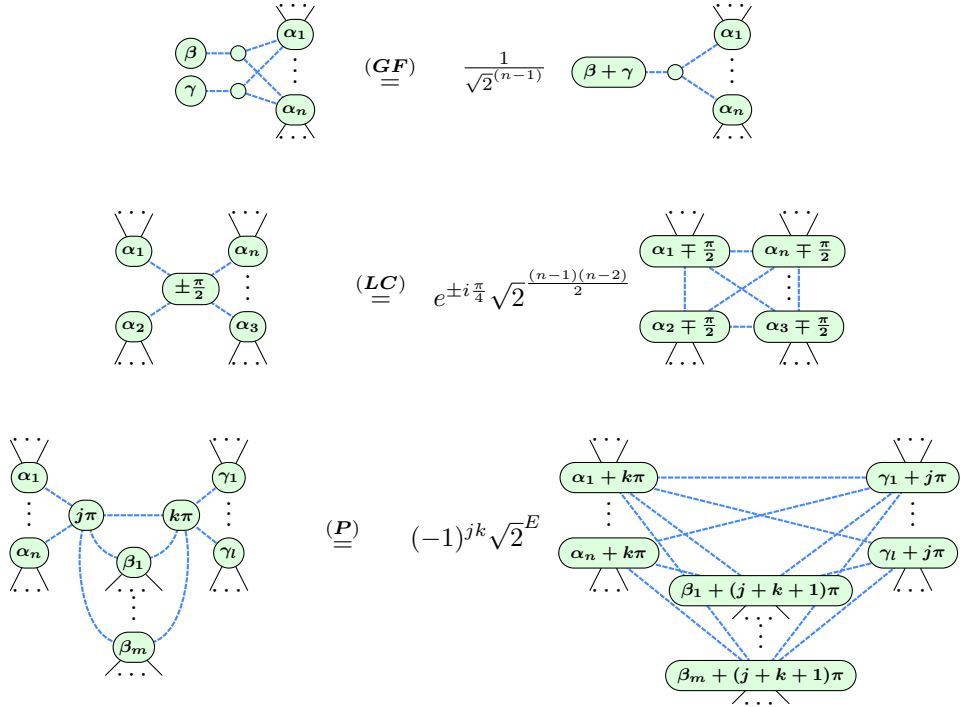

\ctikzfig{figures/deriv-rules}
\caption{Three useful rewriting rules \cite{Kissinger-2022}, derivable from the basic set of Figure \ref{fig:zxrules}, where $\alpha,\beta,\gamma\in[0,2\pi)$, $j,k\in\{0,1\}$, and $E=(n-1)m+(l-1)m+(n-1)(l-1)$.}
\label{fig:derivrules}
\end{figure}

\section{Kissinger et al. Decompositions}
\label{app:cats}

The ZX-calculus based classical simulation approach against which we compare is the method of Kissinger et al. \cite{Kissinger2022cat}, which includes the following `cat' decompositions:

\begin{equation}
    \input{cats.tikz}
\end{equation}

plus the state of the art for a generally applicable decomposition:

\begin{equation}
    \label{eq:magic-5-decomp}
    \input{magic-state-5-decomp.tikz}
\end{equation}

Respectively, these decompositions, from top to bottom, achieve:

\begin{itemize}
    \item $\alpha_{\ket{\text{cat}_\text{3}}} = 1/3$
    \item $\alpha_{\ket{\text{cat}_\text{4}}} = 1/2$
    \item $\alpha_{\ket{\text{cat}_\text{5}}} \approx 0.317$
    \item $\alpha_{\ket{\text{cat}_\text{6}}} \approx 0.264$
    \item $\alpha_{\text{magic}_5} \approx 0.396$
\end{itemize}

\section{Circuit Classes}
\label{app:classes}

As outlined in Section \ref{sec:results}, we benchmark our results for four classes of circuit. The specifics of each are outlined ahead in this appendix, following our general computational setup.

For all experiments, we have a timeout of 90 seconds per circuit simulated.
All $T$-counts shown are the $T$-count \textit{after initial Clifford simplification}.
We filter for circuits that have a $T$-count  $>10$ after initial simplification, 
since simulating below such low $T$-count is trivial.
All $\log_2$ values are taken in base 2, 
so graphs showing a $\log$ scale with a best fit line are showing an exponential fit $y = c2^{mx}$.
When comparing with the default \texttt{quizx} \cite{quizx} implementation,
it uses the greedy approach on \catstate{n} states \cite{Kissinger2022cat}.
We show the results for \texttt{single}, corresponding to Decompositions \ref{decom:lonephase} and \ref{decom:multicat3},
and \texttt{single+paired}, corresponding to using all $4$ decompositions introduced in Section \ref{sec:decomps}.
For each circuit, we randomly plug an output state $\bra{\vec{x}}$ where $\vec{x}\in \{0,1\}^n$
and decompose from there.

\subsection{Clifford+T circuits}

We examine the combination of random Clifford+T circuits with CCZ gates, as benchmarked in Koch et al. \cite{koch2023} In this approach, random circuits for $q\in\{50,100\}$ qubits are generated using T, CCZ, and Clifford gates (including CNOT, CZ, Hadamard, and S). Specifically, gates are sampled randomly from this set, with T and CCZ gates each having a 5\% sampling probability. We then evaluate our method’s performance in comparison to the QuiZX implementation in \cite{quizx}, which represents CCZ (controlled-CZ) gates as:

\begin{equation}
    \label{eq:7t-ccz} 
    \input{ccz-box.tikz} := \input{ccz.tikz} \ =\ \sqrt{2}^5\ \input{7t-ccz.tikz}
\end{equation} 

It is derived by simplifying the `textbook' CCZ or Toffoli representation using CNOT and T gates (see Nielsen and Chuang \cite{Nielsen-Chuang-2010}, Section 4.3).

Such circuits we label in our experiments as CCZ circuits. We also analyze fully random Clifford+T circuits constructed via exponentiated Paulis \cite{Kissinger-2022} (not including CCZ gates).

\vspace{2cm} %temp?
\subsection{IQP circuits}

Instantaneous Quantum Polynomial (IQP) circuits were used in the benchmark by \cite{Codsi2022} 
as a potential model for demonstrating quantum supremacy \cite{Bremner-2017}. 
These circuits are feasible for quantum computers and represent a class of commuting quantum computations challenging to simulate classically, 
even under constraints like sparsity and noise \cite{Bremner-2017}. 
While some question IQP circuits’ suitability for demonstrating quantum supremacy \cite{codsi2023,Codsi2022}, 
our heuristic approach yields performance comparable to prior results.

An IQP circuit is defined as follows:

\begin{defi}[IQP circuit \cite{Bremner-2016,codsi2023}]
    An IQP circuit is a circuit composed of Hadamard gates, $CZ(\frac{k\pi}{2})$-gates,
    and $T^m$-gates, where $k,m\in \mathbb{Z}$.
    Thus, the following \autoref{fig:def-IQP} is a simplified IQP circuit.
\begin{figure}[h!]
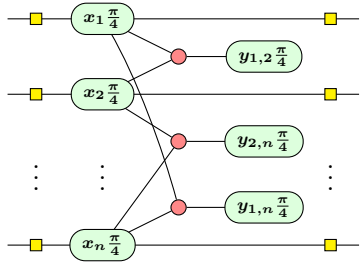

    \ctikzfig{def-IQP}
    \caption{An IQP circuit as a ZX-diagram, for $x_i,y_{i,j}\in\mathbb{Z}$, 
    as defined in Codsi and van de Wetering \cite{codsi2023}.}
    \label{fig:def-IQP}
\end{figure}
\end{defi}

Given a $n$-qubit IQP circuit, the heuristic approach to stabilizer decomposition will give $O(2^n)$ terms. Because of this, the $\beta$ efficiency is better for comparison purposes.

\begin{defi}[$\beta$ efficiency]
    \label{def:beta-efficiency}
    Suppose we have ZX-diagram in reduced gadget form, 
    which has $n$ $T$-like spiders not part of phase gadgets,
    and $\tilde{O}(n^k)$ phase gadgets, for $k\in \mathbb{R}_{> 1}$.
    Let $p$ be the number of terms in the stabilizer decomposition $D$.
    We define the $\beta$ \textit{efficiency} as
    \begin{equation}
        \beta(D) = \frac{p}{n}
    \end{equation}
\end{defi}

\clearpage

\subsection{Modified hidden shift circuits}

Hidden shift circuits have served as benchmarks since their introduction in Bravyi and Gosset \cite{Bravyi-2016bss}, initially for the BSS decomposition benchmark and later for the \catstate{n} decomposition \cite{Bravyi-2019,Kissinger-2022,Kissinger2022cat,Peres-2023}. However, QuiZX advancements have simplified these circuit decompositions, with conjectures suggesting they are simulable in polynomial time \cite{Codsi2022}. Consequently, Koch et al. \cite{koch2023} proposed the \textit{modified hidden shift} circuits.

The $T$-count in hidden shift circuits originates solely from $T$-like spiders in CCZ gates. We define the CCZ gate as in Equation \ref{eq:7t-ccz} \cite{Kissinger-2022}.

While original hidden shift circuits utilized CCZ gates, the modified circuits introduced controlled swap gates instead. 
Adding CNOT and Hadamard gates to create this controlled swap gate made the circuits complex again.

The controlled swap (Friedkin) gate is defined as: 
\begin{equation} 
\input{cswap-box.tikz} := \input{cswap.tikz} 
\end{equation} 

\section{Derivations}
\label{app:derivs}

The dynamic decompositions introduced in Section \ref{sec:decomps} may be derived via the rewriting rules of Figures \ref{fig:zxrules} and \ref{fig:derivrules}, as shown ahead.

\newpage

\noindent\textbf{Proof.} The derivation of Decomposition \ref{decom:lonephase} follows:
\begin{equation}
    \input{lone-phase-deriv.tikz}
\end{equation}

\noindent\textbf{Proof.} The derivation of Decomposition \ref{decom:multicat3} follows:
\begin{equation}
    \input{multi-cat3-deriv.tikz}
\end{equation}

\noindent\textbf{Proof.} The derivation of Decomposition \ref{decom:d3} follows:
\begin{equation}
    \input{pair-eq-cut.tikz}
\end{equation}

\noindent\textbf{Proof.} The derivation of Decomposition \ref{decom:d4} follows:
\begin{equation}
    \input{pair-eq-phase.tikz}
\end{equation}

Note that each of these decompositions remain applicable with any set of T-spiders ($\frac{\pi}{4}$ phase spiders) being instead T-like ($\frac{n\pi}{4}$ for odd $n$). Further note that each of these decompositions are designed for ZX-diagrams in reduced gadget form, but can, if desired, be generalized with just minor modification. In Decomposition \ref{decom:multicat3}, for instance, the $1$-legged spiders may be generalized to $n$-legged spiders, and the cut $\frac{\pi}{4}$ phase to an $\alpha\in\mathbb{R}$ phase, like so:

\begin{equation}
    \label{eq:cat-3-cut-general}
    \input{multi-cat3-general.tikz}
\end{equation}

Lastly, note that we use `$\approx$' in this context to mean `equal up to a constant scalar factor', which is often neglected for brevity (but can be deduced from the rewriting rules).

\section{Additional Dynamic Decompositions}

One additional decomposition, alluded to in the work of Codsi \cite{Codsi2022} and formalized in the work of Ahmad \cite{ahmad2024}, applies to a set of phase gadgets of near-identical children. In such cases, cutting the spiders which belong to the symmetric difference facilitates an instance of gadget fusion. The following example illustrates this:

\begin{equation}
    \input{phase-gadget-heur-ex.tikz}
\end{equation}

A second such decomposition is a variation on the cutting decomposition. Instead of cutting a vertex directly, one may separate its edges into two arbitrarily proportioned sets. Under certain circumstances, this can be more effective than traditional vertex cutting, as the following example highlights:

\begin{equation}
    \input{pg-pair.tikz}
\end{equation}

Experimentally, we found these extra decompositions were rarely helpful in practice, particularly given the additional computational overhead to check for their applicability. As such, we note these as of theoretical interest but neglected them from our decomposition strategy in our experiments.

%\begin{thebibliography}{000} %for 3 digits
%\begin{thebibliography}{00}  %for 2 digits
%\begin{thebibliography}{0}   %for 1 digit

% \begin{thebibliography}{0}

% \end{thebibliography}

\bibliographystyle{ws-ijmpc}
\bibliography{ws-sample}

\end{document}

%% file: defs.tex
\usepackage{amsmath,amsthm,amssymb}
\usepackage{tikzit}
\usepackage{tikzscale}
\usepackage{keycommand}
\usepackage{pgfplots}
\usepackage{hyperref}
\usepackage{relsize}
\usepackage{xcolor}
\usepackage{algorithm}
\usepackage{algpseudocode}
\usepackage{titlesec}
\usepackage{subcaption}

\graphicspath{{.}{./figures/}}

\input{quantum.tikzstyles}
\input{dots.tikzdefs}

\titlespacing*{\chapter}{0pt}{-40pt}{40pt}

\newtheorem{decom}{Decomposition}
\newtheorem{defi}{Definition}

\newtheorem*{theorem*}{Theorem}
\newtheorem*{corollary*}{Corollary}
\newtheorem*{lemma*}{Lemma}
\newtheorem*{proposition*}{Proposition}

\newcommand{\C}{\ensuremath{\mathbb{C}}}
\newcommand{\eqrule}[1]{\ensuremath{\stackrel{#1}{=}}}

\newcommand{\bra}[1]{\ensuremath{\left\langle #1 \right|}}
\newcommand{\ket}[1]{\ensuremath{\left|  #1 \right\rangle}}

\newcommand{\catstate}[1]{\ket{\text{cat}_{#1}}}

\newcommand\cbox[1]{\vcenter{\hbox{#1}}}

\newkeycommand{\boxmap}[style=box][1]{\,\tikz{\node[style=\commandkey{style}] (x) {$#1$};\draw(-1.25,0)--(x)--(1.25,0);}\,}

\newkeycommand{\pointmap}[style=ket][1]{\,\tikz{\node[style=\commandkey{style}] (x) at (-0.3,0) {$#1$};\node [style=none] (3) at (0.9, 0) {};\node [style=none] (3) at (-0.9, 0) {};\draw (x)--(0.7,0);}\,}
\newkeycommand{\point}[style=ket,estyle=none][1]{\,\tikz{\node[style=\commandkey{style}] (x) at (-0.3,0) {$#1$};\draw [\commandkey{estyle}] (x)--(0.7,0);}\,}
\newkeycommand{\longpoint}[style=ket,estyle=][1]{\,\tikz{\node[style=\commandkey{style}] (x) at (-0.3,0) {$#1$};\draw [\commandkey{estyle}] (x)--(1.2,0);}\,}

\newkeycommand{\copointmap}[style=bra][1]{\,\tikz{\node[style=\commandkey{style}] (x) at (0.3,0) {$#1$};\draw (x)--(-0.7,0);}\,}

\newkeycommand{\twopoint}[style=ket,estyle=][1]{\,\tikz{\node[style=\commandkey{style}] (x) at (-0.3,0) {$#1$};\draw [\commandkey{estyle}] (x)+(0.25,0.2)--(0.7,0.2);\draw [\commandkey{estyle}] (x)+(0.25,-0.2)--(0.7,-0.2);}\,}

\newkeycommand{\twopointketbra}[style1=bra,style2=ket,style3=ket][3]{\,%
\begin{tikzpicture}
  \begin{pgfonlayer}{nodelayer}
    \node [style=none] (0) at (0, -1.75) {};
    \node [style=none] (1) at (-0.75, 1.75) {};
    \node [style=\commandkey{style1}] (2) at (0, -0.75) {$#1$};
    \node [style=\commandkey{style2}] (3) at (-0.75, 0.75) {$#2$};
    \node [style=\commandkey{style3}] (4) at (0.75, 0.75) {$#3$};
    \node [style=none] (5) at (0.75, 1.75) {};
  \end{pgfonlayer}
  \begin{pgfonlayer}{edgelayer}
    \draw (0.center) to (2);
    \draw (1.center) to (3);
    \draw (4) to (5.center);
  \end{pgfonlayer}
\end{tikzpicture}\,}

\newkeycommand{\twotwoketbra}[style1=bra,style2=bra,style3=ket,style4=ket][4]{\,%
\begin{tikzpicture}
\begin{pgfonlayer}{nodelayer}
    \node [style=none] (0) at (-1.75, -0.5) {};
    \node [style=none] (1) at (-1.75, 0.5) {};
    \node [style=\commandkey{style1}] (2) at (-0.75, -0.5) {$#1$};
    \node [style=\commandkey{style2}] (3) at (-0.75, 0.5) {$#2$};
    \node [style=\commandkey{style3}] (4) at (0.75, -0.5) {$#3$};
    \node [style=\commandkey{style4}] (5) at (0.75, 0.5) {$#4$};
    \node [style=none] (6) at (1.75, -0.5) {};
    \node [style=none] (7) at (1.75, 0.5) {};
\end{pgfonlayer}
\begin{pgfonlayer}{edgelayer}
    \draw (0.center) to (2);
    \draw (1.center) to (3);
    \draw (4) to (6.center);
    \draw (5) to (7.center);
\end{pgfonlayer}
\end{tikzpicture}\,}

\newkeycommand{\pointbraket}[style1=ket,style2=bra,estyle=][2]{\,%
\begin{tikzpicture}
  \begin{pgfonlayer}{nodelayer}
    \node [style=\commandkey{style1}] (0) at (-0.625, 0) {$#1$};
    \node [style=\commandkey{style2}] (1) at (0.625, 0) {$#2$};
  \end{pgfonlayer}
  \begin{pgfonlayer}{edgelayer}
    \draw [\commandkey{estyle}] (0) to (1);
  \end{pgfonlayer}
\end{tikzpicture}\,}

\newkeycommand{\phase}[style=Z phase dot][1]{\,\begin{tikzpicture}
    \begin{pgfonlayer}{nodelayer}
        \node [style=none] (0) at (1, 0) {};
        \node [style=\commandkey{style}] (2) at (0, 0) {$#1$}; 
        \node [style=none] (3) at (-1, 0) {};
    \end{pgfonlayer}
    \begin{pgfonlayer}{edgelayer}
        \draw (2) to (0.center);
        \draw (3.center) to (2);
    \end{pgfonlayer}
\end{tikzpicture}\,}

\newcommand{\magicspider}{\point[style=Z phase dot]{\frac{\pi}{4}}}

\makeatletter
\newcommand{\boxshape}[3]{%
\pgfdeclareshape{#1}{
\inheritsavedanchors[from=rectangle] % this is nearly a rectangle
\inheritanchorborder[from=rectangle]
\inheritanchor[from=rectangle]{center}
\inheritanchor[from=rectangle]{north}
\inheritanchor[from=rectangle]{south}
\inheritanchor[from=rectangle]{west}
\inheritanchor[from=rectangle]{east}
\backgroundpath{% this is new
\southwest \pgf@xa=\pgf@x \pgf@ya=\pgf@y
\northeast \pgf@xb=\pgf@x \pgf@yb=\pgf@y

\@tempdima=#2
\@tempdimb=#3

\pgfpathmoveto{\pgfpoint{\pgf@xa - 5pt + \@tempdima}{\pgf@ya}}
\pgfpathlineto{\pgfpoint{\pgf@xa - 5pt - \@tempdima}{\pgf@yb}}
\pgfpathlineto{\pgfpoint{\pgf@xb + 5pt + \@tempdimb}{\pgf@yb}}
\pgfpathlineto{\pgfpoint{\pgf@xb + 5pt - \@tempdimb}{\pgf@ya}}
\pgfpathlineto{\pgfpoint{\pgf@xa - 5pt + \@tempdima}{\pgf@ya}}
\pgfpathclose
}
}}

\boxshape{NEbox}{0pt}{5pt}
\boxshape{SEbox}{0pt}{-5pt}
\boxshape{NWbox}{5pt}{0pt}
\boxshape{SWbox}{-5pt}{0pt}
\boxshape{EBox}{-3pt}{3pt}
\boxshape{WBox}{3pt}{-3pt}
\makeatother

\makeatletter

\pgfdeclareshape{cornerpoint}{
\inheritsavedanchors[from=rectangle] % this is nearly a rectangle
\inheritanchorborder[from=rectangle]
\inheritanchor[from=rectangle]{center}
\inheritanchor[from=rectangle]{north}
\inheritanchor[from=rectangle]{south}
\inheritanchor[from=rectangle]{west}
\inheritanchor[from=rectangle]{east}
\backgroundpath{% this is new
\southwest \pgf@xa=\pgf@x \pgf@ya=\pgf@y
\northeast \pgf@xb=\pgf@x \pgf@yb=\pgf@y

\pgfmathsetmacro{\pgf@shorten@left}{\pgfkeysvalueof{/tikz/shorten left}}
\pgfmathsetmacro{\pgf@shorten@right}{\pgfkeysvalueof{/tikz/shorten right}}

\pgfpathmoveto{\pgfpoint{\pgf@xa - 5pt}{0.5 * (\pgf@ya + \pgf@yb)}}
\pgfpathlineto{\pgfpoint{\pgf@xb - 1.5 * \pgf@shorten@left}{\pgf@ya - 8pt + \pgf@shorten@left}}
\pgfpathlineto{\pgfpoint{\pgf@xb}{\pgf@ya - 8pt + \pgf@shorten@left}}
\pgfpathlineto{\pgfpoint{\pgf@xb}{\pgf@yb + 8pt - \pgf@shorten@right}}
\pgfpathlineto{\pgfpoint{\pgf@xb - 1.5 * \pgf@shorten@right}{\pgf@yb + 8pt - \pgf@shorten@right}}
\pgfpathclose
}
}

\pgfdeclareshape{cornercopoint}{
\inheritsavedanchors[from=rectangle] % this is nearly a rectangle
\inheritanchorborder[from=rectangle]
\inheritanchor[from=rectangle]{center}
\inheritanchor[from=rectangle]{north}
\inheritanchor[from=rectangle]{south}
\inheritanchor[from=rectangle]{west}
\inheritanchor[from=rectangle]{east}
\backgroundpath{% this is new
\southwest \pgf@xa=\pgf@x \pgf@ya=\pgf@y
\northeast \pgf@xb=\pgf@x \pgf@yb=\pgf@y

\pgfmathsetmacro{\pgf@shorten@left}{\pgfkeysvalueof{/tikz/shorten left}}
\pgfmathsetmacro{\pgf@shorten@right}{\pgfkeysvalueof{/tikz/shorten right}}

\pgfpathmoveto{\pgfpoint{\pgf@xb + 5pt}{0.5 * (\pgf@ya + \pgf@yb)}}
\pgfpathlineto{\pgfpoint{\pgf@xa + 1.5 * \pgf@shorten@left}{\pgf@ya - 8pt + \pgf@shorten@left}}
\pgfpathlineto{\pgfpoint{\pgf@xa}{\pgf@ya - 8pt + \pgf@shorten@left}}
\pgfpathlineto{\pgfpoint{\pgf@xa}{\pgf@yb + 8pt - \pgf@shorten@right}}
\pgfpathlineto{\pgfpoint{\pgf@xa + 1.5 * \pgf@shorten@right}{\pgf@yb + 8pt - \pgf@shorten@right}}
\pgfpathclose
}
}

\makeatother

\pgfkeyssetvalue{/tikz/shorten left}{0pt}
\pgfkeyssetvalue{/tikz/shorten right}{0pt}

\newkeycommand{\kpoint}[style=kpoint,estyle=][1]{\,\tikz{\node[style=\commandkey{style}] (x) at (-0.05,0) {$#1$};\draw [\commandkey{estyle}] (x)--(1.05,0);}\,}
\newkeycommand{\kpointbraket}[style1=kpoint,style2=kpoint adjoint,estyle=][2]{\,%
\begin{tikzpicture}
  \begin{pgfonlayer}{nodelayer}
    \node [style=\commandkey{style1}] (0) at (-0.735, 0) {$#1$};
    \node [style=\commandkey{style2}] (1) at (0.735, 0) {$#2$};
  \end{pgfonlayer}
  \begin{pgfonlayer}{edgelayer}
    \draw [\commandkey{estyle}] (0) to (1);
  \end{pgfonlayer}
\end{tikzpicture}\,}

\newsavebox\magicspidercd
\begin{lrbox}{\magicspidercd}
  \magicspider
\end{lrbox}

\newsavebox\amplitudecd
\begin{lrbox}{\amplitudecd}
  \input{amplitude.tikz}
\end{lrbox}

\newkeycommand{\catfour}[style=none]{\,\tikz{\node[style=\commandkey{style}] (x) at (0,0.5) {\catstate{4}};
\node[style=none] (x) at (0,0) {};}\,}

\newsavebox\catstatecd
\begin{lrbox}{\catstatecd}
\catfour
\end{lrbox}

%% file: quantum.tikzstyles
\tikzstyle{gate}=[shape=rectangle, text height=1.5ex, text depth=0.25ex, yshift=0.5mm, fill=white, draw=black, minimum height=5mm, yshift=-0.5mm, minimum width=5mm, font={\small}, tikzit category=circuit]
\tikzstyle{big gate}=[shape=rectangle, text height=1.5ex, text depth=0.25ex, yshift=0.5mm, fill=white, draw=black, minimum height=10mm, yshift=-0.5mm, minimum width=5mm, font={\small}, tikzit category=circuit]
\tikzstyle{Z dot}=[inner sep=0mm, minimum size=2mm, shape=circle, draw=black, fill={rgb,255: red,221; green,255; blue,221}, tikzit category=zx]
\tikzstyle{Z phase dot}=[minimum size=5mm, font={\footnotesize\boldmath}, shape=rectangle, rounded corners=2mm, inner sep=1mm, outer sep=-0.5mm, scale=0.8, tikzit shape=rectangle, draw=black, fill={rgb,255: red,221; green,255; blue,221}, tikzit draw=blue, tikzit category=zx]
\tikzstyle{X dot}=[Z dot, shape=circle, draw=black, fill={rgb,255: red,255; green,136; blue,136}, tikzit category=zx]
\tikzstyle{X phase dot}=[Z phase dot, tikzit shape=rectangle, tikzit draw=blue, fill={rgb,255: red,255; green,136; blue,136}, font={\footnotesize\boldmath}, tikzit category=zx]
\tikzstyle{hadamard}=[fill=yellow, draw=black, shape=rectangle, inner sep=0.6mm, minimum height=1.5mm, minimum width=1.5mm, tikzit category=zx]
\tikzstyle{paulibox}=[fill={rgb,255: red,221; green,221; blue,255}, draw=black, shape=rectangle, inner sep=0.6mm, minimum height=5mm, minimum width=5mm, font={\footnotesize}, text height=1.5ex, text depth=0.25ex, tikzit category=zx]
\tikzstyle{vertex}=[inner sep=0mm, minimum size=1mm, shape=circle, draw=black, fill=black, tikzit category=misc]
\tikzstyle{vertex set}=[inner sep=0mm, minimum size=1mm, shape=circle, draw=black, fill=white, font={\footnotesize\boldmath}, tikzit category=misc]
\tikzstyle{small black dot}=[fill=black, draw=black, shape=circle, inner sep=0pt, minimum width=1.2mm, tikzit category=circuit]
\tikzstyle{cnot ctrl}=[fill=black, draw=black, shape=circle, inner sep=0pt, minimum width=1.2mm, tikzit category=circuit]
\tikzstyle{cnot targ}=[fill=white, draw=white, shape=circle, tikzit category=circuit, label={center:$\oplus$}, inner sep=0pt, minimum width=2.1mm, tikzit fill={rgb,255: red,102; green,204; blue,255}, tikzit draw=black]
\tikzstyle{ket}=[fill=white, draw=black, shape=regular polygon, regular polygon sides=3, regular polygon rotate=-30, scale=0.7, inner sep=1pt, tikzit category=circuit, tikzit shape=rectangle, tikzit fill=green]
\tikzstyle{bra}=[fill=white, draw=black, shape=regular polygon, regular polygon sides=3, regular polygon rotate=30, scale=0.7, inner sep=1pt, tikzit category=circuit, tikzit shape=rectangle, tikzit fill=red]
\tikzstyle{scalar}=[shape=rectangle, text height=1.5ex, text depth=0.25ex, yshift=0.5mm, fill=white, draw=black, minimum height=5mm, yshift=-0.5mm, minimum width=5mm, font={\small}]
\tikzstyle{clabel}=[fill=white, draw=none, shape=rectangle, tikzit fill={rgb,255: red,56; green,255; blue,242}, font={\footnotesize}, inner sep=1pt, tikzit category=labels]
\tikzstyle{empty diagram}=[draw={gray!40!white}, dashed, shape=rectangle, minimum width=1cm, minimum height=1cm, tikzit category=misc]
\tikzstyle{amap}=[fill=white, draw=black, shape=NEbox, tikzit category=asymmetric, tikzit fill=yellow, tikzit shape=rectangle]
\tikzstyle{amap conj}=[fill=white, draw=black, shape=NWbox, tikzit category=asymmetric, tikzit fill=green, tikzit shape=rectangle]
\tikzstyle{amap adj}=[fill=white, draw=black, shape=SEbox, tikzit category=asymmetric, tikzit fill=red, tikzit shape=rectangle]
\tikzstyle{amap trans}=[fill=white, draw=black, shape=SWbox, tikzit category=asymmetric, tikzit fill=orange, tikzit shape=rectangle]
\tikzstyle{astate}=[fill=white, draw=black, shape=NEtriangle, tikzit category=asymmetric, tikzit shape=circle, tikzit fill=yellow]
\tikzstyle{astate conj}=[fill=white, draw=black, shape=NWtriangle, tikzit category=asymmetric, tikzit shape=circle, tikzit fill=green]
\tikzstyle{astate adj}=[fill=white, draw=black, shape=SEtriangle, tikzit category=asymmetric, tikzit shape=circle, tikzit fill=red]
\tikzstyle{astate trans}=[fill=white, draw=black, shape=SWtriangle, tikzit category=asymmetric, tikzit shape=circle, tikzit fill=orange]

\tikzstyle{gray ket}=[ket, fill=gray!40!white]
\tikzstyle{gray bra}=[bra, fill=gray!40!white]

\tikzstyle{box}=[draw,shape=rectangle,inner sep=2pt,minimum height=6mm,minimum width=6mm,fill=white]

\tikzstyle{kpoint common}=[draw,fill=white,inner sep=1pt,minimum height=4mm]
\tikzstyle{kpoint}=[shape=cornerpoint,shorten left=5pt,kpoint common]
\tikzstyle{kpoint adjoint}=[shape=cornercopoint,shorten left=5pt,kpoint common]
\tikzstyle{kpoint conjugate}=[shape=cornerpoint,shorten right=5pt,kpoint common]
\tikzstyle{kpoint transpose}=[shape=cornercopoint,shorten right=5pt,kpoint common]

\tikzstyle{hadamard edge}=[-, dashed, dash pattern=on 2pt off 0.5pt, thick, draw={rgb,255: red,68; green,136; blue,255}]
\tikzstyle{box edge}=[-, dashed, dash pattern=on 2pt off 0.5pt, thick, draw={rgb,255: red,203; green,192; blue,225}]
\tikzstyle{brace edge}=[-, tikzit draw=blue, decorate, decoration={brace,amplitude=1mm,raise=-1mm}]
\tikzstyle{diredge}=[->]
\tikzstyle{double edge}=[-, double, shorten <=-1mm, shorten >=-1mm, double distance=2pt]
\tikzstyle{gray edge}=[-, {gray!60!white}]
\tikzstyle{pointer edge}=[->, very thick, gray]
\tikzstyle{boldedge}=[-, line width=1.6pt, shorten <=-0.17mm, shorten >=-0.17mm]
\tikzstyle{bidir edge}=[<->, very thick, draw={rgb,255: red,191; green,191; blue,191}]
\tikzstyle{separator edge}=[-, dashed, dash pattern=on 2pt off 0.5pt, thick, draw={rgb,255: red,153; green,153; blue,153}]

%% file: figures/amplitude.tikz
\begin{tikzpicture}
	\begin{pgfonlayer}{nodelayer}
		\node [style=none] (0) at (-2, 1.25) {};
		\node [style=none] (1) at (-2, -1.25) {};
		\node [style=none] (2) at (2, -1.25) {};
		\node [style=none] (3) at (2, 1.25) {};
		\node [style=none] (7) at (2.5, 0.25) {$\vdots$};
		\node [style=none] (8) at (2, 1) {};
		\node [style=none] (10) at (2, -1) {};
		\node [style=none] (11) at (0, 0) {$V$};
		\node [style=none] (14) at (-2, 1) {};
		\node [style=none] (16) at (-2, -1) {};
		\node [style=none] (17) at (-2.5, 0.25) {$\vdots$};
		\node [style=X dot] (18) at (-3, 1) {};
		\node [style=X dot] (20) at (-3, -1) {};
		\node [style=X phase dot] (21) at (3, 1) {$x_1\pi$};
		\node [style=X phase dot] (23) at (3, -1) {$x_n\pi$};
	\end{pgfonlayer}
	\begin{pgfonlayer}{edgelayer}
		\draw (0.center) to (3.center);
		\draw (3.center) to (2.center);
		\draw (2.center) to (1.center);
		\draw (1.center) to (0.center);
		\draw (18) to (14.center);
		\draw (20) to (16.center);
		\draw (8.center) to (21);
		\draw (10.center) to (23);
	\end{pgfonlayer}
\end{tikzpicture}

%% file: figures/cut.tikz
\begin{tikzpicture}
	\begin{pgfonlayer}{nodelayer}
		\node [style=Z phase dot] (0) at (0, 0) {$\frac{\pi}{4}$};
		\node [style=none] (1) at (1.5, 0) {};
		\node [style=none] (2) at (2.5, 0) {$=$};
		\node [style=none] (3) at (4, 0) {$\frac{1}{\sqrt{2}}\big($};
		\node [style=none] (4) at (11.5, 0) {\big)};
		\node [style=X dot] (5) at (5, 0) {};
		\node [style=none] (6) at (6.5, 0) {};
		\node [style=X phase dot] (7) at (9.75, 0) {$\pi$};
		\node [style=none] (8) at (11.25, 0) {};
		\node [style=none] (9) at (8.75, 0) {$e^{i\frac{\pi}{4}}$};
		\node [style=none] (10) at (7.5, 0) {$+$};
		\node [style=none] (11) at (2.5, 0.75) {\CutRule};
	\end{pgfonlayer}
	\begin{pgfonlayer}{edgelayer}
		\draw (0) to (1.center);
		\draw (5) to (6.center);
		\draw (7) to (8.center);
	\end{pgfonlayer}
\end{tikzpicture}

%% file: figures/lone-phase.tikz
\begin{tikzpicture}
	\begin{pgfonlayer}{nodelayer}
		\node [style=Z phase dot] (0) at (0, 0) {$\frac{\pi}{4}$};
		\node [style=Z phase dot] (1) at (-1, 1.5) {$\frac{\pi}{4}$};
		\node [style=Z phase dot] (2) at (1, 1.5) {$\frac{\pi}{4}$};
		\node [style=none] (3) at (0, 1.5) {$\cdots$};
		\node [style=none] (4) at (-1, -1.5) {};
		\node [style=none] (5) at (1, -1.5) {};
		\node [style=none] (6) at (0, -1.25) {$\cdots$};
		\node [style=none] (21) at (-1, 2.25) {\textcolor{lightgray}{\footnotesize $x_1$}};
		\node [style=none] (22) at (1, 2.25) {\textcolor{lightgray}{\footnotesize $x_n$}};
		\node [style=none] (25) at (3, 0) {$\approx$};
		\node [style=none] (37) at (5, -0.25) {$\sum\limits_{a\in\{0,1\}}$};
		\node [style=none] (38) at (9.25, 1) {$e^{ia\frac{\pi}{4}}\left(1+e^{i\pi(\frac{1}{4}+a)}\right)^n$};
		\node [style=none] (39) at (8, -1.5) {};
		\node [style=none] (40) at (10.5, -1.5) {};
		\node [style=none] (41) at (9.25, -1.25) {$\cdots$};
		\node [style=Z phase dot] (42) at (8, -0.5) {$a\pi$};
		\node [style=Z phase dot] (43) at (10.5, -0.5) {$a\pi$};
	\end{pgfonlayer}
	\begin{pgfonlayer}{edgelayer}
		\draw [style=hadamard edge] (2) to (0);
		\draw [style=hadamard edge] (0) to (1);
		\draw (0) to (4.center);
		\draw (0) to (5.center);
		\draw [style=hadamard edge] (42) to (39.center);
		\draw [style=hadamard edge] (43) to (40.center);
	\end{pgfonlayer}
\end{tikzpicture}

%% file: figures/multi-cat3.tikz
\begin{tikzpicture}
	\begin{pgfonlayer}{nodelayer}
		\node [style=Z dot] (0) at (-5.25, 1.25) {};
		\node [style=Z dot] (1) at (-3.25, 1.25) {};
		\node [style=Z phase dot] (2) at (-5.25, 2.25) {$\frac{\pi}{4}$};
		\node [style=Z phase dot] (3) at (-3.25, 2.25) {$\frac{\pi}{4}$};
		\node [style=Z phase dot] (4) at (-6.25, 0) {$\frac{\pi}{4}$};
		\node [style=Z phase dot] (5) at (-4.25, 0) {$\frac{\pi}{4}$};
		\node [style=Z phase dot] (6) at (-2.25, 0) {$\frac{\pi}{4}$};
		\node [style=none] (7) at (-6.75, -1) {};
		\node [style=none] (8) at (-6.25, -0.75) {$...$};
		\node [style=none] (9) at (-5.75, -1) {};
		\node [style=none] (10) at (-4.75, -1) {};
		\node [style=none] (11) at (-4.25, -0.75) {$...$};
		\node [style=none] (12) at (-3.75, -1) {};
		\node [style=none] (13) at (-2.75, -1) {};
		\node [style=none] (14) at (-2.25, -0.75) {$...$};
		\node [style=none] (15) at (-1.75, -1) {};
		\node [style=Z dot] (20) at (-7.25, 1.25) {};
		\node [style=Z phase dot] (21) at (-7.25, 2.25) {$\frac{\pi}{4}$};
		\node [style=Z phase dot] (22) at (-8.25, 0) {$\frac{\pi}{4}$};
		\node [style=none] (23) at (-8.75, -1) {};
		\node [style=none] (24) at (-8.25, -0.75) {$...$};
		\node [style=none] (25) at (-7.75, -1) {};
		\node [style=none] (89) at (-4.25, 2) {$...$};
		\node [style=none] (93) at (-3.25, 0) {$...$};
		\node [style=none] (214) at (-7.25, 3) {\textcolor{lightgray}{\footnotesize $x_1$}};
		\node [style=none] (215) at (-5.25, 3) {\textcolor{lightgray}{\footnotesize $x_2$}};
		\node [style=none] (216) at (-3.25, 3) {\textcolor{lightgray}{\footnotesize $x_n$}};
		\node [style=none] (217) at (-7, -0.25) {\textcolor{lightgray}{\footnotesize $y_1$}};
		\node [style=none] (218) at (-5, -0.25) {\textcolor{lightgray}{\footnotesize $y_2$}};
		\node [style=none] (219) at (-1.5, -0.25) {\textcolor{lightgray}{\footnotesize $y_n$}};
		\node [style=none] (220) at (-9, -0.25) {\textcolor{lightgray}{\footnotesize $v$}};
		\node [style=none] (221) at (0, 0.75) {$\approx$};
		\node [style=Z phase dot] (222) at (8, 0.75) {$(1-a)\frac{\pi}{2}$};
		\node [style=Z phase dot] (223) at (11, 0.75) {$(1-a)\frac{\pi}{2}$};
		\node [style=Z phase dot] (224) at (14.5, 0.75) {$(1-a)\frac{\pi}{2}$};
		\node [style=none] (225) at (7.5, -0.25) {};
		\node [style=none] (226) at (8, 0) {$...$};
		\node [style=none] (227) at (8.5, -0.25) {};
		\node [style=none] (228) at (10.5, -0.25) {};
		\node [style=none] (229) at (11, 0) {$...$};
		\node [style=none] (230) at (11.5, -0.25) {};
		\node [style=none] (231) at (14, -0.25) {};
		\node [style=none] (232) at (14.5, 0) {$...$};
		\node [style=none] (233) at (15, -0.25) {};
		\node [style=none] (234) at (4.5, -0.25) {};
		\node [style=none] (235) at (5.25, 0) {$...$};
		\node [style=none] (236) at (6, -0.25) {};
		\node [style=Z phase dot] (237) at (4.5, 0.75) {$a\pi$};
		\node [style=Z phase dot] (238) at (6, 0.75) {$a\pi$};
		\node [style=none] (239) at (2, 0.75) {$\mathlarger{\sum}$};
		\node [style=none] (240) at (5.25, 2) {$e^{ia(n+1)\frac{\pi}{4}}$};
		\node [style=none] (241) at (2, -0.25) {$\scriptstyle{a\in\{0,1\}}$};
		\node [style=none] (242) at (12.75, 0.75) {$...$};
		\node [style=none] (244) at (8, 1.5) {\textcolor{lightgray}{\footnotesize $y_1$}};
		\node [style=none] (245) at (11, 1.5) {\textcolor{lightgray}{\footnotesize $y_2$}};
		\node [style=none] (246) at (14.5, 1.5) {\textcolor{lightgray}{\footnotesize $y_n$}};
	\end{pgfonlayer}
	\begin{pgfonlayer}{edgelayer}
		\draw [style=hadamard edge] (2) to (0);
		\draw [style=hadamard edge] (0) to (5);
		\draw [style=hadamard edge] (3) to (1);
		\draw [style=hadamard edge] (1) to (6);
		\draw (4) to (7.center);
		\draw (4) to (9.center);
		\draw (5) to (10.center);
		\draw (5) to (12.center);
		\draw (6) to (13.center);
		\draw (6) to (15.center);
		\draw [style=hadamard edge] (21) to (20);
		\draw [style=hadamard edge] (4) to (20);
		\draw (22) to (23.center);
		\draw (22) to (25.center);
		\draw [style=hadamard edge] (20) to (22);
		\draw [style=hadamard edge] (22) to (0);
		\draw [style=hadamard edge] (22) to (1);
		\draw (222) to (225.center);
		\draw (222) to (227.center);
		\draw (223) to (228.center);
		\draw (223) to (230.center);
		\draw (224) to (231.center);
		\draw (224) to (233.center);
		\draw [style=hadamard edge] (234.center) to (237);
		\draw [style=hadamard edge] (236.center) to (238);
	\end{pgfonlayer}
\end{tikzpicture}

%% file: figures/pair-decomp.tikz
\begin{tikzpicture}
	\begin{pgfonlayer}{nodelayer}
		\node [style=Z phase dot] (25) at (7.5, 1) {$(1-a)\frac{\pi}{2}$};
		\node [style=Z phase dot] (26) at (11.5, 1) {$(1-a)\frac{\pi}{2}$};
		\node [style=none] (27) at (9.5, 1) {$...$};
		\node [style=none] (28) at (6.5, 2) {};
		\node [style=none] (29) at (8.25, 2) {};
		\node [style=none] (30) at (7.5, 1.75) {$...$};
		\node [style=none] (31) at (10.75, 2) {};
		\node [style=none] (32) at (12.5, 2) {};
		\node [style=none] (33) at (11.5, 1.75) {$...$};
		\node [style=none] (34) at (9.5, 1.75) {$...$};
		\node [style=none] (35) at (0, 0) {$\approx$};
		\node [style=none] (37) at (1.75, 0) {$\mathlarger{\sum}$};
		\node [style=none] (38) at (1.75, -1) {$\scriptstyle{a\in\{0,1\}}$};
		\node [style=none] (39) at (6.5, 0.25) {\textcolor{lightgray}{\footnotesize $x_1$}};
		\node [style=none] (40) at (12.5, 0.25) {\textcolor{lightgray}{\footnotesize $x_n$}};
		\node [style=none] (41) at (4.25, 0.25) {$e^{ia(n+1)\frac{\pi}{4}}$};
		\node [style=Z phase dot] (42) at (9.5, -0.5) {$(1-a)\frac{\pi}{2}$};
		\node [style=Z phase dot] (43) at (10.25, -1.75) {$a\pi$};
		\node [style=none] (44) at (7.75, -2.25) {$...$};
		\node [style=Z phase dot] (45) at (11.75, -1.75) {$a\pi$};
		\node [style=none] (46) at (6.5, -2.75) {};
		\node [style=none] (47) at (8.5, -2.75) {};
		\node [style=none] (48) at (10.5, -2.75) {};
		\node [style=none] (49) at (12.5, -2.75) {};
		\node [style=none] (50) at (11.25, -2.25) {$...$};
		\node [style=none] (51) at (-1, 3) {};
		\node [style=Z phase dot] (52) at (-6.5, -2) {$\frac{\pi}{4}$};
		\node [style=Z phase dot] (53) at (-3.5, -2) {$\frac{\pi}{4}$};
		\node [style=Z dot] (54) at (-6.5, 0) {};
		\node [style=Z dot] (55) at (-3.5, 0) {};
		\node [style=none] (56) at (-5, 0) {$...$};
		\node [style=Z phase dot] (57) at (-8, 2) {$\frac{\pi}{4}$};
		\node [style=Z phase dot] (58) at (-6, 2) {$\frac{\pi}{4}$};
		\node [style=Z phase dot] (59) at (-4, 2) {$\frac{\pi}{4}$};
		\node [style=Z phase dot] (60) at (-2, 2) {$\frac{\pi}{4}$};
		\node [style=none] (61) at (-5, 2) {$...$};
		\node [style=none] (62) at (-9, 3) {};
		\node [style=none] (63) at (-7.25, 3) {};
		\node [style=none] (64) at (-8, 2.75) {$...$};
		\node [style=none] (65) at (-2.75, 3) {};
		\node [style=none] (67) at (-2, 2.75) {$...$};
		\node [style=none] (68) at (-4, -3) {};
		\node [style=none] (69) at (-2, -3) {};
		\node [style=none] (70) at (-3.25, -2.75) {$...$};
		\node [style=none] (71) at (-8, -3) {};
		\node [style=none] (72) at (-6, -3) {};
		\node [style=none] (73) at (-6.75, -2.75) {$...$};
		\node [style=none] (74) at (-5, 2.75) {$...$};
		\node [style=none] (75) at (-7.5, -2) {\textcolor{lightgray}{\footnotesize $z_1$}};
		\node [style=none] (76) at (-2.5, -2) {\textcolor{lightgray}{\footnotesize $z_2$}};
		\node [style=none] (77) at (-8.5, 1.25) {\textcolor{lightgray}{\footnotesize $x_1$}};
		\node [style=none] (78) at (-1.5, 1.25) {\textcolor{lightgray}{\footnotesize $x_n$}};
		\node [style=none] (79) at (-4, 2.75) {\textcolor{lightgray}{\footnotesize $y_n$}};
		\node [style=none] (80) at (-6, 2.75) {\textcolor{lightgray}{\footnotesize $y_1$}};
	\end{pgfonlayer}
	\begin{pgfonlayer}{edgelayer}
		\draw (29.center) to (25);
		\draw (25) to (28.center);
		\draw (31.center) to (26);
		\draw (26) to (32.center);
		\draw [bend right] (42) to (46.center);
		\draw [bend right=15] (42) to (47.center);
		\draw [style=hadamard edge, bend left=15, looseness=0.75] (42) to (43);
		\draw [style=hadamard edge, bend left=15, looseness=0.50] (43) to (48.center);
		\draw [style=hadamard edge, bend left=15, looseness=0.75] (42) to (45);
		\draw [style=hadamard edge, bend left=15, looseness=0.75] (45) to (49.center);
		\draw [style=hadamard edge] (55) to (52);
		\draw [style=hadamard edge] (54) to (53);
		\draw [style=hadamard edge] (53) to (55);
		\draw [style=hadamard edge] (54) to (52);
		\draw [style=hadamard edge] (54) to (57);
		\draw [style=hadamard edge] (54) to (58);
		\draw [style=hadamard edge] (59) to (55);
		\draw [style=hadamard edge] (55) to (60);
		\draw (63.center) to (57);
		\draw (57) to (62.center);
		\draw (65.center) to (60);
		\draw (52) to (71.center);
		\draw (52) to (72.center);
		\draw (53) to (68.center);
		\draw (53) to (69.center);
		\draw (60) to (51.center);
	\end{pgfonlayer}
\end{tikzpicture}

%% file: figures/pair-phase-decomp.tikz
\begin{tikzpicture}
	\begin{pgfonlayer}{nodelayer}
		\node [style=Z phase dot] (119) at (-7.5, 2.75) {$\frac{\pi}{4}$};
		\node [style=Z phase dot] (120) at (-4.5, 2.75) {$\frac{\pi}{4}$};
		\node [style=Z phase dot] (124) at (-9, 5) {$\frac{\pi}{4}$};
		\node [style=Z phase dot] (125) at (-7, 5) {$\frac{\pi}{4}$};
		\node [style=Z phase dot] (126) at (-5, 5) {$\frac{\pi}{4}$};
		\node [style=Z phase dot] (127) at (-3, 5) {$\frac{\pi}{4}$};
		\node [style=none] (128) at (-6, 5) {$...$};
		\node [style=none] (135) at (-5, 1.75) {};
		\node [style=none] (136) at (-3, 1.75) {};
		\node [style=none] (137) at (-4.25, 2) {$...$};
		\node [style=none] (138) at (-9, 1.75) {};
		\node [style=none] (139) at (-7, 1.75) {};
		\node [style=none] (140) at (-7.75, 2) {$...$};
		\node [style=none] (141) at (-8.5, 2.75) {\textcolor{lightgray}{\footnotesize $y_1$}};
		\node [style=none] (142) at (-3.5, 2.75) {\textcolor{lightgray}{\footnotesize $y_2$}};
		\node [style=none] (158) at (-0.25, 3.75) {$\approx$};
		\node [style=none] (160) at (2, 3.75) {$\mathlarger{\sum}$};
		\node [style=none] (161) at (2, 2.75) {$\scriptstyle{a\in\{0,1\}}$};
		\node [style=Z phase dot] (164) at (7, 4) {$(1-a)\frac{\pi}{2}$};
		\node [style=Z phase dot] (166) at (7.75, 2.75) {$a\pi$};
		\node [style=none] (167) at (5.25, 2.25) {$...$};
		\node [style=none] (228) at (7, 5.5) {$\left(1+e^{i\pi(\frac{1}{4}+a)}\right)^n$};
		\node [style=Z phase dot] (229) at (9.25, 2.75) {$a\pi$};
		\node [style=none] (231) at (4, 1.75) {};
		\node [style=none] (232) at (6, 1.75) {};
		\node [style=none] (233) at (8, 1.75) {};
		\node [style=none] (234) at (10, 1.75) {};
		\node [style=none] (235) at (8.75, 2.25) {$...$};
		\node [style=none] (247) at (-3, 5.75) {\textcolor{lightgray}{\footnotesize $x_n$}};
		\node [style=none] (248) at (-9, 5.75) {\textcolor{lightgray}{\footnotesize $x_1$}};
		\node [style=none] (249) at (-7, 5.75) {\textcolor{lightgray}{\footnotesize $x_2$}};
		\node [style=none] (250) at (-5, 5.75) {\textcolor{lightgray}{\footnotesize $x_{n-1}$}};
	\end{pgfonlayer}
	\begin{pgfonlayer}{edgelayer}
		\draw (119) to (138.center);
		\draw (119) to (139.center);
		\draw (120) to (135.center);
		\draw (120) to (136.center);
		\draw [style=hadamard edge] (124) to (120);
		\draw [style=hadamard edge] (125) to (119);
		\draw [style=hadamard edge] (126) to (119);
		\draw [style=hadamard edge] (127) to (119);
		\draw [style=hadamard edge] (124) to (119);
		\draw [style=hadamard edge] (125) to (120);
		\draw [style=hadamard edge] (126) to (120);
		\draw [style=hadamard edge] (127) to (120);
		\draw [bend right] (164) to (231.center);
		\draw [bend right=15] (164) to (232.center);
		\draw [style=hadamard edge, bend left=15, looseness=0.75] (164) to (166);
		\draw [style=hadamard edge, bend left=15, looseness=0.50] (166) to (233.center);
		\draw [style=hadamard edge, bend left=15, looseness=0.75] (164) to (229);
		\draw [style=hadamard edge, bend left=15, looseness=0.75] (229) to (234.center);
	\end{pgfonlayer}
\end{tikzpicture}

%% file: figures/cats.tikz
\begin{tikzpicture}
	\begin{pgfonlayer}{nodelayer}
		\node [style=none] (16) at (2, 0) {$=$};
		\node [style=none] (17) at (3.75, 0) {$\frac{1}{2}$};
		\node [style=Z dot] (18) at (12.5, 0) {};
		\node [style=Z dot] (19) at (13.5, 2) {};
		\node [style=Z dot] (20) at (13.5, 1) {};
		\node [style=Z dot] (21) at (13.5, 0) {};
		\node [style=Z dot] (22) at (13.5, -1) {};
		\node [style=Z dot] (23) at (13.5, -2) {};
		\node [style=Z dot] (24) at (11.5, 0) {};
		\node [style=none] (25) at (8.5, 0) {$+$};
		\node [style=none] (26) at (10, 0) {$\frac{ie^{i\pi/4}}{\sqrt{2}}$};
		\node [style=none] (27) at (14.25, 2) {};
		\node [style=none] (28) at (14.25, 1) {};
		\node [style=none] (29) at (14.25, 0) {};
		\node [style=none] (30) at (14.25, -1) {};
		\node [style=none] (31) at (14.25, -2) {};
		\node [style=Z dot] (32) at (18.5, 0) {};
		\node [style=Z phase dot] (33) at (19.5, 2) {$\frac{\pi}{2}$};
		\node [style=Z phase dot] (34) at (19.5, 1) {$\frac{\pi}{2}$};
		\node [style=Z phase dot] (35) at (19.5, 0) {$\frac{\pi}{2}$};
		\node [style=Z phase dot] (36) at (19.5, -1) {$\frac{\pi}{2}$};
		\node [style=Z phase dot] (37) at (19.5, -2) {$\frac{\pi}{2}$};
		\node [style=Z dot] (38) at (17.5, 0) {};
		\node [style=none] (39) at (15, 0) {$-$};
		\node [style=none] (41) at (20.5, 2) {};
		\node [style=none] (42) at (20.5, 1) {};
		\node [style=none] (43) at (20.5, 0) {};
		\node [style=none] (44) at (20.5, -1) {};
		\node [style=none] (45) at (20.5, -2) {};
		\node [style=Z dot] (46) at (-1, 0) {};
		\node [style=Z phase dot] (47) at (0, 2) {$\frac{\pi}{4}$};
		\node [style=Z phase dot] (48) at (0, 1) {$\frac{\pi}{4}$};
		\node [style=Z phase dot] (49) at (0, 0) {$\frac{\pi}{4}$};
		\node [style=Z phase dot] (50) at (0, -1) {$\frac{\pi}{4}$};
		\node [style=Z phase dot] (51) at (0, -2) {$\frac{\pi}{4}$};
		\node [style=none] (53) at (1, 2) {};
		\node [style=none] (54) at (1, 1) {};
		\node [style=none] (55) at (1, 0) {};
		\node [style=none] (56) at (1, -1) {};
		\node [style=none] (57) at (1, -2) {};
		\node [style=Z phase dot] (58) at (6.75, 0) {$\,-\frac{\pi}{2}\,$};
		\node [style=none] (59) at (7.75, 2) {};
		\node [style=none] (60) at (7.75, 1) {};
		\node [style=none] (61) at (7.75, 0) {};
		\node [style=none] (62) at (7.75, -1) {};
		\node [style=none] (63) at (7.75, -2) {};
		\node [style=Z dot] (64) at (5.25, 0) {};
		\node [style=none] (65) at (16.25, 0) {$\frac{e^{i\pi/4}}{\sqrt{2}}$};
		\node [style=none] (66) at (2, -6.5) {$=$};
		\node [style=none] (120) at (3.75, -6.5) {$\frac{1}{2}$};
		\node [style=Z dot] (129) at (12.5, -6.5) {};
		\node [style=Z dot] (130) at (13.5, -4) {};
		\node [style=Z dot] (131) at (13.5, -5) {};
		\node [style=Z dot] (132) at (13.5, -6) {};
		\node [style=Z dot] (133) at (13.5, -8) {};
		\node [style=Z dot] (134) at (13.5, -9) {};
		\node [style=Z dot] (135) at (11.5, -6.5) {};
		\node [style=none] (136) at (8.5, -6.5) {$+$};
		\node [style=none] (137) at (10, -6.5) {$\frac{ie^{i\pi/4}}{\sqrt{2}}$};
		\node [style=none] (138) at (14.25, -4) {};
		\node [style=none] (139) at (14.25, -5) {};
		\node [style=none] (140) at (14.25, -6) {};
		\node [style=none] (141) at (14.25, -8) {};
		\node [style=none] (142) at (14.25, -9) {};
		\node [style=Z dot] (143) at (13.5, -7) {};
		\node [style=none] (144) at (14.25, -7) {};
		\node [style=Z dot] (145) at (18.5, -6.5) {};
		\node [style=Z phase dot] (146) at (19.5, -4) {$\frac{\pi}{2}$};
		\node [style=Z phase dot] (147) at (19.5, -5) {$\frac{\pi}{2}$};
		\node [style=Z phase dot] (148) at (19.5, -6) {$\frac{\pi}{2}$};
		\node [style=Z phase dot] (149) at (19.5, -8) {$\frac{\pi}{2}$};
		\node [style=Z phase dot] (150) at (19.5, -9) {$\frac{\pi}{2}$};
		\node [style=Z dot] (151) at (17.5, -6.5) {};
		\node [style=none] (152) at (15, -6.5) {$-$};
		\node [style=none] (153) at (20.5, -4) {};
		\node [style=none] (154) at (20.5, -5) {};
		\node [style=none] (155) at (20.5, -6) {};
		\node [style=none] (156) at (20.5, -8) {};
		\node [style=none] (157) at (20.5, -9) {};
		\node [style=none] (158) at (16.25, -6.5) {$\frac{e^{i\pi/4}}{\sqrt{2}}$};
		\node [style=Z phase dot] (159) at (19.5, -7) {$\frac{\pi}{2}$};
		\node [style=none] (160) at (20.5, -7) {};
		\node [style=Z dot] (161) at (-1, -6.5) {};
		\node [style=Z phase dot] (162) at (0, -4) {$\frac{\pi}{4}$};
		\node [style=Z phase dot] (163) at (0, -5) {$\frac{\pi}{4}$};
		\node [style=Z phase dot] (164) at (0, -6) {$\frac{\pi}{4}$};
		\node [style=Z phase dot] (165) at (0, -8) {$\frac{\pi}{4}$};
		\node [style=Z phase dot] (166) at (0, -9) {$\frac{\pi}{4}$};
		\node [style=none] (167) at (1, -4) {};
		\node [style=none] (168) at (1, -5) {};
		\node [style=none] (169) at (1, -6) {};
		\node [style=none] (170) at (1, -8) {};
		\node [style=none] (171) at (1, -9) {};
		\node [style=Z phase dot] (172) at (0, -7) {$\frac{\pi}{4}$};
		\node [style=none] (173) at (1, -7) {};
		\node [style=Z phase dot] (174) at (6.75, -6.5) {$\,-\frac{\pi}{2}\,$};
		\node [style=none] (175) at (7.75, -4) {};
		\node [style=none] (176) at (7.75, -5) {};
		\node [style=none] (177) at (7.75, -6) {};
		\node [style=none] (178) at (7.75, -8) {};
		\node [style=none] (179) at (7.75, -9) {};
		\node [style=Z dot] (180) at (5.25, -6.5) {};
		\node [style=none] (181) at (7.75, -7) {};
		\node [style=none] (182) at (2, 5.5) {$=$};
		\node [style=Z dot] (183) at (-1, 5.5) {};
		\node [style=Z phase dot] (185) at (0, 7) {$\frac{\pi}{4}$};
		\node [style=Z phase dot] (186) at (0, 6) {$\frac{\pi}{4}$};
		\node [style=Z phase dot] (187) at (0, 4) {$\frac{\pi}{4}$};
		\node [style=none] (190) at (1, 7) {};
		\node [style=none] (191) at (1, 6) {};
		\node [style=none] (192) at (1, 4) {};
		\node [style=Z phase dot] (194) at (0, 5) {$\frac{\pi}{4}$};
		\node [style=none] (195) at (1, 5) {};
		\node [style=none] (196) at (3.75, 5.5) {$\frac{e^{-i\pi/4}}{\sqrt{2}}$};
		\node [style=Z phase dot] (197) at (6.75, 5.5) {$\,-\frac{\pi}{2}\,$};
		\node [style=none] (199) at (7.75, 7) {};
		\node [style=none] (200) at (7.75, 6) {};
		\node [style=none] (201) at (7.75, 4) {};
		\node [style=Z dot] (203) at (5.25, 5.5) {};
		\node [style=none] (204) at (7.75, 5) {};
		\node [style=Z dot] (205) at (12.5, 5.5) {};
		\node [style=Z dot] (207) at (13.5, 7) {};
		\node [style=Z dot] (208) at (13.5, 6) {};
		\node [style=Z dot] (209) at (13.5, 4) {};
		\node [style=Z dot] (211) at (11.5, 5.5) {};
		\node [style=none] (212) at (8.5, 5.5) {$+$};
		\node [style=none] (213) at (10, 5.5) {$i$};
		\node [style=none] (215) at (14.25, 7) {};
		\node [style=none] (216) at (14.25, 6) {};
		\node [style=none] (217) at (14.25, 4) {};
		\node [style=Z dot] (219) at (13.5, 5) {};
		\node [style=none] (220) at (14.25, 5) {};
		\node [style=none] (221) at (2, 10) {$=$};
		\node [style=none] (222) at (3.75, 10) {$\frac{e^{-i\pi/4}}{\sqrt{2}}$};
		\node [style=Z dot] (223) at (12.5, 10) {};
		\node [style=Z dot] (225) at (13.5, 11) {};
		\node [style=Z dot] (226) at (13.5, 10) {};
		\node [style=Z dot] (227) at (13.5, 9) {};
		\node [style=Z dot] (229) at (11.5, 10) {};
		\node [style=none] (230) at (8.5, 10) {$+$};
		\node [style=none] (231) at (10, 10) {$i$};
		\node [style=none] (233) at (14.25, 11) {};
		\node [style=none] (234) at (14.25, 10) {};
		\node [style=none] (235) at (14.25, 9) {};
		\node [style=Z dot] (237) at (-1, 10) {};
		\node [style=Z phase dot] (239) at (0, 11) {$\frac{\pi}{4}$};
		\node [style=Z phase dot] (240) at (0, 10) {$\frac{\pi}{4}$};
		\node [style=Z phase dot] (241) at (0, 9) {$\frac{\pi}{4}$};
		\node [style=none] (244) at (1, 11) {};
		\node [style=none] (245) at (1, 10) {};
		\node [style=none] (246) at (1, 9) {};
		\node [style=Z phase dot] (248) at (6.75, 10) {$\,-\frac{\pi}{2}\,$};
		\node [style=none] (250) at (7.75, 11) {};
		\node [style=none] (251) at (7.75, 10) {};
		\node [style=none] (252) at (7.75, 9) {};
		\node [style=Z dot] (254) at (5.25, 10) {};
	\end{pgfonlayer}
	\begin{pgfonlayer}{edgelayer}
		\draw [style=hadamard edge] (18) to (21);
		\draw [style=hadamard edge, in=180, out=75, looseness=1.25] (18) to (20);
		\draw [style=hadamard edge, in=180, out=90] (18) to (19);
		\draw [style=hadamard edge, in=-180, out=-75, looseness=1.25] (18) to (22);
		\draw [style=hadamard edge, in=180, out=-90] (18) to (23);
		\draw (24) to (18);
		\draw (19) to (27.center);
		\draw (20) to (28.center);
		\draw (21) to (29.center);
		\draw (22) to (30.center);
		\draw (23) to (31.center);
		\draw [style=hadamard edge] (32) to (35);
		\draw [style=hadamard edge, in=180, out=75, looseness=1.25] (32) to (34);
		\draw [style=hadamard edge, in=180, out=90] (32) to (33);
		\draw [style=hadamard edge, in=-180, out=-75, looseness=1.25] (32) to (36);
		\draw [style=hadamard edge, in=180, out=-90] (32) to (37);
		\draw (38) to (32);
		\draw (33) to (41.center);
		\draw (34) to (42.center);
		\draw (35) to (43.center);
		\draw (36) to (44.center);
		\draw (37) to (45.center);
		\draw [style=hadamard edge] (46) to (49);
		\draw [style=hadamard edge, in=180, out=75, looseness=1.25] (46) to (48);
		\draw [style=hadamard edge, in=180, out=90] (46) to (47);
		\draw [style=hadamard edge, in=-180, out=-75, looseness=1.25] (46) to (50);
		\draw [style=hadamard edge, in=180, out=-90] (46) to (51);
		\draw (47) to (53.center);
		\draw (48) to (54.center);
		\draw (49) to (55.center);
		\draw (50) to (56.center);
		\draw (51) to (57.center);
		\draw (58) to (61.center);
		\draw [in=180, out=75, looseness=1.25] (58) to (60.center);
		\draw [in=180, out=90] (58) to (59.center);
		\draw [in=-180, out=-75, looseness=1.25] (58) to (62.center);
		\draw [in=180, out=-90] (58) to (63.center);
		\draw [style=hadamard edge] (64) to (58);
		\draw [style=hadamard edge, in=-180, out=60, looseness=1.25] (129) to (132);
		\draw [style=hadamard edge, in=180, out=75, looseness=1.25] (129) to (131);
		\draw [style=hadamard edge, in=180, out=90] (129) to (130);
		\draw [style=hadamard edge, in=-180, out=-75, looseness=1.25] (129) to (133);
		\draw [style=hadamard edge, in=180, out=-90] (129) to (134);
		\draw (135) to (129);
		\draw (130) to (138.center);
		\draw (131) to (139.center);
		\draw (132) to (140.center);
		\draw (133) to (141.center);
		\draw (134) to (142.center);
		\draw (143) to (144.center);
		\draw [style=hadamard edge, in=-180, out=-60, looseness=1.25] (129) to (143);
		\draw [style=hadamard edge, in=-180, out=60, looseness=1.25] (145) to (148);
		\draw [style=hadamard edge, in=180, out=75, looseness=1.25] (145) to (147);
		\draw [style=hadamard edge, in=180, out=90] (145) to (146);
		\draw [style=hadamard edge, in=-180, out=-75, looseness=1.25] (145) to (149);
		\draw [style=hadamard edge, in=180, out=-90] (145) to (150);
		\draw (151) to (145);
		\draw (146) to (153.center);
		\draw (147) to (154.center);
		\draw (148) to (155.center);
		\draw (149) to (156.center);
		\draw (150) to (157.center);
		\draw (159) to (160.center);
		\draw [style=hadamard edge, in=180, out=-60, looseness=1.25] (145) to (159);
		\draw [style=hadamard edge, in=-180, out=60, looseness=1.25] (161) to (164);
		\draw [style=hadamard edge, in=180, out=75, looseness=1.25] (161) to (163);
		\draw [style=hadamard edge, in=180, out=90] (161) to (162);
		\draw [style=hadamard edge, in=-180, out=-75, looseness=1.25] (161) to (165);
		\draw [style=hadamard edge, in=180, out=-90] (161) to (166);
		\draw (162) to (167.center);
		\draw (163) to (168.center);
		\draw (164) to (169.center);
		\draw (165) to (170.center);
		\draw (166) to (171.center);
		\draw (172) to (173.center);
		\draw [style=hadamard edge, in=180, out=-60, looseness=1.25] (161) to (172);
		\draw [in=-180, out=60, looseness=1.25] (174) to (177.center);
		\draw [in=180, out=75, looseness=1.25] (174) to (176.center);
		\draw [in=180, out=90] (174) to (175.center);
		\draw [in=-180, out=-75, looseness=1.25] (174) to (178.center);
		\draw [in=180, out=-90] (174) to (179.center);
		\draw (180) to (174);
		\draw [in=180, out=-60, looseness=1.25] (174) to (181.center);
		\draw [style=hadamard edge, in=-180, out=60, looseness=1.25] (183) to (186);
		\draw [style=hadamard edge, in=180, out=90, looseness=1.25] (183) to (185);
		\draw [style=hadamard edge, in=-180, out=-90, looseness=1.25] (183) to (187);
		\draw (185) to (190.center);
		\draw (186) to (191.center);
		\draw (187) to (192.center);
		\draw (194) to (195.center);
		\draw [style=hadamard edge, in=180, out=-60, looseness=1.25] (183) to (194);
		\draw [in=-180, out=60, looseness=1.25] (197) to (200.center);
		\draw [in=180, out=90, looseness=1.25] (197) to (199.center);
		\draw [in=-180, out=-90, looseness=1.25] (197) to (201.center);
		\draw (203) to (197);
		\draw [in=180, out=-60, looseness=1.25] (197) to (204.center);
		\draw [style=hadamard edge, in=-180, out=60, looseness=1.25] (205) to (208);
		\draw [style=hadamard edge, in=180, out=90, looseness=1.25] (205) to (207);
		\draw [style=hadamard edge, in=-180, out=-90, looseness=1.25] (205) to (209);
		\draw (211) to (205);
		\draw (207) to (215.center);
		\draw (208) to (216.center);
		\draw (209) to (217.center);
		\draw (219) to (220.center);
		\draw [style=hadamard edge, in=-180, out=-60, looseness=1.25] (205) to (219);
		\draw [style=hadamard edge] (223) to (226);
		\draw [style=hadamard edge, in=180, out=90, looseness=1.25] (223) to (225);
		\draw [style=hadamard edge, in=-180, out=-90, looseness=1.25] (223) to (227);
		\draw (229) to (223);
		\draw (225) to (233.center);
		\draw (226) to (234.center);
		\draw (227) to (235.center);
		\draw [style=hadamard edge] (237) to (240);
		\draw [style=hadamard edge, in=180, out=90, looseness=1.25] (237) to (239);
		\draw [style=hadamard edge, in=-180, out=-90, looseness=1.25] (237) to (241);
		\draw (239) to (244.center);
		\draw (240) to (245.center);
		\draw (241) to (246.center);
		\draw (248) to (251.center);
		\draw [in=180, out=90, looseness=1.25] (248) to (250.center);
		\draw [in=-180, out=-90, looseness=1.25] (248) to (252.center);
		\draw [style=hadamard edge] (254) to (248);
	\end{pgfonlayer}
\end{tikzpicture}

%% file: figures/magic-state-5-decomp.tikz
\begin{tikzpicture}[scale=2]
	\begin{pgfonlayer}{nodelayer}
		\node [style=none] (98) at (0.05, 0) {$=$};
		\node [style=none] (134) at (2.062, -1.525) {};
		\node [style=none] (135) at (2.062, -0.9) {};
		\node [style=Z phase dot] (136) at (1.175, 0.025) {$~-\frac\pi2~$};
		\node [style=none] (137) at (2.062, -0.275) {};
		\node [style=none] (138) at (2.062, 0.35) {};
		\node [style=none] (139) at (2.062, 0.975) {};
		\node [style=Z phase dot] (140) at (2.062, 1.6) {$\ -\frac\pi4~$};
		\node [style=none] (141) at (3.3, 0) {$+~2\sqrt{2}i e^{i\pi/4}$};
		\node [style=Z dot] (142) at (5.312, -1.5) {};
		\node [style=Z dot] (143) at (5.312, -0.875) {};
		\node [style=Z dot] (144) at (4.675, 0.05) {};
		\node [style=Z dot] (145) at (5.312, -0.25) {};
		\node [style=Z dot] (146) at (5.312, 0.375) {};
		\node [style=Z dot] (147) at (5.312, 1) {};
		\node [style=Z phase dot] (148) at (5.312, 1.625) {$\ -\frac\pi4~$};
		\node [style=none] (149) at (6.975, 0) {$-\ 2\sqrt{2} e^{i\pi/4}$};
		\node [style=none] (150) at (5.687, -1.5) {};
		\node [style=none] (151) at (5.687, -0.875) {};
		\node [style=none] (152) at (5.687, -0.25) {};
		\node [style=none] (153) at (5.687, 0.375) {};
		\node [style=none] (154) at (5.687, 1) {};
		\node [style=none] (155) at (0.55, 0) {$2$};
		\node [style=Z phase dot] (156) at (8.987, -1.5) {$\frac\pi2$};
		\node [style=Z phase dot] (157) at (8.987, -0.875) {$\frac\pi2$};
		\node [style=Z dot] (158) at (8.1, 0.05) {};
		\node [style=Z phase dot] (159) at (8.987, -0.25) {$\frac\pi2$};
		\node [style=Z phase dot] (160) at (8.987, 0.375) {$\frac\pi2$};
		\node [style=Z phase dot] (161) at (8.987, 1) {$\frac\pi2$};
		\node [style=Z phase dot] (162) at (8.987, 1.625) {$\frac\pi4$};
		\node [style=none] (163) at (9.487, -1.5) {};
		\node [style=none] (164) at (9.487, -0.875) {};
		\node [style=none] (165) at (9.487, -0.25) {};
		\node [style=none] (166) at (9.487, 0.375) {};
		\node [style=none] (167) at (9.487, 1) {};
		\node [style=Z phase dot] (168) at (-1.013, -1.25) {$\frac\pi4$};
		\node [style=Z phase dot] (169) at (-1.013, -0.625) {$\frac\pi4$};
		\node [style=Z phase dot] (170) at (-1.013, 0) {$\frac\pi4$};
		\node [style=Z phase dot] (171) at (-1.013, 0.625) {$\frac\pi4$};
		\node [style=Z phase dot] (172) at (-1.013, 1.25) {$\frac\pi4$};
		\node [style=none] (173) at (-0.513, -1.25) {};
		\node [style=none] (174) at (-0.513, -0.625) {};
		\node [style=none] (175) at (-0.513, 0) {};
		\node [style=none] (176) at (-0.513, 0.625) {};
		\node [style=none] (177) at (-0.513, 1.25) {};
	\end{pgfonlayer}
	\begin{pgfonlayer}{edgelayer}
		\draw [in=165, out=-90] (136) to (134.center);
		\draw [in=165, out=-75] (136) to (135.center);
		\draw [in=180, out=-45] (136) to (137.center);
		\draw [in=45, out=180] (138.center) to (136);
		\draw [in=-165, out=75] (136) to (139.center);
		\draw [in=-165, out=90] (136) to (140);
		\draw [style=hadamard edge, in=165, out=-90] (144) to (142);
		\draw [style=hadamard edge, in=165, out=-75] (144) to (143);
		\draw [style=hadamard edge, in=180, out=-45] (144) to (145);
		\draw [style=hadamard edge, in=45, out=180] (146) to (144);
		\draw [style=hadamard edge, in=-165, out=75] (144) to (147);
		\draw [style=hadamard edge, in=-165, out=90] (144) to (148);
		\draw (142) to (150.center);
		\draw (143) to (151.center);
		\draw (145) to (152.center);
		\draw (146) to (153.center);
		\draw (147) to (154.center);
		\draw [style=hadamard edge, in=165, out=-90] (158) to (156);
		\draw [style=hadamard edge, in=165, out=-75] (158) to (157);
		\draw [style=hadamard edge, in=180, out=-45] (158) to (159);
		\draw [style=hadamard edge, in=45, out=180] (160) to (158);
		\draw [style=hadamard edge, in=-165, out=75] (158) to (161);
		\draw [style=hadamard edge, in=-165, out=90] (158) to (162);
		\draw (156) to (163.center);
		\draw (157) to (164.center);
		\draw (159) to (165.center);
		\draw (160) to (166.center);
		\draw (161) to (167.center);
		\draw (168) to (173.center);
		\draw (169) to (174.center);
		\draw (170) to (175.center);
		\draw (171) to (176.center);
		\draw (172) to (177.center);
	\end{pgfonlayer}
\end{tikzpicture}

%% file: figures/ccz-box.tikz
\begin{tikzpicture}
	\begin{pgfonlayer}{nodelayer}
		\node [style=big gate] (0) at (0, 0) {CCZ};
		\node [style=none] (3) at (1.5, 0.75) {};
		\node [style=none] (4) at (1.5, 0) {};
		\node [style=none] (5) at (1.5, -0.75) {};
		\node [style=none] (6) at (-1.5, -0.75) {};
		\node [style=none] (7) at (-1.5, 0) {};
		\node [style=none] (8) at (-1.5, 0.75) {};
	\end{pgfonlayer}
	\begin{pgfonlayer}{edgelayer}
		\draw (4.center) to (7.center);
		\draw (6.center) to (5.center);
		\draw (8.center) to (3.center);
	\end{pgfonlayer}
\end{tikzpicture}

%% file: figures/ccz.tikz
\begin{tikzpicture}
	\begin{pgfonlayer}{nodelayer}
		\node [style=none] (0) at (-1, 0) {};
		\node [style=none] (1) at (-1, 1.5) {};
		\node [style=none] (2) at (-1, -1.5) {};
		\node [style=cnot ctrl] (3) at (0, 0) {};
		\node [style=cnot ctrl] (4) at (0, 1.5) {};
		\node [style=cnot ctrl] (5) at (0, -1.5) {};
		\node [style=none] (6) at (1, 0) {};
		\node [style=none] (7) at (1, 1.5) {};
		\node [style=none] (8) at (1, -1.5) {};
	\end{pgfonlayer}
	\begin{pgfonlayer}{edgelayer}
		\draw (4) to (3);
		\draw (3) to (5);
		\draw (0.center) to (6.center);
		\draw (1.center) to (7.center);
		\draw (2.center) to (8.center);
	\end{pgfonlayer}
\end{tikzpicture}

%% file: figures/7t-ccz.tikz
\begin{tikzpicture}
	\begin{pgfonlayer}{nodelayer}
		\node [style=Z phase dot] (3) at (0, -2) {$\frac\pi4$};
		\node [style=none] (4) at (-2, 2) {};
		\node [style=none] (5) at (-2, 0) {};
		\node [style=none] (6) at (-2, -2) {};
		\node [style=none] (7) at (2, 2) {};
		\node [style=none] (8) at (2, 0) {};
		\node [style=none] (9) at (2, -2) {};
		\node [style=Z dot] (10) at (1, -1) {};
		\node [style=Z dot] (11) at (-1, 1) {};
		\node [style=Z dot] (12) at (1, 1) {};
		\node [style=Z phase dot] (13) at (-2, 1) {-$\frac\pi4$};
		\node [style=Z phase dot] (14) at (2, -1) {-$\frac\pi4$};
		\node [style=Z phase dot] (15) at (2, 1) {$\frac\pi4$};
		\node [style=Z phase dot] (16) at (0, 0) {$\frac\pi4$};
		\node [style=Z phase dot] (17) at (0, 2) {$\frac\pi4$};
		\node [style=Z dot] (18) at (-1, -1) {};
		\node [style=Z phase dot] (19) at (-2, -1) {-$\frac\pi4$};
	\end{pgfonlayer}
	\begin{pgfonlayer}{edgelayer}
		\draw (6.center) to (3);
		\draw (3) to (9.center);
		\draw [style=hadamard edge] (10) to (14);
		\draw [style=hadamard edge] (11) to (13);
		\draw [style=hadamard edge] (12) to (15);
		\draw (5.center) to (16);
		\draw (16) to (8.center);
		\draw [style=hadamard edge] (16) to (12);
		\draw [style=hadamard edge] (16) to (10);
		\draw (4.center) to (17);
		\draw (17) to (7.center);
		\draw [style=hadamard edge] (17) to (12);
		\draw [style=hadamard edge] (17) to (11);
		\draw [style=hadamard edge] (18) to (19);
		\draw [style=hadamard edge] (12) to (3);
		\draw [style=hadamard edge] (3) to (10);
		\draw [style=hadamard edge] (11) to (16);
		\draw [style=hadamard edge] (18) to (3);
		\draw [style=hadamard edge] (18) to (17);
	\end{pgfonlayer}
\end{tikzpicture}

%% file: figures/cswap-box.tikz
\begin{tikzpicture}
	\begin{pgfonlayer}{nodelayer}
		\node [style=big gate] (0) at (0, 0) {CSwap};
		\node [style=none] (3) at (2.5, 0.75) {};
		\node [style=none] (4) at (2.5, 0) {};
		\node [style=none] (5) at (2.5, -0.75) {};
		\node [style=none] (6) at (-2.5, -0.75) {};
		\node [style=none] (7) at (-2.5, 0) {};
		\node [style=none] (8) at (-2.5, 0.75) {};
	\end{pgfonlayer}
	\begin{pgfonlayer}{edgelayer}
		\draw (4.center) to (7.center);
		\draw (6.center) to (5.center);
		\draw (8.center) to (3.center);
	\end{pgfonlayer}
\end{tikzpicture}

%% file: figures/cswap.tikz
\begin{tikzpicture}
	\begin{pgfonlayer}{nodelayer}
		\node [style=big gate] (0) at (0, 0) {CCZ};
		\node [style=hadamard] (1) at (-1.5, -0.75) {};
		\node [style=hadamard] (2) at (1.5, -0.75) {};
		\node [style=Z dot] (3) at (-2.5, -0.75) {};
		\node [style=Z dot] (4) at (2.5, -0.75) {};
		\node [style=X dot] (5) at (2.5, 0) {};
		\node [style=X dot] (6) at (-2.5, 0) {};
		\node [style=none] (7) at (-3.5, 0.75) {};
		\node [style=none] (8) at (-3.5, 0) {};
		\node [style=none] (9) at (-3.5, -0.75) {};
		\node [style=none] (10) at (3.5, 0.75) {};
		\node [style=none] (11) at (3.5, 0) {};
		\node [style=none] (12) at (3.5, -0.75) {};
	\end{pgfonlayer}
	\begin{pgfonlayer}{edgelayer}
		\draw (7.center) to (10.center);
		\draw (11.center) to (8.center);
		\draw (9.center) to (12.center);
		\draw (5) to (4);
		\draw (6) to (3);
	\end{pgfonlayer}
\end{tikzpicture}

%% file: figures/lone-phase-deriv.tikz
\begin{tikzpicture}
	\begin{pgfonlayer}{nodelayer}
		\node [style=Z phase dot] (0) at (0, 0) {$\frac{\pi}{4}$};
		\node [style=Z phase dot] (1) at (-1, 1.5) {$\frac{\pi}{4}$};
		\node [style=Z phase dot] (2) at (1, 1.5) {$\frac{\pi}{4}$};
		\node [style=none] (3) at (0, 1.5) {$\cdots$};
		\node [style=none] (4) at (-1, -1.5) {};
		\node [style=none] (5) at (1, -1.5) {};
		\node [style=none] (6) at (0, -1.25) {$\cdots$};
		\node [style=none] (7) at (3, 0) {$\approx$};
		\node [style=none] (9) at (3, 1) {\CutRule};
		\node [style=none] (10) at (5, -0.25) {$\sum\limits_{a\in\{0,1\}}$};
		\node [style=none] (11) at (6.75, 0.25) {$e^{ia\frac{\pi}{4}}$};
		\node [style=Z phase dot] (13) at (8.5, 1.5) {$\frac{\pi}{4}$};
		\node [style=Z phase dot] (14) at (12, 1.5) {$\frac{\pi}{4}$};
		\node [style=none] (15) at (10.25, 1.5) {$\cdots$};
		\node [style=none] (16) at (8.5, -1.5) {};
		\node [style=none] (17) at (12, -1.5) {};
		\node [style=none] (18) at (10.25, -1.25) {$\cdots$};
		\node [style=Z phase dot] (19) at (9.25, -0.75) {$a\pi$};
		\node [style=Z phase dot] (20) at (11.25, -0.75) {$a\pi$};
		\node [style=none] (21) at (-1, 2.25) {\textcolor{lightgray}{\footnotesize $x_1$}};
		\node [style=none] (22) at (1, 2.25) {\textcolor{lightgray}{\footnotesize $x_n$}};
		\node [style=none] (23) at (8.5, 2.25) {\textcolor{lightgray}{\footnotesize $x_1$}};
		\node [style=none] (24) at (12, 2.25) {\textcolor{lightgray}{\footnotesize $x_n$}};
		\node [style=none] (25) at (3, -10.5) {$\approx$};
		\node [style=none] (26) at (3, -9.5) {\ScalarRuleA};
		\node [style=none] (27) at (5, -10.75) {$\sum\limits_{a\in\{0,1\}}$};
		\node [style=none] (28) at (9.25, -9.5) {$e^{ia\frac{\pi}{4}}\left(1+e^{i\pi(\frac{1}{4}+a)}\right)^n$};
		\node [style=none] (32) at (8, -12) {};
		\node [style=none] (33) at (10.5, -12) {};
		\node [style=none] (34) at (9.25, -11.75) {$\cdots$};
		\node [style=Z phase dot] (35) at (8, -11) {$a\pi$};
		\node [style=Z phase dot] (36) at (10.5, -11) {$a\pi$};
		\node [style=none] (37) at (3, -5.75) {$\approx$};
		\node [style=none] (38) at (3, -4.75) {\FusionRule};
		\node [style=none] (39) at (5, -6) {$\sum\limits_{a\in\{0,1\}}$};
		\node [style=none] (40) at (6.75, -5.5) {$e^{ia\frac{\pi}{4}}$};
		\node [style=Z phase dot] (41) at (8.5, -4.25) {$\frac{\pi}{4}+a\pi$};
		\node [style=Z phase dot] (42) at (12, -4.25) {$\frac{\pi}{4}+a\pi$};
		\node [style=none] (43) at (10.25, -4.25) {$\cdots$};
		\node [style=none] (44) at (8.5, -7.25) {};
		\node [style=none] (45) at (12, -7.25) {};
		\node [style=none] (46) at (10.25, -7) {$\cdots$};
		\node [style=Z phase dot] (47) at (8.5, -6.25) {$a\pi$};
		\node [style=Z phase dot] (48) at (12, -6.25) {$a\pi$};
		\node [style=none] (49) at (8.5, -3.5) {\textcolor{lightgray}{\footnotesize $x_1$}};
		\node [style=none] (50) at (12, -3.5) {\textcolor{lightgray}{\footnotesize $x_n$}};
		\node [style=Z phase dot] (51) at (9.25, 0.5) {$a\pi$};
		\node [style=Z phase dot] (52) at (11.25, 0.5) {$a\pi$};
	\end{pgfonlayer}
	\begin{pgfonlayer}{edgelayer}
		\draw [style=hadamard edge] (2) to (0);
		\draw [style=hadamard edge] (0) to (1);
		\draw (0) to (4.center);
		\draw (0) to (5.center);
		\draw [style=hadamard edge] (19) to (16.center);
		\draw [style=hadamard edge] (20) to (17.center);
		\draw [style=hadamard edge] (35) to (32.center);
		\draw [style=hadamard edge] (36) to (33.center);
		\draw [style=hadamard edge] (47) to (44.center);
		\draw [style=hadamard edge] (48) to (45.center);
		\draw (51) to (13);
		\draw (52) to (14);
	\end{pgfonlayer}
\end{tikzpicture}

%% file: figures/multi-cat3-deriv.tikz
\begin{tikzpicture}
	\begin{pgfonlayer}{nodelayer}
		\node [style=Z dot] (0) at (-5.25, 1.25) {};
		\node [style=Z dot] (1) at (-3.25, 1.25) {};
		\node [style=Z phase dot] (2) at (-5.25, 2.25) {$\frac{\pi}{4}$};
		\node [style=Z phase dot] (3) at (-3.25, 2.25) {$\frac{\pi}{4}$};
		\node [style=Z phase dot] (4) at (-6.25, 0) {$\frac{\pi}{4}$};
		\node [style=Z phase dot] (5) at (-4.25, 0) {$\frac{\pi}{4}$};
		\node [style=Z phase dot] (6) at (-2.25, 0) {$\frac{\pi}{4}$};
		\node [style=none] (7) at (-6.75, -1) {};
		\node [style=none] (8) at (-6.25, -0.75) {$...$};
		\node [style=none] (9) at (-5.75, -1) {};
		\node [style=none] (10) at (-4.75, -1) {};
		\node [style=none] (11) at (-4.25, -0.75) {$...$};
		\node [style=none] (12) at (-3.75, -1) {};
		\node [style=none] (13) at (-2.75, -1) {};
		\node [style=none] (14) at (-2.25, -0.75) {$...$};
		\node [style=none] (15) at (-1.75, -1) {};
		\node [style=Z dot] (20) at (-7.25, 1.25) {};
		\node [style=Z phase dot] (21) at (-7.25, 2.25) {$\frac{\pi}{4}$};
		\node [style=Z phase dot] (22) at (-8.25, 0) {$\frac{\pi}{4}$};
		\node [style=none] (23) at (-8.75, -1) {};
		\node [style=none] (24) at (-8.25, -0.75) {$...$};
		\node [style=none] (25) at (-7.75, -1) {};
		\node [style=none] (53) at (0, 0.75) {$\approx$};
		\node [style=none] (86) at (2, 0.75) {$\mathlarger{\sum}$};
		\node [style=none] (88) at (2, -0.25) {$\scriptstyle{a\in\{0,1\}}$};
		\node [style=none] (89) at (-4.25, 2) {$...$};
		\node [style=none] (93) at (-3.25, 0) {$...$};
		\node [style=none] (94) at (0, -16) {$\approx$};
		\node [style=Z phase dot] (95) at (8, -16) {$(1-a)\frac{\pi}{2}$};
		\node [style=Z phase dot] (96) at (11, -16) {$(1-a)\frac{\pi}{2}$};
		\node [style=Z phase dot] (97) at (14.5, -16) {$(1-a)\frac{\pi}{2}$};
		\node [style=none] (98) at (7.5, -17) {};
		\node [style=none] (99) at (8, -16.75) {$...$};
		\node [style=none] (100) at (8.5, -17) {};
		\node [style=none] (101) at (10.5, -17) {};
		\node [style=none] (102) at (11, -16.75) {$...$};
		\node [style=none] (103) at (11.5, -17) {};
		\node [style=none] (104) at (14, -17) {};
		\node [style=none] (105) at (14.5, -16.75) {$...$};
		\node [style=none] (106) at (15, -17) {};
		\node [style=none] (107) at (4.5, -17) {};
		\node [style=none] (108) at (5.25, -16.75) {$...$};
		\node [style=none] (109) at (6, -17) {};
		\node [style=Z phase dot] (110) at (4.5, -16) {$a\pi$};
		\node [style=Z phase dot] (111) at (6, -16) {$a\pi$};
		\node [style=none] (112) at (2, -16) {$\mathlarger{\sum}$};
		\node [style=none] (113) at (5.25, -14.75) {$e^{ia(n+1)\frac{\pi}{4}}$};
		\node [style=none] (114) at (2, -17) {$\scriptstyle{a\in\{0,1\}}$};
		\node [style=none] (115) at (12.75, -16) {$...$};
		\node [style=Z dot] (116) at (9.75, 1.75) {};
		\node [style=Z dot] (117) at (11.75, 1.75) {};
		\node [style=Z phase dot] (118) at (9.75, 2.75) {$\frac{\pi}{4}$};
		\node [style=Z phase dot] (119) at (11.75, 2.75) {$\frac{\pi}{4}$};
		\node [style=Z phase dot] (120) at (8.75, 0) {$\frac{\pi}{4}$};
		\node [style=Z phase dot] (121) at (10.75, 0) {$\frac{\pi}{4}$};
		\node [style=Z phase dot] (122) at (12.75, 0) {$\frac{\pi}{4}$};
		\node [style=none] (123) at (8.25, -1) {};
		\node [style=none] (124) at (8.75, -0.75) {$...$};
		\node [style=none] (125) at (9.25, -1) {};
		\node [style=none] (126) at (10.25, -1) {};
		\node [style=none] (127) at (10.75, -0.75) {$...$};
		\node [style=none] (128) at (11.25, -1) {};
		\node [style=none] (129) at (12.25, -1) {};
		\node [style=none] (130) at (12.75, -0.75) {$...$};
		\node [style=none] (131) at (13.25, -1) {};
		\node [style=Z dot] (132) at (7.75, 1.75) {};
		\node [style=Z phase dot] (133) at (7.75, 2.75) {$\frac{\pi}{4}$};
		\node [style=none] (138) at (10.75, 2.5) {$...$};
		\node [style=none] (139) at (11.75, 0) {$...$};
		\node [style=none] (140) at (5.5, -1) {};
		\node [style=none] (141) at (6.25, -0.75) {$...$};
		\node [style=none] (142) at (7, -1) {};
		\node [style=Z phase dot] (143) at (5.5, 0) {$a\pi$};
		\node [style=Z phase dot] (144) at (7, 0) {$a\pi$};
		\node [style=Z phase dot] (145) at (9, 1) {$a\pi$};
		\node [style=Z phase dot] (146) at (11, 1) {$a\pi$};
		\node [style=Z phase dot] (147) at (7, 1) {$a\pi$};
		\node [style=none] (148) at (0, -5) {$\approx$};
		\node [style=none] (149) at (2, -5) {$\mathlarger{\sum}$};
		\node [style=none] (150) at (2, -6) {$\scriptstyle{a\in\{0,1\}}$};
		\node [style=Z phase dot] (151) at (9.75, -4.5) {$a\pi$};
		\node [style=Z phase dot] (152) at (11.75, -4.5) {$a\pi$};
		\node [style=Z phase dot] (153) at (9.75, -3.25) {$\frac{\pi}{4}$};
		\node [style=Z phase dot] (154) at (11.75, -3.25) {$\frac{\pi}{4}$};
		\node [style=Z phase dot] (155) at (8.75, -5.75) {$\frac{\pi}{4}$};
		\node [style=Z phase dot] (156) at (10.75, -5.75) {$\frac{\pi}{4}$};
		\node [style=Z phase dot] (157) at (12.75, -5.75) {$\frac{\pi}{4}$};
		\node [style=none] (158) at (8.25, -6.75) {};
		\node [style=none] (159) at (8.75, -6.5) {$...$};
		\node [style=none] (160) at (9.25, -6.75) {};
		\node [style=none] (161) at (10.25, -6.75) {};
		\node [style=none] (162) at (10.75, -6.5) {$...$};
		\node [style=none] (163) at (11.25, -6.75) {};
		\node [style=none] (164) at (12.25, -6.75) {};
		\node [style=none] (165) at (12.75, -6.5) {$...$};
		\node [style=none] (166) at (13.25, -6.75) {};
		\node [style=Z phase dot] (167) at (7.75, -4.5) {$a\pi$};
		\node [style=Z phase dot] (168) at (7.75, -3.25) {$\frac{\pi}{4}$};
		\node [style=none] (169) at (10.75, -3.5) {$...$};
		\node [style=none] (170) at (11.75, -5.75) {$...$};
		\node [style=none] (171) at (5.5, -6.75) {};
		\node [style=none] (172) at (6.25, -6.5) {$...$};
		\node [style=none] (173) at (7, -6.75) {};
		\node [style=Z phase dot] (174) at (5.5, -5.75) {$a\pi$};
		\node [style=Z phase dot] (175) at (7, -5.75) {$a\pi$};
		\node [style=none] (179) at (0, 1.75) {\CutRule};
		\node [style=none] (180) at (0, -4) {\FusionRule};
		\node [style=none] (181) at (0, -10.75) {$\approx$};
		\node [style=none] (182) at (2, -10.75) {$\mathlarger{\sum}$};
		\node [style=none] (183) at (2, -11.75) {$\scriptstyle{a\in\{0,1\}}$};
		\node [style=Z phase dot] (186) at (10.5, -9.75) {$\frac{\pi}{4}-a\frac{\pi}{2}$};
		\node [style=Z phase dot] (187) at (13.5, -9.25) {$\frac{\pi}{4}-a\frac{\pi}{2}$};
		\node [style=Z phase dot] (188) at (9, -11.5) {$\frac{\pi}{4}$};
		\node [style=Z phase dot] (189) at (11.5, -11.5) {$\frac{\pi}{4}$};
		\node [style=Z phase dot] (190) at (14.5, -11.5) {$\frac{\pi}{4}$};
		\node [style=none] (191) at (8.5, -12.5) {};
		\node [style=none] (192) at (9, -12.25) {$...$};
		\node [style=none] (193) at (9.5, -12.5) {};
		\node [style=none] (194) at (11, -12.5) {};
		\node [style=none] (195) at (11.5, -12.25) {$...$};
		\node [style=none] (196) at (12, -12.5) {};
		\node [style=none] (197) at (14, -12.5) {};
		\node [style=none] (198) at (14.5, -12.25) {$...$};
		\node [style=none] (199) at (15, -12.5) {};
		\node [style=Z phase dot] (201) at (8, -10.25) {$\frac{\pi}{4}-a\frac{\pi}{2}$};
		\node [style=none] (203) at (13, -11.5) {$...$};
		\node [style=none] (204) at (5.5, -12.5) {};
		\node [style=none] (205) at (6.25, -12.25) {$...$};
		\node [style=none] (206) at (7, -12.5) {};
		\node [style=Z phase dot] (207) at (5.5, -11.5) {$a\pi$};
		\node [style=Z phase dot] (208) at (7, -11.5) {$a\pi$};
		\node [style=none] (209) at (0, -9.75) {\PiRule};
		\node [style=none] (210) at (0, -15) {\FusionRule};
		\node [style=none] (211) at (4, 0.75) {$e^{ia\frac{\pi}{4}}$};
		\node [style=none] (212) at (5, -10.25) {$e^{ia(n+1)\frac{\pi}{4}}$};
		\node [style=none] (214) at (-7.25, 3) {\textcolor{lightgray}{\footnotesize $x_1$}};
		\node [style=none] (215) at (-5.25, 3) {\textcolor{lightgray}{\footnotesize $x_2$}};
		\node [style=none] (216) at (-3.25, 3) {\textcolor{lightgray}{\footnotesize $x_n$}};
		\node [style=none] (217) at (-7, -0.25) {\textcolor{lightgray}{\footnotesize $y_1$}};
		\node [style=none] (218) at (-5, -0.25) {\textcolor{lightgray}{\footnotesize $y_2$}};
		\node [style=none] (219) at (-1.5, -0.25) {\textcolor{lightgray}{\footnotesize $y_n$}};
		\node [style=none] (220) at (-9, -0.25) {\textcolor{lightgray}{\footnotesize $v$}};
		\node [style=none] (221) at (7.75, 3.5) {\textcolor{lightgray}{\footnotesize $x_1$}};
		\node [style=none] (222) at (9.75, 3.5) {\textcolor{lightgray}{\footnotesize $x_2$}};
		\node [style=none] (223) at (11.75, 3.5) {\textcolor{lightgray}{\footnotesize $x_n$}};
		\node [style=none] (224) at (8, -0.25) {\textcolor{lightgray}{\footnotesize $y_1$}};
		\node [style=none] (225) at (10, -0.25) {\textcolor{lightgray}{\footnotesize $y_2$}};
		\node [style=none] (226) at (13.5, -0.25) {\textcolor{lightgray}{\footnotesize $y_n$}};
		\node [style=none] (227) at (7.75, -2.5) {\textcolor{lightgray}{\footnotesize $x_1$}};
		\node [style=none] (228) at (9.75, -2.5) {\textcolor{lightgray}{\footnotesize $x_2$}};
		\node [style=none] (229) at (11.75, -2.5) {\textcolor{lightgray}{\footnotesize $x_n$}};
		\node [style=none] (230) at (8, -6.25) {\textcolor{lightgray}{\footnotesize $y_1$}};
		\node [style=none] (231) at (10, -6.25) {\textcolor{lightgray}{\footnotesize $y_2$}};
		\node [style=none] (232) at (13.5, -6.25) {\textcolor{lightgray}{\footnotesize $y_n$}};
		\node [style=none] (233) at (4, -5) {$e^{ia\frac{\pi}{4}}$};
		\node [style=none] (234) at (9.75, -12) {\textcolor{lightgray}{\footnotesize $y_1$}};
		\node [style=none] (235) at (12.25, -12) {\textcolor{lightgray}{\footnotesize $y_2$}};
		\node [style=none] (236) at (15.25, -12) {\textcolor{lightgray}{\footnotesize $y_n$}};
		\node [style=none] (237) at (8, -15.25) {\textcolor{lightgray}{\footnotesize $y_1$}};
		\node [style=none] (238) at (11, -15.25) {\textcolor{lightgray}{\footnotesize $y_2$}};
		\node [style=none] (239) at (14.5, -15.25) {\textcolor{lightgray}{\footnotesize $y_n$}};
		\node [style=none] (240) at (8, -9.5) {\textcolor{lightgray}{\footnotesize $x_1$}};
		\node [style=none] (241) at (10.5, -9) {\textcolor{lightgray}{\footnotesize $x_2$}};
		\node [style=none] (242) at (13.5, -8.5) {\textcolor{lightgray}{\footnotesize $x_n$}};
	\end{pgfonlayer}
	\begin{pgfonlayer}{edgelayer}
		\draw [style=hadamard edge] (2) to (0);
		\draw [style=hadamard edge] (0) to (5);
		\draw [style=hadamard edge] (3) to (1);
		\draw [style=hadamard edge] (1) to (6);
		\draw (4) to (7.center);
		\draw (4) to (9.center);
		\draw (5) to (10.center);
		\draw (5) to (12.center);
		\draw (6) to (13.center);
		\draw (6) to (15.center);
		\draw [style=hadamard edge] (21) to (20);
		\draw [style=hadamard edge] (4) to (20);
		\draw (22) to (23.center);
		\draw (22) to (25.center);
		\draw [style=hadamard edge] (20) to (22);
		\draw [style=hadamard edge] (22) to (0);
		\draw [style=hadamard edge] (22) to (1);
		\draw (95) to (98.center);
		\draw (95) to (100.center);
		\draw (96) to (101.center);
		\draw (96) to (103.center);
		\draw (97) to (104.center);
		\draw (97) to (106.center);
		\draw [style=hadamard edge] (107.center) to (110);
		\draw [style=hadamard edge] (109.center) to (111);
		\draw [style=hadamard edge] (118) to (116);
		\draw [style=hadamard edge] (116) to (121);
		\draw [style=hadamard edge] (119) to (117);
		\draw [style=hadamard edge] (117) to (122);
		\draw (120) to (123.center);
		\draw (120) to (125.center);
		\draw (121) to (126.center);
		\draw (121) to (128.center);
		\draw (122) to (129.center);
		\draw (122) to (131.center);
		\draw [style=hadamard edge] (133) to (132);
		\draw [style=hadamard edge] (120) to (132);
		\draw [style=hadamard edge] (140.center) to (143);
		\draw [style=hadamard edge] (142.center) to (144);
		\draw (132) to (147);
		\draw (116) to (145);
		\draw (117) to (146);
		\draw [style=hadamard edge] (153) to (151);
		\draw [style=hadamard edge] (151) to (156);
		\draw [style=hadamard edge] (154) to (152);
		\draw [style=hadamard edge] (152) to (157);
		\draw (155) to (158.center);
		\draw (155) to (160.center);
		\draw (156) to (161.center);
		\draw (156) to (163.center);
		\draw (157) to (164.center);
		\draw (157) to (166.center);
		\draw [style=hadamard edge] (168) to (167);
		\draw [style=hadamard edge] (155) to (167);
		\draw [style=hadamard edge] (171.center) to (174);
		\draw [style=hadamard edge] (173.center) to (175);
		\draw (188) to (191.center);
		\draw (188) to (193.center);
		\draw (189) to (194.center);
		\draw (189) to (196.center);
		\draw (190) to (197.center);
		\draw (190) to (199.center);
		\draw [style=hadamard edge] (204.center) to (207);
		\draw [style=hadamard edge] (206.center) to (208);
		\draw [bend right=15, looseness=0.75] (186) to (189);
		\draw [bend right=15, looseness=0.75] (187) to (190);
		\draw [bend right=15, looseness=0.75] (201) to (188);
	\end{pgfonlayer}
\end{tikzpicture}

%% file: figures/pair-eq-cut.tikz
\begin{tikzpicture}
	\begin{pgfonlayer}{nodelayer}
		\node [style=Z phase dot] (25) at (5.25, 7) {$\frac{\pi}{4}$};
		\node [style=Z phase dot] (26) at (8.25, 7) {$\frac{\pi}{4}$};
		\node [style=Z dot] (27) at (5.25, 9.75) {};
		\node [style=Z dot] (28) at (8.25, 9.75) {};
		\node [style=none] (29) at (6.75, 9.75) {$...$};
		\node [style=Z phase dot] (30) at (3.75, 11) {$\frac{\pi}{4}$};
		\node [style=Z phase dot] (31) at (5.75, 11) {$\frac{\pi}{4}$};
		\node [style=Z phase dot] (32) at (7.75, 11) {$\frac{\pi}{4}$};
		\node [style=Z phase dot] (33) at (9.75, 11) {$\frac{\pi}{4}$};
		\node [style=none] (34) at (6.75, 11) {$...$};
		\node [style=none] (35) at (2.75, 12) {};
		\node [style=none] (36) at (4.5, 12) {};
		\node [style=none] (37) at (3.75, 11.75) {$...$};
		\node [style=none] (38) at (9, 12) {};
		\node [style=none] (39) at (10.75, 12) {};
		\node [style=none] (40) at (9.75, 11.75) {$...$};
		\node [style=none] (41) at (7.75, 6) {};
		\node [style=none] (42) at (9.75, 6) {};
		\node [style=none] (43) at (8.5, 6.25) {$...$};
		\node [style=none] (44) at (3.75, 6) {};
		\node [style=none] (45) at (5.75, 6) {};
		\node [style=none] (46) at (5, 6.25) {$...$};
		\node [style=none] (49) at (6.75, 11.75) {$...$};
		\node [style=Z dot] (50) at (6.75, 8.75) {};
		\node [style=Z dot] (51) at (6.75, 7.75) {};
		\node [style=Z phase dot] (55) at (5.5, -2) {$\frac{\pi}{4}$};
		\node [style=Z phase dot] (56) at (8.5, -2) {$\frac{\pi}{4}$};
		\node [style=none] (59) at (7, 0.5) {$...$};
		\node [style=Z phase dot] (60) at (4, 1.75) {$\frac{\pi}{4}$};
		\node [style=Z phase dot] (61) at (6, 1.75) {$\frac{\pi}{4}$};
		\node [style=Z phase dot] (62) at (8, 1.75) {$\frac{\pi}{4}$};
		\node [style=Z phase dot] (63) at (10, 1.75) {$\frac{\pi}{4}$};
		\node [style=none] (64) at (7, 1.75) {$...$};
		\node [style=none] (65) at (3, 2.75) {};
		\node [style=none] (66) at (4.75, 2.75) {};
		\node [style=none] (67) at (4, 2.5) {$...$};
		\node [style=none] (68) at (9.25, 2.75) {};
		\node [style=none] (69) at (11, 2.75) {};
		\node [style=none] (70) at (10, 2.5) {$...$};
		\node [style=none] (71) at (8, -3) {};
		\node [style=none] (72) at (10, -3) {};
		\node [style=none] (73) at (8.75, -2.75) {$...$};
		\node [style=none] (74) at (4, -3) {};
		\node [style=none] (75) at (6, -3) {};
		\node [style=none] (76) at (5.25, -2.75) {$...$};
		\node [style=none] (79) at (7, 2.5) {$...$};
		\node [style=Z dot] (80) at (5.5, 0.5) {};
		\node [style=Z dot] (81) at (8.5, 0.5) {};
		\node [style=Z dot] (82) at (7, -1.25) {};
		\node [style=none] (84) at (0, -0.5) {$\approx$};
		\node [style=none] (85) at (0, 0.5) {\CutRule};
		\node [style=none] (86) at (1.75, -0.5) {$\mathlarger{\sum}$};
		\node [style=none] (88) at (1.75, -1.5) {$\scriptstyle{a\in\{0,1\}}$};
		\node [style=none] (115) at (0, -17.25) {$\approx$};
		\node [style=none] (116) at (0, -16.25) {\PiRule};
		\node [style=none] (117) at (1.75, -17.25) {$\mathlarger{\sum}$};
		\node [style=none] (118) at (1.75, -18.25) {$\scriptstyle{a\in\{0,1\}}$};
		\node [style=none] (119) at (4.25, 7) {\textcolor{lightgray}{\footnotesize $z_1$}};
		\node [style=none] (120) at (9.25, 7) {\textcolor{lightgray}{\footnotesize $z_2$}};
		\node [style=none] (121) at (7.5, 8.5) {\textcolor{lightgray}{\footnotesize $v$}};
		\node [style=none] (122) at (0, 8.75) {$\approx$};
		\node [style=Z phase dot] (126) at (-7.5, 7.25) {$\frac{\pi}{4}$};
		\node [style=Z phase dot] (127) at (-4.5, 7.25) {$\frac{\pi}{4}$};
		\node [style=Z dot] (128) at (-7.5, 9.25) {};
		\node [style=Z dot] (129) at (-4.5, 9.25) {};
		\node [style=none] (130) at (-6, 9.25) {$...$};
		\node [style=Z phase dot] (131) at (-9, 11.25) {$\frac{\pi}{4}$};
		\node [style=Z phase dot] (132) at (-7, 11.25) {$\frac{\pi}{4}$};
		\node [style=Z phase dot] (133) at (-5, 11.25) {$\frac{\pi}{4}$};
		\node [style=Z phase dot] (134) at (-3, 11.25) {$\frac{\pi}{4}$};
		\node [style=none] (135) at (-6, 11.25) {$...$};
		\node [style=none] (136) at (-10, 12.25) {};
		\node [style=none] (137) at (-8.25, 12.25) {};
		\node [style=none] (138) at (-9, 12) {$...$};
		\node [style=none] (139) at (-3.75, 12.25) {};
		\node [style=none] (140) at (-2, 12.25) {};
		\node [style=none] (141) at (-3, 12) {$...$};
		\node [style=none] (142) at (-5, 6.25) {};
		\node [style=none] (143) at (-3, 6.25) {};
		\node [style=none] (144) at (-4.25, 6.5) {$...$};
		\node [style=none] (145) at (-9, 6.25) {};
		\node [style=none] (146) at (-7, 6.25) {};
		\node [style=none] (147) at (-7.75, 6.5) {$...$};
		\node [style=none] (150) at (-6, 12) {$...$};
		\node [style=none] (152) at (0, 9.75) {\PivotRule};
		\node [style=none] (153) at (-8.5, 7.25) {\textcolor{lightgray}{\footnotesize $z_1$}};
		\node [style=none] (154) at (-3.5, 7.25) {\textcolor{lightgray}{\footnotesize $z_2$}};
		\node [style=none] (155) at (-9.5, 10.5) {\textcolor{lightgray}{\footnotesize $x_1$}};
		\node [style=none] (156) at (-2.5, 10.5) {\textcolor{lightgray}{\footnotesize $x_n$}};
		\node [style=none] (157) at (-5, 12) {\textcolor{lightgray}{\footnotesize $y_n$}};
		\node [style=none] (158) at (-7, 12) {\textcolor{lightgray}{\footnotesize $y_1$}};
		\node [style=none] (159) at (3.25, 10.25) {\textcolor{lightgray}{\footnotesize $x_1$}};
		\node [style=none] (160) at (10.25, 10.25) {\textcolor{lightgray}{\footnotesize $x_n$}};
		\node [style=none] (161) at (7.75, 11.75) {\textcolor{lightgray}{\footnotesize $y_n$}};
		\node [style=none] (162) at (5.75, 11.75) {\textcolor{lightgray}{\footnotesize $y_1$}};
		\node [style=none] (163) at (3.5, 1) {\textcolor{lightgray}{\footnotesize $x_1$}};
		\node [style=none] (164) at (10.5, 1) {\textcolor{lightgray}{\footnotesize $x_n$}};
		\node [style=none] (165) at (8, 2.5) {\textcolor{lightgray}{\footnotesize $y_n$}};
		\node [style=none] (166) at (6, 2.5) {\textcolor{lightgray}{\footnotesize $y_1$}};
		\node [style=none] (167) at (4.5, -2) {\textcolor{lightgray}{\footnotesize $z_1$}};
		\node [style=none] (168) at (9.5, -2) {\textcolor{lightgray}{\footnotesize $z_2$}};
		\node [style=Z phase dot] (173) at (7.5, -24.75) {$(1-a)\frac{\pi}{2}$};
		\node [style=Z phase dot] (174) at (11.5, -24.75) {$(1-a)\frac{\pi}{2}$};
		\node [style=none] (175) at (9.5, -24.75) {$...$};
		\node [style=none] (176) at (6.5, -23.75) {};
		\node [style=none] (177) at (8.25, -23.75) {};
		\node [style=none] (178) at (7.5, -24) {$...$};
		\node [style=none] (179) at (10.75, -23.75) {};
		\node [style=none] (180) at (12.5, -23.75) {};
		\node [style=none] (181) at (11.5, -24) {$...$};
		\node [style=none] (185) at (9.5, -24) {$...$};
		\node [style=none] (186) at (0, -25.75) {$\approx$};
		\node [style=none] (187) at (0, -24.75) {\FusionRule};
		\node [style=none] (188) at (1.75, -25.75) {$\mathlarger{\sum}$};
		\node [style=none] (189) at (1.75, -26.75) {$\scriptstyle{a\in\{0,1\}}$};
		\node [style=none] (190) at (6.5, -25.5) {\textcolor{lightgray}{\footnotesize $x_1$}};
		\node [style=none] (191) at (12.5, -25.5) {\textcolor{lightgray}{\footnotesize $x_n$}};
		\node [style=none] (192) at (4.25, -25.5) {$e^{ia(n+1)\frac{\pi}{4}}$};
		\node [style=Z phase dot] (196) at (5, -15) {$\frac{\pi}{4}$};
		\node [style=Z phase dot] (197) at (7, -15) {$\frac{\pi}{4}-a\frac{\pi}{2}$};
		\node [style=Z phase dot] (198) at (10, -15) {$\frac{\pi}{4}-a\frac{\pi}{2}$};
		\node [style=Z phase dot] (199) at (12, -15) {$\frac{\pi}{4}$};
		\node [style=none] (200) at (8.5, -15) {$...$};
		\node [style=none] (201) at (4, -14) {};
		\node [style=none] (202) at (5.75, -14) {};
		\node [style=none] (203) at (5, -14.25) {$...$};
		\node [style=none] (204) at (11.25, -14) {};
		\node [style=none] (205) at (13, -14) {};
		\node [style=none] (206) at (12, -14.25) {$...$};
		\node [style=none] (217) at (4.5, -15.75) {\textcolor{lightgray}{\footnotesize $x_1$}};
		\node [style=none] (218) at (12.5, -15.75) {\textcolor{lightgray}{\footnotesize $x_n$}};
		\node [style=none] (219) at (10, -14.25) {\textcolor{lightgray}{\footnotesize $y_n$}};
		\node [style=none] (220) at (7, -14.25) {\textcolor{lightgray}{\footnotesize $y_1$}};
		\node [style=none] (223) at (4.5, -17) {$e^{ia(n+1)\frac{\pi}{4}}$};
		\node [style=none] (224) at (6.25, -17.5) {\textcolor{lightgray}{\footnotesize $z_1$}};
		\node [style=none] (225) at (11.5, -17.5) {\textcolor{lightgray}{\footnotesize $z_2$}};
		\node [style=Z phase dot] (226) at (10, -18) {$\frac{\pi}{4}-a\frac{\pi}{2}$};
		\node [style=Z phase dot] (227) at (10.25, -19.25) {$a\pi$};
		\node [style=Z phase dot] (228) at (11.75, -19.25) {$a\pi$};
		\node [style=none] (229) at (10.5, -20.25) {};
		\node [style=none] (230) at (12.5, -20.25) {};
		\node [style=none] (231) at (11, -19.75) {$...$};
		\node [style=Z phase dot] (232) at (7, -18) {$\frac{\pi}{4}$};
		\node [style=none] (233) at (6, -19.75) {$...$};
		\node [style=none] (234) at (4.5, -20.25) {};
		\node [style=none] (235) at (6.5, -20.25) {};
		\node [style=Z phase dot] (236) at (9.5, -26.25) {$(1-a)\frac{\pi}{2}$};
		\node [style=Z phase dot] (237) at (10.25, -27.5) {$a\pi$};
		\node [style=none] (238) at (7.75, -28) {$...$};
		\node [style=Z phase dot] (239) at (11.75, -27.5) {$a\pi$};
		\node [style=none] (240) at (6.5, -28.5) {};
		\node [style=none] (241) at (8.5, -28.5) {};
		\node [style=none] (242) at (10.5, -28.5) {};
		\node [style=none] (243) at (12.5, -28.5) {};
		\node [style=none] (244) at (11.25, -28) {$...$};
		\node [style=none] (245) at (0, -9) {$\approx$};
		\node [style=none] (246) at (0, -8) {\FusionRule};
		\node [style=none] (247) at (1.75, -9) {$\mathlarger{\sum}$};
		\node [style=none] (248) at (1.75, -10) {$\scriptstyle{a\in\{0,1\}}$};
		\node [style=Z phase dot] (279) at (5.5, -0.5) {$a\pi$};
		\node [style=Z phase dot] (280) at (8.5, -0.5) {$a\pi$};
		\node [style=Z phase dot] (281) at (7, -0.25) {$a\pi$};
		\node [style=Z phase dot] (282) at (5.5, -10.5) {$\frac{\pi}{4}$};
		\node [style=Z phase dot] (283) at (8.5, -10.5) {$\frac{\pi}{4}$};
		\node [style=none] (284) at (7, -8) {$...$};
		\node [style=Z phase dot] (285) at (4, -6.75) {$\frac{\pi}{4}$};
		\node [style=Z phase dot] (286) at (6, -6.75) {$\frac{\pi}{4}$};
		\node [style=Z phase dot] (287) at (8, -6.75) {$\frac{\pi}{4}$};
		\node [style=Z phase dot] (288) at (10, -6.75) {$\frac{\pi}{4}$};
		\node [style=none] (289) at (7, -6.75) {$...$};
		\node [style=none] (290) at (3, -5.75) {};
		\node [style=none] (291) at (4.75, -5.75) {};
		\node [style=none] (292) at (4, -6) {$...$};
		\node [style=none] (293) at (9.25, -5.75) {};
		\node [style=none] (294) at (11, -5.75) {};
		\node [style=none] (295) at (10, -6) {$...$};
		\node [style=none] (296) at (8, -11.5) {};
		\node [style=none] (297) at (10, -11.5) {};
		\node [style=none] (298) at (8.75, -11.25) {$...$};
		\node [style=none] (299) at (4, -11.5) {};
		\node [style=none] (300) at (6, -11.5) {};
		\node [style=none] (301) at (5.25, -11.25) {$...$};
		\node [style=none] (302) at (7, -6) {$...$};
		\node [style=Z phase dot] (303) at (5.5, -8) {$a\pi$};
		\node [style=Z phase dot] (304) at (8.5, -8) {$a\pi$};
		\node [style=Z phase dot] (305) at (7, -9.75) {$a\pi$};
		\node [style=none] (306) at (3.5, -7.5) {\textcolor{lightgray}{\footnotesize $x_1$}};
		\node [style=none] (307) at (10.5, -7.5) {\textcolor{lightgray}{\footnotesize $x_n$}};
		\node [style=none] (308) at (8, -6) {\textcolor{lightgray}{\footnotesize $y_n$}};
		\node [style=none] (309) at (6, -6) {\textcolor{lightgray}{\footnotesize $y_1$}};
		\node [style=none] (310) at (4.5, -10.5) {\textcolor{lightgray}{\footnotesize $z_1$}};
		\node [style=none] (311) at (9.5, -10.5) {\textcolor{lightgray}{\footnotesize $z_2$}};
	\end{pgfonlayer}
	\begin{pgfonlayer}{edgelayer}
		\draw [style=hadamard edge] (27) to (30);
		\draw [style=hadamard edge] (27) to (31);
		\draw [style=hadamard edge] (32) to (28);
		\draw [style=hadamard edge] (28) to (33);
		\draw (36.center) to (30);
		\draw (30) to (35.center);
		\draw (38.center) to (33);
		\draw (33) to (39.center);
		\draw (25) to (44.center);
		\draw (25) to (45.center);
		\draw (26) to (41.center);
		\draw (26) to (42.center);
		\draw [style=hadamard edge] (27) to (50);
		\draw [style=hadamard edge] (50) to (28);
		\draw [style=hadamard edge] (51) to (50);
		\draw [style=hadamard edge] (51) to (25);
		\draw [style=hadamard edge] (51) to (26);
		\draw (66.center) to (60);
		\draw (60) to (65.center);
		\draw (68.center) to (63);
		\draw (63) to (69.center);
		\draw (55) to (74.center);
		\draw (55) to (75.center);
		\draw (56) to (71.center);
		\draw (56) to (72.center);
		\draw [style=hadamard edge] (82) to (55);
		\draw [style=hadamard edge] (82) to (56);
		\draw [style=hadamard edge] (60) to (80);
		\draw [style=hadamard edge] (80) to (61);
		\draw [style=hadamard edge] (62) to (81);
		\draw [style=hadamard edge] (81) to (63);
		\draw [style=hadamard edge] (129) to (126);
		\draw [style=hadamard edge] (128) to (127);
		\draw [style=hadamard edge] (127) to (129);
		\draw [style=hadamard edge] (128) to (126);
		\draw [style=hadamard edge] (128) to (131);
		\draw [style=hadamard edge] (128) to (132);
		\draw [style=hadamard edge] (133) to (129);
		\draw [style=hadamard edge] (129) to (134);
		\draw (137.center) to (131);
		\draw (131) to (136.center);
		\draw (139.center) to (134);
		\draw (134) to (140.center);
		\draw (126) to (145.center);
		\draw (126) to (146.center);
		\draw (127) to (142.center);
		\draw (127) to (143.center);
		\draw (177.center) to (173);
		\draw (173) to (176.center);
		\draw (179.center) to (174);
		\draw (174) to (180.center);
		\draw (202.center) to (196);
		\draw (196) to (201.center);
		\draw (204.center) to (199);
		\draw (199) to (205.center);
		\draw [bend right=60, looseness=1.25] (196) to (197);
		\draw [bend right=60, looseness=1.25] (198) to (199);
		\draw (232) to (226);
		\draw [style=hadamard edge] (227) to (229.center);
		\draw [style=hadamard edge] (228) to (230.center);
		\draw [style=hadamard edge] (226) to (227);
		\draw [style=hadamard edge, bend left=15] (226) to (228);
		\draw [bend right, looseness=0.75] (232) to (234.center);
		\draw (232) to (235.center);
		\draw [bend right] (236) to (240.center);
		\draw [bend right=15] (236) to (241.center);
		\draw [style=hadamard edge, bend left=15, looseness=0.75] (236) to (237);
		\draw [style=hadamard edge, bend left=15, looseness=0.50] (237) to (242.center);
		\draw [style=hadamard edge, bend left=15, looseness=0.75] (236) to (239);
		\draw [style=hadamard edge, bend left=15, looseness=0.75] (239) to (243.center);
		\draw (279) to (80);
		\draw (280) to (81);
		\draw (281) to (82);
		\draw (291.center) to (285);
		\draw (285) to (290.center);
		\draw (293.center) to (288);
		\draw (288) to (294.center);
		\draw (282) to (299.center);
		\draw (282) to (300.center);
		\draw (283) to (296.center);
		\draw (283) to (297.center);
		\draw [style=hadamard edge] (305) to (282);
		\draw [style=hadamard edge] (305) to (283);
		\draw [style=hadamard edge] (285) to (303);
		\draw [style=hadamard edge] (303) to (286);
		\draw [style=hadamard edge] (287) to (304);
		\draw [style=hadamard edge] (304) to (288);
	\end{pgfonlayer}
\end{tikzpicture}

%% file: figures/pair-eq-phase.tikz
\begin{tikzpicture}
	\begin{pgfonlayer}{nodelayer}
		\node [style=Z phase dot] (25) at (4.75, 1.75) {$\frac{\pi}{4}$};
		\node [style=Z phase dot] (26) at (7.75, 1.75) {$\frac{\pi}{4}$};
		\node [style=Z phase dot] (30) at (3.25, 5.25) {$\frac{\pi}{4}$};
		\node [style=Z phase dot] (31) at (5.25, 5.25) {$\frac{\pi}{4}$};
		\node [style=Z phase dot] (32) at (7.25, 5.25) {$\frac{\pi}{4}$};
		\node [style=Z phase dot] (33) at (9.25, 5.25) {$\frac{\pi}{4}$};
		\node [style=none] (34) at (6.25, 5.25) {$...$};
		\node [style=none] (41) at (7.25, 0.75) {};
		\node [style=none] (42) at (9.25, 0.75) {};
		\node [style=none] (43) at (8, 1) {$...$};
		\node [style=none] (44) at (3.25, 0.75) {};
		\node [style=none] (45) at (5.25, 0.75) {};
		\node [style=none] (46) at (4.5, 1) {$...$};
		\node [style=none] (47) at (3.75, 1.75) {\textcolor{lightgray}{\footnotesize $y_1$}};
		\node [style=none] (48) at (8.75, 1.75) {\textcolor{lightgray}{\footnotesize $y_2$}};
		\node [style=Z dot] (50) at (6.25, 3.5) {};
		\node [style=Z dot] (51) at (6.25, 2.5) {};
		\node [style=none] (54) at (7.25, 3.5) {\textcolor{lightgray}{\footnotesize $v$}};
		\node [style=none] (59) at (6.25, -2.5) {$...$};
		\node [style=Z phase dot] (60) at (3.25, -2) {$\frac{\pi}{4}$};
		\node [style=Z phase dot] (61) at (5.25, -2) {$\frac{\pi}{4}$};
		\node [style=Z phase dot] (62) at (7.25, -2) {$\frac{\pi}{4}$};
		\node [style=Z phase dot] (63) at (9.25, -2) {$\frac{\pi}{4}$};
		\node [style=Z phase dot] (80) at (4, -3.25) {$a\pi$};
		\node [style=Z phase dot] (81) at (7.25, -3.25) {$a\pi$};
		\node [style=none] (84) at (0, -3.75) {$\approx$};
		\node [style=none] (85) at (0, -2.75) {\CutRule};
		\node [style=none] (86) at (1.5, -3.75) {$\mathlarger{\sum}$};
		\node [style=none] (88) at (1.5, -4.75) {$\scriptstyle{a\in\{0,1\}}$};
		\node [style=Z phase dot] (119) at (-7.75, 2.75) {$\frac{\pi}{4}$};
		\node [style=Z phase dot] (120) at (-4.75, 2.75) {$\frac{\pi}{4}$};
		\node [style=Z phase dot] (124) at (-9.25, 5) {$\frac{\pi}{4}$};
		\node [style=Z phase dot] (125) at (-7.25, 5) {$\frac{\pi}{4}$};
		\node [style=Z phase dot] (126) at (-5.25, 5) {$\frac{\pi}{4}$};
		\node [style=Z phase dot] (127) at (-3.25, 5) {$\frac{\pi}{4}$};
		\node [style=none] (128) at (-6.25, 5) {$...$};
		\node [style=none] (135) at (-5.25, 1.75) {};
		\node [style=none] (136) at (-3.25, 1.75) {};
		\node [style=none] (137) at (-4.5, 2) {$...$};
		\node [style=none] (138) at (-9.25, 1.75) {};
		\node [style=none] (139) at (-7.25, 1.75) {};
		\node [style=none] (140) at (-8, 2) {$...$};
		\node [style=none] (141) at (-8.75, 2.75) {\textcolor{lightgray}{\footnotesize $y_1$}};
		\node [style=none] (142) at (-3.75, 2.75) {\textcolor{lightgray}{\footnotesize $y_2$}};
		\node [style=none] (143) at (0, 3.25) {$\approx$};
		\node [style=none] (144) at (0, 4.25) {\PivotRule};
		\node [style=Z phase dot] (145) at (5.25, -3.25) {$a\pi$};
		\node [style=Z phase dot] (146) at (8.75, -3.25) {$a\pi$};
		\node [style=none] (158) at (0, -30) {$\approx$};
		\node [style=none] (159) at (0, -29) {\FusionRule};
		\node [style=none] (160) at (2, -30) {$\mathlarger{\sum}$};
		\node [style=none] (161) at (2, -31) {$\scriptstyle{a\in\{0,1\}}$};
		\node [style=Z phase dot] (164) at (7, -30.5) {$(1-a)\frac{\pi}{2}$};
		\node [style=Z phase dot] (166) at (7.75, -31.75) {$a\pi$};
		\node [style=none] (167) at (5.25, -32.25) {$...$};
		\node [style=Z phase dot] (168) at (4.75, -5.25) {$\frac{\pi}{4}$};
		\node [style=Z phase dot] (169) at (7.75, -5.25) {$\frac{\pi}{4}$};
		\node [style=none] (170) at (7.25, -6.25) {};
		\node [style=none] (171) at (9.25, -6.25) {};
		\node [style=none] (172) at (8, -6) {$...$};
		\node [style=none] (173) at (3.25, -6.25) {};
		\node [style=none] (174) at (5.25, -6.25) {};
		\node [style=none] (175) at (4.5, -6) {$...$};
		\node [style=none] (176) at (3.75, -5.25) {\textcolor{lightgray}{\footnotesize $y_1$}};
		\node [style=none] (177) at (8.75, -5.25) {\textcolor{lightgray}{\footnotesize $y_2$}};
		\node [style=Z dot] (178) at (6.25, -5) {};
		\node [style=Z phase dot] (179) at (6.25, -4.25) {$a\pi$};
		\node [style=none] (180) at (7, -9.75) {$...$};
		\node [style=Z phase dot] (181) at (3, -9.75) {$\frac{\pi}{4}+a\pi$};
		\node [style=Z phase dot] (182) at (5.5, -9.75) {$\frac{\pi}{4}+a\pi$};
		\node [style=Z phase dot] (183) at (8.5, -9.75) {$\frac{\pi}{4}+a\pi$};
		\node [style=Z phase dot] (184) at (11, -9.75) {$\frac{\pi}{4}+a\pi$};
		\node [style=none] (187) at (0, -11.25) {$\approx$};
		\node [style=none] (188) at (0, -10.25) {\FusionRule};
		\node [style=none] (189) at (1.5, -11.25) {$\mathlarger{\sum}$};
		\node [style=none] (190) at (1.5, -12.25) {$\scriptstyle{a\in\{0,1\}}$};
		\node [style=Z phase dot] (193) at (5.5, -12) {$\frac{\pi}{4}$};
		\node [style=Z phase dot] (194) at (8.5, -12) {$\frac{\pi}{4}$};
		\node [style=none] (195) at (8, -13) {};
		\node [style=none] (196) at (10, -13) {};
		\node [style=none] (197) at (8.75, -12.75) {$...$};
		\node [style=none] (198) at (4, -13) {};
		\node [style=none] (199) at (6, -13) {};
		\node [style=none] (200) at (5.25, -12.75) {$...$};
		\node [style=none] (201) at (4.5, -12) {\textcolor{lightgray}{\footnotesize $y_1$}};
		\node [style=none] (202) at (9.5, -12) {\textcolor{lightgray}{\footnotesize $y_2$}};
		\node [style=Z phase dot] (203) at (7, -11.5) {$a\pi$};
		\node [style=none] (209) at (0, -17.5) {$\approx$};
		\node [style=none] (210) at (0, -16.5) {\ScalarRuleA};
		\node [style=none] (211) at (2, -17.5) {$\mathlarger{\sum}$};
		\node [style=none] (212) at (2, -18.5) {$\scriptstyle{a\in\{0,1\}}$};
		\node [style=none] (227) at (7, -16.5) {$\left(1+e^{i\pi(\frac{1}{4}+a)}\right)^n$};
		\node [style=none] (228) at (7, -29) {$e^{ia\frac{\pi}{4}} \left(1+e^{i\pi(\frac{1}{4}+a)}\right)^n$};
		\node [style=Z phase dot] (229) at (9.25, -31.75) {$a\pi$};
		\node [style=none] (231) at (4, -32.75) {};
		\node [style=none] (232) at (6, -32.75) {};
		\node [style=none] (233) at (8, -32.75) {};
		\node [style=none] (234) at (10, -32.75) {};
		\node [style=none] (235) at (8.75, -32.25) {$...$};
		\node [style=none] (247) at (-3.25, 5.75) {\textcolor{lightgray}{\footnotesize $x_n$}};
		\node [style=none] (248) at (-9.25, 5.75) {\textcolor{lightgray}{\footnotesize $x_1$}};
		\node [style=none] (249) at (-7.25, 5.75) {\textcolor{lightgray}{\footnotesize $x_2$}};
		\node [style=none] (250) at (-5.25, 5.75) {\textcolor{lightgray}{\footnotesize $x_{n-1}$}};
		\node [style=none] (251) at (9.25, 6) {\textcolor{lightgray}{\footnotesize $x_n$}};
		\node [style=none] (252) at (3.25, 6) {\textcolor{lightgray}{\footnotesize $x_1$}};
		\node [style=none] (253) at (5.25, 6) {\textcolor{lightgray}{\footnotesize $x_2$}};
		\node [style=none] (254) at (7.25, 6) {\textcolor{lightgray}{\footnotesize $x_{n-1}$}};
		\node [style=none] (255) at (9.25, -1.25) {\textcolor{lightgray}{\footnotesize $x_n$}};
		\node [style=none] (256) at (3.25, -1.25) {\textcolor{lightgray}{\footnotesize $x_1$}};
		\node [style=none] (257) at (5.25, -1.25) {\textcolor{lightgray}{\footnotesize $x_2$}};
		\node [style=none] (258) at (7.25, -1.25) {\textcolor{lightgray}{\footnotesize $x_{n-1}$}};
		\node [style=none] (259) at (11, -9) {\textcolor{lightgray}{\footnotesize $x_n$}};
		\node [style=none] (260) at (3, -9) {\textcolor{lightgray}{\footnotesize $x_1$}};
		\node [style=none] (261) at (5.5, -9) {\textcolor{lightgray}{\footnotesize $x_2$}};
		\node [style=none] (262) at (8.5, -9) {\textcolor{lightgray}{\footnotesize $x_{n-1}$}};
		\node [style=none] (264) at (0, -24) {$\approx$};
		\node [style=none] (265) at (0, -23) {\PiRule};
		\node [style=none] (266) at (2, -24) {$\mathlarger{\sum}$};
		\node [style=none] (267) at (2, -25) {$\scriptstyle{a\in\{0,1\}}$};
		\node [style=none] (268) at (5.25, -23.5) {\textcolor{lightgray}{\footnotesize $y_1$}};
		\node [style=none] (269) at (9.5, -23.5) {\textcolor{lightgray}{\footnotesize $y_2$}};
		\node [style=none] (270) at (7, -22.5) {$e^{ia\frac{\pi}{4}} \left(1+e^{i\pi(\frac{1}{4}+a)}\right)^n$};
		\node [style=Z phase dot] (271) at (8, -24) {$\frac{\pi}{4}-a\frac{\pi}{2}$};
		\node [style=Z phase dot] (272) at (8.25, -25.25) {$a\pi$};
		\node [style=Z phase dot] (273) at (9.75, -25.25) {$a\pi$};
		\node [style=none] (274) at (8.5, -26.25) {};
		\node [style=none] (275) at (10.5, -26.25) {};
		\node [style=none] (276) at (9, -25.75) {$...$};
		\node [style=Z phase dot] (277) at (6, -24) {$\frac{\pi}{4}$};
		\node [style=none] (278) at (5, -25.75) {$...$};
		\node [style=none] (279) at (3.5, -26.25) {};
		\node [style=none] (280) at (5.5, -26.25) {};
		\node [style=Z phase dot] (281) at (5.5, -18.25) {$\frac{\pi}{4}$};
		\node [style=Z phase dot] (282) at (8.5, -18.25) {$\frac{\pi}{4}$};
		\node [style=none] (283) at (8, -19.25) {};
		\node [style=none] (284) at (10, -19.25) {};
		\node [style=none] (285) at (8.75, -19) {$...$};
		\node [style=none] (286) at (4, -19.25) {};
		\node [style=none] (287) at (6, -19.25) {};
		\node [style=none] (288) at (5.25, -19) {$...$};
		\node [style=none] (289) at (4.5, -18.25) {\textcolor{lightgray}{\footnotesize $y_1$}};
		\node [style=none] (290) at (9.5, -18.25) {\textcolor{lightgray}{\footnotesize $y_2$}};
		\node [style=Z phase dot] (291) at (7, -17.75) {$a\pi$};
	\end{pgfonlayer}
	\begin{pgfonlayer}{edgelayer}
		\draw (25) to (44.center);
		\draw (25) to (45.center);
		\draw (26) to (41.center);
		\draw (26) to (42.center);
		\draw [style=hadamard edge] (51) to (50);
		\draw [style=hadamard edge] (51) to (25);
		\draw [style=hadamard edge] (51) to (26);
		\draw (60) to (80);
		\draw (62) to (81);
		\draw (119) to (138.center);
		\draw (119) to (139.center);
		\draw (120) to (135.center);
		\draw (120) to (136.center);
		\draw [style=hadamard edge] (124) to (120);
		\draw [style=hadamard edge] (125) to (119);
		\draw [style=hadamard edge] (126) to (119);
		\draw [style=hadamard edge] (127) to (119);
		\draw [style=hadamard edge] (124) to (119);
		\draw [style=hadamard edge] (125) to (120);
		\draw [style=hadamard edge] (126) to (120);
		\draw [style=hadamard edge] (127) to (120);
		\draw [style=hadamard edge] (30) to (50);
		\draw [style=hadamard edge] (31) to (50);
		\draw [style=hadamard edge] (32) to (50);
		\draw [style=hadamard edge] (33) to (50);
		\draw (61) to (145);
		\draw (63) to (146);
		\draw (168) to (173.center);
		\draw (168) to (174.center);
		\draw (169) to (170.center);
		\draw (169) to (171.center);
		\draw [style=hadamard edge] (178) to (168);
		\draw [style=hadamard edge] (178) to (169);
		\draw (179) to (178);
		\draw (193) to (198.center);
		\draw (193) to (199.center);
		\draw (194) to (195.center);
		\draw (194) to (196.center);
		\draw [style=hadamard edge] (203) to (193);
		\draw [style=hadamard edge] (203) to (194);
		\draw [bend right] (164) to (231.center);
		\draw [bend right=15] (164) to (232.center);
		\draw [style=hadamard edge, bend left=15, looseness=0.75] (164) to (166);
		\draw [style=hadamard edge, bend left=15, looseness=0.50] (166) to (233.center);
		\draw [style=hadamard edge, bend left=15, looseness=0.75] (164) to (229);
		\draw [style=hadamard edge, bend left=15, looseness=0.75] (229) to (234.center);
		\draw (277) to (271);
		\draw [style=hadamard edge] (272) to (274.center);
		\draw [style=hadamard edge] (273) to (275.center);
		\draw [style=hadamard edge] (271) to (272);
		\draw [style=hadamard edge, bend left=15] (271) to (273);
		\draw [bend right, looseness=0.75] (277) to (279.center);
		\draw (277) to (280.center);
		\draw (281) to (286.center);
		\draw (281) to (287.center);
		\draw (282) to (283.center);
		\draw (282) to (284.center);
		\draw [style=hadamard edge] (291) to (281);
		\draw [style=hadamard edge] (291) to (282);
	\end{pgfonlayer}
\end{tikzpicture}

%% file: figures/multi-cat3-general.tikz
\begin{tikzpicture}
	\begin{pgfonlayer}{nodelayer}
		\node [style=Z dot] (0) at (-8, 1.25) {};
		\node [style=Z dot] (1) at (-6, 1.25) {};
		\node [style=Z phase dot] (2) at (-8, 2.25) {$\frac{\pi}{4}$};
		\node [style=Z phase dot] (3) at (-6, 2.25) {$\frac{\pi}{4}$};
		\node [style=Z phase dot] (4) at (-9, 0) {$\frac{\pi}{4}$};
		\node [style=Z phase dot] (5) at (-7, 0) {$\frac{\pi}{4}$};
		\node [style=Z phase dot] (6) at (-5, 0) {$\frac{\pi}{4}$};
		\node [style=none] (7) at (-9.5, -1) {};
		\node [style=none] (8) at (-9, -0.75) {$...$};
		\node [style=none] (9) at (-8.5, -1) {};
		\node [style=none] (10) at (-7.5, -1) {};
		\node [style=none] (11) at (-7, -0.75) {$...$};
		\node [style=none] (12) at (-6.5, -1) {};
		\node [style=none] (13) at (-5.5, -1) {};
		\node [style=none] (14) at (-5, -0.75) {$...$};
		\node [style=none] (15) at (-4.5, -1) {};
		\node [style=Z dot] (20) at (-10, 1.25) {};
		\node [style=Z phase dot] (21) at (-10, 2.25) {$\frac{\pi}{4}$};
		\node [style=Z phase dot] (22) at (-11, 0) {$\alpha$};
		\node [style=none] (23) at (-11.5, -1) {};
		\node [style=none] (24) at (-11, -0.75) {$...$};
		\node [style=none] (25) at (-10.5, -1) {};
		\node [style=none] (53) at (-3.25, 0) {$\approx$};
		\node [style=Z phase dot] (59) at (3.75, 0) {$(1-a)\frac{\pi}{2}$};
		\node [style=Z phase dot] (60) at (6.75, 0) {$(1-a)\frac{\pi}{2}$};
		\node [style=Z phase dot] (61) at (10.25, 0) {$(1-a)\frac{\pi}{2}$};
		\node [style=none] (62) at (3.25, -1) {};
		\node [style=none] (63) at (3.75, -0.75) {$...$};
		\node [style=none] (64) at (4.25, -1) {};
		\node [style=none] (65) at (6.25, -1) {};
		\node [style=none] (66) at (6.75, -0.75) {$...$};
		\node [style=none] (67) at (7.25, -1) {};
		\node [style=none] (68) at (9.75, -1) {};
		\node [style=none] (69) at (10.25, -0.75) {$...$};
		\node [style=none] (70) at (10.75, -1) {};
		\node [style=none] (73) at (0.25, -1) {};
		\node [style=none] (74) at (1, -0.75) {$...$};
		\node [style=none] (75) at (1.75, -1) {};
		\node [style=Z phase dot] (79) at (0.25, 0) {$a\pi$};
		\node [style=Z phase dot] (80) at (1.75, 0) {$a\pi$};
		\node [style=none] (86) at (-1.75, 0) {$\mathlarger{\sum}$};
		\node [style=none] (87) at (5.5, 3) {$e^{ia(\alpha+\frac{n\pi}{4})}$};
		\node [style=none] (88) at (-1.75, -1) {$\scriptstyle{a\in\{0,1\}}$};
		\node [style=none] (89) at (-7, 2.25) {$...$};
		\node [style=none] (91) at (8.5, 0) {$...$};
		\node [style=none] (93) at (-6, 0) {$...$};
		\node [style=none] (94) at (-10.5, 3.25) {};
		\node [style=none] (95) at (-9.5, 3.25) {};
		\node [style=none] (96) at (-8.5, 3.25) {};
		\node [style=none] (97) at (-7.5, 3.25) {};
		\node [style=none] (98) at (-6.5, 3.25) {};
		\node [style=none] (99) at (-5.5, 3.25) {};
		\node [style=Z phase dot] (100) at (3, 1.25) {$a\pi$};
		\node [style=Z phase dot] (101) at (4.5, 1.25) {$a\pi$};
		\node [style=none] (106) at (3.75, 0.75) {$...$};
		\node [style=none] (109) at (-10, 3) {$...$};
		\node [style=none] (110) at (-8, 3) {$...$};
		\node [style=none] (111) at (-6, 3) {$...$};
		\node [style=none] (112) at (3, 2) {};
		\node [style=none] (113) at (4.5, 2) {};
		\node [style=Z phase dot] (114) at (6, 1.25) {$a\pi$};
		\node [style=Z phase dot] (115) at (7.5, 1.25) {$a\pi$};
		\node [style=none] (116) at (6.75, 0.75) {$...$};
		\node [style=none] (117) at (6, 2) {};
		\node [style=none] (118) at (7.5, 2) {};
		\node [style=Z phase dot] (119) at (9.5, 1.25) {$a\pi$};
		\node [style=Z phase dot] (120) at (11, 1.25) {$a\pi$};
		\node [style=none] (121) at (10.25, 0.75) {$...$};
		\node [style=none] (122) at (9.5, 2) {};
		\node [style=none] (123) at (11, 2) {};
	\end{pgfonlayer}
	\begin{pgfonlayer}{edgelayer}
		\draw [style=hadamard edge] (2) to (0);
		\draw [style=hadamard edge] (0) to (5);
		\draw [style=hadamard edge] (3) to (1);
		\draw [style=hadamard edge] (1) to (6);
		\draw (4) to (7.center);
		\draw (4) to (9.center);
		\draw (5) to (10.center);
		\draw (5) to (12.center);
		\draw (6) to (13.center);
		\draw (6) to (15.center);
		\draw [style=hadamard edge] (21) to (20);
		\draw [style=hadamard edge] (4) to (20);
		\draw (22) to (23.center);
		\draw (22) to (25.center);
		\draw [style=hadamard edge] (20) to (22);
		\draw [style=hadamard edge] (22) to (0);
		\draw [style=hadamard edge] (22) to (1);
		\draw (59) to (62.center);
		\draw (59) to (64.center);
		\draw (60) to (65.center);
		\draw (60) to (67.center);
		\draw (61) to (68.center);
		\draw (61) to (70.center);
		\draw [style=hadamard edge] (73.center) to (79);
		\draw [style=hadamard edge] (75.center) to (80);
		\draw (94.center) to (21);
		\draw (21) to (95.center);
		\draw (96.center) to (2);
		\draw (2) to (97.center);
		\draw (98.center) to (3);
		\draw (3) to (99.center);
		\draw [style=hadamard edge] (100) to (59);
		\draw [style=hadamard edge] (59) to (101);
		\draw [style=hadamard edge] (112.center) to (100);
		\draw [style=hadamard edge] (101) to (113.center);
		\draw [style=hadamard edge] (117.center) to (114);
		\draw [style=hadamard edge] (115) to (118.center);
		\draw [style=hadamard edge] (122.center) to (119);
		\draw [style=hadamard edge] (120) to (123.center);
		\draw [style=hadamard edge] (114) to (60);
		\draw [style=hadamard edge] (60) to (115);
		\draw [style=hadamard edge] (119) to (61);
		\draw [style=hadamard edge] (61) to (120);
	\end{pgfonlayer}
\end{tikzpicture}

%% file: figures/phase-gadget-heur-ex.tikz
\begin{tikzpicture}
	\begin{pgfonlayer}{nodelayer}
		\node [style=Z dot] (0) at (-7, 1.5) {};
		\node [style=Z dot] (1) at (-4, 1.5) {};
		\node [style=Z phase dot] (2) at (-7, 3) {$\frac{\pi}{4}$};
		\node [style=Z phase dot] (3) at (-4, 3) {$\frac{\pi}{4}$};
		\node [style=Z phase dot] (4) at (-8.5, 0) {$\frac{\pi}{4}$};
		\node [style=Z phase dot] (5) at (-5.5, 0) {$\frac{\pi}{4}$};
		\node [style=Z phase dot] (6) at (-2.5, 0) {$\frac{\pi}{4}$};
		\node [style=none] (7) at (-9, -1) {};
		\node [style=none] (8) at (-8.5, -0.75) {$...$};
		\node [style=none] (9) at (-8, -1) {};
		\node [style=none] (10) at (-6, -1) {};
		\node [style=none] (11) at (-5.5, -0.75) {$...$};
		\node [style=none] (12) at (-5, -1) {};
		\node [style=none] (13) at (-3, -1) {};
		\node [style=none] (14) at (-2.5, -0.75) {$...$};
		\node [style=none] (15) at (-2, -1) {};
		\node [style=Z dot] (18) at (5.25, 1.5) {};
		\node [style=Z phase dot] (20) at (5.25, 3) {$\frac{\pi}{4}$};
		\node [style=Z phase dot] (21) at (8.25, 3) {$\frac{\pi}{4}$};
		\node [style=Z phase dot] (22) at (5.25, 0) {$\frac{\pi}{4}$};
		\node [style=Z phase dot] (23) at (8.25, 0) {$\frac{\pi}{4}$};
		\node [style=none] (25) at (4.75, -1) {};
		\node [style=none] (26) at (5.25, -0.75) {$...$};
		\node [style=none] (27) at (5.75, -1) {};
		\node [style=none] (28) at (7.75, -1) {};
		\node [style=none] (29) at (8.25, -0.75) {$...$};
		\node [style=none] (30) at (8.75, -1) {};
		\node [style=none] (31) at (10.75, -1) {};
		\node [style=none] (32) at (11.5, -0.75) {$...$};
		\node [style=none] (33) at (12.25, -1) {};
		\node [style=X phase dot] (35) at (10.75, 0) {$a\pi$};
		\node [style=X phase dot] (36) at (12.25, 0) {$a\pi$};
		\node [style=Z dot] (37) at (5.25, -4.5) {};
		\node [style=Z dot] (38) at (8.25, -4.5) {};
		\node [style=Z phase dot] (39) at (5.25, -3) {$\frac{\pi}{4}$};
		\node [style=Z phase dot] (40) at (8.25, -3) {$(-1)^a \frac{\pi}{4}$};
		\node [style=Z phase dot] (41) at (5.25, -6) {$\frac{\pi}{4}$};
		\node [style=Z phase dot] (42) at (8.25, -6) {$\frac{\pi}{4}$};
		\node [style=none] (43) at (4.75, -7) {};
		\node [style=none] (44) at (5.25, -6.75) {$...$};
		\node [style=none] (45) at (5.75, -7) {};
		\node [style=none] (46) at (7.75, -7) {};
		\node [style=none] (47) at (8.25, -6.75) {$...$};
		\node [style=none] (48) at (8.75, -7) {};
		\node [style=none] (49) at (10.75, -7) {};
		\node [style=none] (50) at (11.5, -6.75) {$...$};
		\node [style=none] (51) at (12.25, -7) {};
		\node [style=X phase dot] (53) at (10.75, -6) {$a\pi$};
		\node [style=X phase dot] (54) at (12.25, -6) {$a\pi$};
		\node [style=Z phase dot] (55) at (8.25, 1.5) {$a\pi$};
		\node [style=none] (56) at (0, 0) {$\approx$};
		\node [style=none] (57) at (0, 1) {\CutRule};
		\node [style=none] (58) at (0, -6) {$\approx$};
		\node [style=none] (59) at (0, -5) {\SpiderRule};
		\node [style=Z dot] (61) at (6.75, -10.5) {};
		\node [style=Z phase dot] (63) at (6.75, -9) {$(1-a) \frac{\pi}{2}$};
		\node [style=Z phase dot] (64) at (5.25, -12) {$\frac{\pi}{4}$};
		\node [style=Z phase dot] (65) at (8.25, -12) {$\frac{\pi}{4}$};
		\node [style=none] (66) at (4.75, -13) {};
		\node [style=none] (67) at (5.25, -12.75) {$...$};
		\node [style=none] (68) at (5.75, -13) {};
		\node [style=none] (69) at (7.75, -13) {};
		\node [style=none] (70) at (8.25, -12.75) {$...$};
		\node [style=none] (71) at (8.75, -13) {};
		\node [style=none] (72) at (10.75, -13) {};
		\node [style=none] (73) at (11.5, -12.75) {$...$};
		\node [style=none] (74) at (12.25, -13) {};
		\node [style=X phase dot] (76) at (10.75, -12) {$a\pi$};
		\node [style=X phase dot] (77) at (12.25, -12) {$a\pi$};
		\node [style=none] (78) at (0, -12) {$\approx$};
		\node [style=none] (79) at (0, -11) {\GFRule};
		\node [style=none] (80) at (1.5, 0) {$\mathlarger{\sum}$};
		\node [style=none] (81) at (3, 0) {$e^{ia\frac{\pi}{4}}$};
		\node [style=none] (82) at (1.5, -1) {$\scriptstyle{a\in\{0,1\}}$};
		\node [style=none] (83) at (1.5, -6) {$\mathlarger{\sum}$};
		\node [style=none] (84) at (3, -6) {$e^{ia\frac{\pi}{2}}$};
		\node [style=none] (85) at (1.5, -7) {$\scriptstyle{a\in\{0,1\}}$};
		\node [style=none] (86) at (1.5, -12) {$\mathlarger{\sum}$};
		\node [style=none] (87) at (3, -12) {$e^{ia\frac{\pi}{2}}$};
		\node [style=none] (88) at (1.5, -13) {$\scriptstyle{a\in\{0,1\}}$};
	\end{pgfonlayer}
	\begin{pgfonlayer}{edgelayer}
		\draw [style=hadamard edge] (2) to (0);
		\draw [style=hadamard edge] (0) to (4);
		\draw [style=hadamard edge] (0) to (5);
		\draw [style=hadamard edge] (3) to (1);
		\draw [style=hadamard edge] (1) to (6);
		\draw (4) to (7.center);
		\draw (4) to (9.center);
		\draw (5) to (10.center);
		\draw (5) to (12.center);
		\draw (6) to (13.center);
		\draw (6) to (15.center);
		\draw [style=hadamard edge] (4) to (1);
		\draw [style=hadamard edge] (1) to (5);
		\draw [style=hadamard edge] (20) to (18);
		\draw [style=hadamard edge] (18) to (22);
		\draw [style=hadamard edge] (18) to (23);
		\draw (22) to (25.center);
		\draw (22) to (27.center);
		\draw (23) to (28.center);
		\draw (23) to (30.center);
		\draw (31.center) to (35);
		\draw (33.center) to (36);
		\draw [style=hadamard edge] (39) to (37);
		\draw [style=hadamard edge] (37) to (41);
		\draw [style=hadamard edge] (37) to (42);
		\draw [style=hadamard edge] (40) to (38);
		\draw (41) to (43.center);
		\draw (41) to (45.center);
		\draw (42) to (46.center);
		\draw (42) to (48.center);
		\draw [style=hadamard edge] (41) to (38);
		\draw [style=hadamard edge] (38) to (42);
		\draw (49.center) to (53);
		\draw (51.center) to (54);
		\draw [style=hadamard edge] (55) to (23);
		\draw [style=hadamard edge] (55) to (22);
		\draw [style=hadamard edge] (55) to (21);
		\draw [style=hadamard edge] (63) to (61);
		\draw (64) to (66.center);
		\draw (64) to (68.center);
		\draw (65) to (69.center);
		\draw (65) to (71.center);
		\draw [style=hadamard edge] (64) to (61);
		\draw [style=hadamard edge] (61) to (65);
		\draw (72.center) to (76);
		\draw (74.center) to (77);
	\end{pgfonlayer}
\end{tikzpicture}

%% file: figures/pg-pair.tikz
\begin{tikzpicture}
	\begin{pgfonlayer}{nodelayer}
		\node [style=Z dot] (0) at (-10, 1.5) {};
		\node [style=Z dot] (1) at (-7, 1.5) {};
		\node [style=Z phase dot] (2) at (-10, 3) {$\frac{\pi}{4}$};
		\node [style=Z phase dot] (3) at (-7, 3) {$\frac{\pi}{4}$};
		\node [style=Z phase dot] (4) at (-11.5, 0) {$\frac{\pi}{4}$};
		\node [style=Z phase dot] (5) at (-8.5, 0) {$\frac{\pi}{4}$};
		\node [style=Z phase dot] (6) at (-5.5, 0) {$\frac{\pi}{4}$};
		\node [style=none] (7) at (-12, -1) {};
		\node [style=none] (8) at (-11.5, -0.75) {$...$};
		\node [style=none] (9) at (-11, -1) {};
		\node [style=none] (10) at (-9, -1) {};
		\node [style=none] (11) at (-8.5, -0.75) {$...$};
		\node [style=none] (12) at (-8, -1) {};
		\node [style=none] (13) at (-6, -1) {};
		\node [style=none] (14) at (-5.5, -0.75) {$...$};
		\node [style=none] (15) at (-5, -1) {};
		\node [style=Z dot] (18) at (1.25, -5.5) {};
		\node [style=Z phase dot] (20) at (1.25, -4) {$\frac{\pi}{4}$};
		\node [style=Z phase dot] (21) at (4.25, -4) {$\frac{\pi}{4}$};
		\node [style=Z phase dot] (22) at (1.25, -7) {$\frac{\pi}{4}$};
		\node [style=Z phase dot] (23) at (4.25, -7) {$\frac{\pi}{4}$};
		\node [style=none] (25) at (0.75, -8) {};
		\node [style=none] (26) at (1.25, -7.75) {$...$};
		\node [style=none] (27) at (1.75, -8) {};
		\node [style=none] (28) at (3.75, -8) {};
		\node [style=none] (29) at (4.25, -7.75) {$...$};
		\node [style=none] (30) at (4.75, -8) {};
		\node [style=Z phase dot] (55) at (4.25, -5.5) {$a\pi$};
		\node [style=none] (56) at (-4, -6.25) {$\approx$};
		\node [style=none] (57) at (-4, -5.25) {\CutRule};
		\node [style=none] (80) at (-2.5, -6.25) {$\mathlarger{\sum}$};
		\node [style=none] (82) at (-2.5, -7.25) {$\scriptstyle{a\in\{0,1\}}$};
		\node [style=Z phase dot] (89) at (-2.5, 0) {$\frac{\pi}{4}$};
		\node [style=none] (90) at (-3, -1) {};
		\node [style=none] (91) at (-2.5, -0.75) {$...$};
		\node [style=none] (92) at (-2, -1) {};
		\node [style=Z dot] (93) at (4, 1.5) {};
		\node [style=Z dot] (94) at (7, 1.5) {};
		\node [style=Z phase dot] (95) at (4, 3) {$\frac{\pi}{4}$};
		\node [style=Z phase dot] (96) at (7, 3) {$\frac{\pi}{4}$};
		\node [style=Z phase dot] (97) at (2.5, 0) {$\frac{\pi}{4}$};
		\node [style=Z phase dot] (98) at (5.5, 0) {$\frac{\pi}{4}$};
		\node [style=Z phase dot] (99) at (8.5, 0) {$\frac{\pi}{4}$};
		\node [style=none] (100) at (2, -1) {};
		\node [style=none] (101) at (2.5, -0.75) {$...$};
		\node [style=none] (102) at (3, -1) {};
		\node [style=none] (103) at (5, -1) {};
		\node [style=none] (104) at (5.5, -0.75) {$...$};
		\node [style=none] (105) at (6, -1) {};
		\node [style=none] (106) at (8, -1) {};
		\node [style=none] (107) at (8.5, -0.75) {$...$};
		\node [style=none] (108) at (9, -1) {};
		\node [style=Z phase dot] (109) at (11.5, 0) {$\frac{\pi}{4}$};
		\node [style=none] (110) at (11, -1) {};
		\node [style=none] (111) at (11.5, -0.75) {$...$};
		\node [style=none] (112) at (12, -1) {};
		\node [style=Z dot] (113) at (10, 1.5) {};
		\node [style=Z dot] (114) at (8.5, 1.5) {};
		\node [style=none] (115) at (0, 0) {$=$};
		\node [style=none] (116) at (0, 1) {\SpiderRule};
		\node [style=Z phase dot] (117) at (7.25, -7) {$\frac{\pi}{4}$};
		\node [style=none] (118) at (6.75, -8) {};
		\node [style=none] (119) at (7.25, -7.75) {$...$};
		\node [style=none] (120) at (7.75, -8) {};
		\node [style=Z phase dot] (121) at (10.25, -7) {$\frac{\pi}{4}$};
		\node [style=none] (122) at (9.75, -8) {};
		\node [style=none] (123) at (10.25, -7.75) {$...$};
		\node [style=none] (124) at (10.75, -8) {};
		\node [style=Z phase dot] (125) at (8.75, -5.5) {$a\pi$};
		\node [style=Z dot] (126) at (2.75, -12.5) {};
		\node [style=Z phase dot] (127) at (2.75, -11) {$(1-a)\frac{\pi}{2}$};
		\node [style=Z phase dot] (129) at (1.25, -14) {$\frac{\pi}{4}$};
		\node [style=Z phase dot] (130) at (4.25, -14) {$\frac{\pi}{4}$};
		\node [style=none] (131) at (0.75, -15) {};
		\node [style=none] (132) at (1.25, -14.75) {$...$};
		\node [style=none] (133) at (1.75, -15) {};
		\node [style=none] (134) at (3.75, -15) {};
		\node [style=none] (135) at (4.25, -14.75) {$...$};
		\node [style=none] (136) at (4.75, -15) {};
		\node [style=none] (138) at (-4, -13.25) {$\approx$};
		\node [style=none] (139) at (-4, -12.25) { \PiRule \GFRule \SpiderRule};
		\node [style=none] (140) at (-2.5, -13.25) {$\mathlarger{\sum}$};
		\node [style=none] (141) at (-2.5, -14.25) {$\scriptstyle{a\in\{0,1\}}$};
		\node [style=Z phase dot] (146) at (8.75, -13) {$(1-a)\frac{\pi}{2}$};
		\node [style=Z phase dot] (147) at (9.75, -14) {$a\pi$};
		\node [style=none] (148) at (7.25, -14.5) {$...$};
		\node [style=Z phase dot] (149) at (11.25, -14) {$a\pi$};
		\node [style=none] (150) at (6, -15) {};
		\node [style=none] (151) at (8, -15) {};
		\node [style=none] (152) at (10, -15) {};
		\node [style=none] (153) at (12, -15) {};
		\node [style=none] (154) at (10.75, -14.5) {$...$};
		\node [style=none] (155) at (-0.75, -13.25) {$e^{ia\frac{\pi}{2}}$};
	\end{pgfonlayer}
	\begin{pgfonlayer}{edgelayer}
		\draw [style=hadamard edge] (2) to (0);
		\draw [style=hadamard edge] (0) to (4);
		\draw [style=hadamard edge] (0) to (5);
		\draw [style=hadamard edge] (3) to (1);
		\draw [style=hadamard edge] (1) to (6);
		\draw (4) to (7.center);
		\draw (4) to (9.center);
		\draw (5) to (10.center);
		\draw (5) to (12.center);
		\draw (6) to (13.center);
		\draw (6) to (15.center);
		\draw [style=hadamard edge] (4) to (1);
		\draw [style=hadamard edge] (1) to (5);
		\draw [style=hadamard edge] (20) to (18);
		\draw [style=hadamard edge] (18) to (22);
		\draw [style=hadamard edge] (18) to (23);
		\draw (22) to (25.center);
		\draw (22) to (27.center);
		\draw (23) to (28.center);
		\draw (23) to (30.center);
		\draw [style=hadamard edge] (55) to (23);
		\draw [style=hadamard edge] (55) to (22);
		\draw [style=hadamard edge] (55) to (21);
		\draw (89) to (90.center);
		\draw (89) to (92.center);
		\draw [style=hadamard edge] (1) to (89);
		\draw [style=hadamard edge] (95) to (93);
		\draw [style=hadamard edge] (93) to (97);
		\draw [style=hadamard edge] (93) to (98);
		\draw [style=hadamard edge] (96) to (94);
		\draw (97) to (100.center);
		\draw (97) to (102.center);
		\draw (98) to (103.center);
		\draw (98) to (105.center);
		\draw (99) to (106.center);
		\draw (99) to (108.center);
		\draw [style=hadamard edge] (97) to (94);
		\draw [style=hadamard edge] (94) to (98);
		\draw (109) to (110.center);
		\draw (109) to (112.center);
		\draw [style=hadamard edge] (113) to (99);
		\draw [style=hadamard edge] (113) to (109);
		\draw [style=hadamard edge] (94) to (114);
		\draw [style=hadamard edge] (114) to (113);
		\draw (117) to (118.center);
		\draw (117) to (120.center);
		\draw (121) to (122.center);
		\draw (121) to (124.center);
		\draw [style=hadamard edge] (125) to (117);
		\draw [style=hadamard edge] (125) to (121);
		\draw [style=hadamard edge] (127) to (126);
		\draw [style=hadamard edge] (126) to (129);
		\draw [style=hadamard edge] (126) to (130);
		\draw (129) to (131.center);
		\draw (129) to (133.center);
		\draw (130) to (134.center);
		\draw (130) to (136.center);
		\draw [bend right] (146) to (150.center);
		\draw [bend right=15] (146) to (151.center);
		\draw [style=hadamard edge, bend left=15, looseness=0.75] (146) to (147);
		\draw [style=hadamard edge, bend left=15, looseness=0.50] (147) to (152.center);
		\draw [style=hadamard edge, bend left=15, looseness=0.75] (146) to (149);
		\draw [style=hadamard edge, bend left=15, looseness=0.75] (149) to (153.center);
	\end{pgfonlayer}
\end{tikzpicture}

%% file: ws-ijmpc.bbl
\begin{thebibliography}{10}

\bibitem{Bravyi-2016bss}
S.~Bravyi and D.~Gosset, {\em Physical Review Letters} {\bf 116} (June 2016).

\bibitem{Nielsen-Chuang-2010}
M.~A. Nielsen and I.~L. Chuang, {\em Quantum Computation and Quantum Information: 10th Anniversary Edition} (Cambridge University Press, 2010).

\bibitem{Kissinger-2022}
A.~Kissinger and J.~van~de Wetering, {\em Quantum Science and Technology} {\bf 7}, p. 044001 (July 2022).

\bibitem{Kissinger2022cat}
A.~Kissinger, J.~van~de Wetering and R.~Vilmart, Classical simulation of quantum circuits with partial and graphical stabiliser decompositions (Schloss Dagstuhl – Leibniz-Zentrum für Informatik, 2022).

\bibitem{Codsi2022}
J.~Codsi, Cutting-edge graphical stabiliser decompositions for classical simulation of quantum circuits, Master's thesis, University of Oxford  (2022).

\bibitem{Laakkonen2022}
T.~Laakkonen, Graphical stabilizer decompositions for counting problems, Master's thesis, University of Oxford  (2022).

\bibitem{sutcliffe2024-paramzx}
M.~Sutcliffe and A.~Kissinger, Fast classical simulation of quantum circuits via parametric rewriting in the zx-calculus  (2024).

\bibitem{sutcliffe2024part}
M.~Sutcliffe, Smarter k-partitioning of zx-diagrams for improved quantum circuit simulation  (2024).

\bibitem{sutcliffe2024proccut}
M.~Sutcliffe and A.~Kissinger, Procedurally optimised zx-diagram cutting for efficient t-decomposition in classical simulation, in {\em Quantum Physics and Logic 2024\/},  eds. A.~Díaz-Caro and V.~Zamdzhiev, Electronic Proceedings in Theoretical Computer Science, Vol.~406 (Open Publishing Association, 2024).

\bibitem{Coecke-2011}
B.~Coecke and R.~Duncan, {\em New Journal of Physics} {\bf 13}, p. 043016 (April 2011).

\bibitem{coecke2017picturing}
B.~Coecke and A.~Kissinger, {\em Picturing Quantum Processes} (Cambridge University Press, 2017).

\bibitem{weteringWorking}
J.~van~de Wetering, Zx-calculus for the working quantum computer scientist  (2020).

\bibitem{KissingerWetering2024Book}
A.~Kissinger and J.~van~de Wetering, {\em {Picturing Quantum Software: An Introduction to the ZX-Calculus and Quantum Compilation}} (Preprint, 2024).

\bibitem{KissingerTcount}
A.~Kissinger and J.~van~de Wetering, {\em Physical Review A} {\bf 102} (August 2020).

\bibitem{Duncan2020graphtheoretic}
R.~Duncan, A.~Kissinger, S.~Perdrix and J.~van~de Wetering, {\em {Quantum}} {\bf 4}, p. 279 (June 2020).

\bibitem{Kissinger2020}
A.~Kissinger and J.~van~de Wetering, {\em Phys. Rev. A} {\bf 102}, p. 022406 (August 2020).

\bibitem{cowtan2019phase}
A.~Cowtan, S.~Dilkes, R.~Duncan, W.~Simmons and S.~Sivarajah, {\em arXiv preprint arXiv:1906.01734}   (2019).

\bibitem{khesin2023graphical}
A.~B. Khesin, J.~Z. Lu and P.~W. Shor, {\em arXiv preprint arXiv:2301.02356}   (2023).

\bibitem{koch2023}
M.~Koch, R.~Yeung and Q.~Wang, Speedy contraction of zx diagrams with triangles via stabiliser decompositions  (2023).

\bibitem{gottesman1998}
D.~Gottesman, The heisenberg representation of quantum computers  (1998).

\bibitem{movassagh2023hardness}
R.~Movassagh, {\em Nature Physics} {\bf 19}, 1719  (2023).

\bibitem{koziell-pipe2024}
A.~Koziell-Pipe, R.~Yeung and M.~Sutcliffe, Towards faster quantum circuit simulation using graph decompositions, {GNN}s and reinforcement learning, in {\em The 4th Workshop on Mathematical Reasoning and AI at NeurIPS'24\/},  (NeurIPS, 2024).

\bibitem{ahmad2024}
W.~A. Ahmad, Efficient heuristics for classical simulation of quantum circuits using zx-calculus, Master's thesis, University of Oxford  (2024).

\bibitem{sutcliffePhd}
M.~Sutcliffe, Novel methods for classical simulation of quantum circuits via zx-calculus, PhD thesis, University of Oxford, (Oxford, United Kingdom, 2025).

\bibitem{sutcliffeReview}
M.~Sutcliffe, Classically simulating quantum circuits with zx-calculus  (2025).

\bibitem{quizx}
A.~Kissinger, J.~van~de Wetering and R.~Vilmart, Quizx: a quick rust port of pyzx \url{https://github.com/zxcalc/quizx},  (2021).

\bibitem{Bremner-2017}
M.~J. Bremner, A.~Montanaro and D.~J. Shepherd, {\em Quantum} {\bf 1}, p.~8 (April 2017).

\bibitem{codsi2023}
J.~Codsi and J.~van~de Wetering, Classically simulating quantum supremacy iqp circuits through a random graph approach  (2023).

\bibitem{Bremner-2016}
M.~J. Bremner, A.~Montanaro and D.~J. Shepherd, {\em Physical Review Letters} {\bf 117} (August 2016).

\bibitem{Bravyi-2019}
S.~Bravyi, D.~Browne, P.~Calpin, E.~Campbell, D.~Gosset and M.~Howard, {\em Quantum} {\bf 3}, p. 181 (September 2019).

\bibitem{Peres-2023}
F.~C.~R. Peres and E.~F. Galvão, {\em Quantum} {\bf 7}, p. 1126 (October 2023).

\end{thebibliography}
